\documentclass[aps,prl,reprint,superscriptaddress]{revtex4-1}
\usepackage[utf8]{inputenc}
\usepackage{amsfonts}
\usepackage{graphicx}
\graphicspath{{Pictures/}}
\usepackage{bm} 
\usepackage{amssymb} 
\usepackage{amsmath} 
\usepackage{braket} 
\usepackage{natbib} 
\usepackage[colorlinks = true, linkcolor = magenta, urlcolor  = purple, citecolor = magenta, anchorcolor = blue]{hyperref}
\usepackage[T1]{fontenc}
\usepackage{xcolor}
\usepackage{marginnote}
\usepackage[normalem]{ulem}

\newcommand{\dr}{\ensuremath{\mathbf{r}}}
\newcommand{\SB}[1] {{\color{black}#1}}

\begin{document}

\title{Inverse Faraday effect in Mott insulators}

\author{Saikat Banerjee}
\email{saikatb@lanl.gov}
\affiliation{Theoretical Division, T-4, Los Alamos National Laboratory, Los Alamos, New Mexico 87545, USA}
\author{Umesh Kumar}
\affiliation{Theoretical Division, T-4, Los Alamos National Laboratory, Los Alamos, New Mexico 87545, USA}
\author{Shi-Zeng Lin} 
\email{szl@lanl.gov}
\affiliation{Theoretical Division, T-4 and CNLS, Los Alamos National Laboratory, Los Alamos, New Mexico 87545, USA}

\date{\today}
\begin{abstract}
The inverse Faraday effect (IFE), where a static magnetization is induced by circularly polarized light, offers a promising route to ultrafast control of spin states. Here we study the IFE in Mott insulators using the Floquet theory. We find two distinct IFE behavior governed by the inversion symmetry. In the Mott insulators with inversion symmetry, we find that the effective magnetic field induced by the IFE couples ferromagnetically to the neighboring spins. While for the Mott insulators without inversion symmetry, the effective magnetic field due to IFE couples antiferromagnetically to the neighboring spins. We apply the theory to the spin-orbit coupled single- and multi-orbital Hubbard model that is relevant for the Kitaev quantum spin liquid material and demonstrate that the magnetic interactions can be tuned by light.  
\end{abstract}

\maketitle

\textit{Introduction.} \textbf{\---} The optical control and manipulation of the magnetic exchange interaction in quantum materials have always been an important centerpiece in condensed matter physics~\cite{RevModPhys.82.2731,RevModPhys.91.025005}. The origin of such magneto-optical studies dates back to Faraday who discovered that the plane of light polarization rotates due to the intrinsic magnetization in a material~\cite{Faraday_1845}. Almost a century later, it was predicted~\cite{Pitaevskii_1961} and subsequently observed~\cite{PhysRevLett.15.190} that a circularly polarized light can also generate static magnetic moments. This opposite phenomenon is known as the inverse Faraday effect (IFE), which offers a natural pathway to the ultrafast manipulation of magnetic order in quantum materials~\cite{Kimel2005,Lottermoser2004}. Over the past few decades, IFE has remained an active area of research and has been observed in a large class of materials ranging from insulating magnets~\cite{Kimel2005} to non-magnetic metals~\cite{PhysRevLett.120.207207,Gu2010}. 

However, despite significant experimental progress, the microscopic origin of the IFE has remained relatively unclear from a theoretical point of view. Most of the previous attempts in this direction relied on semi-classical analysis~\cite{Pitaevskii_1961,Hertel2006,PhysRevB.79.212412,PhysRevB.74.134430}. Earlier theoretical work by Battiato \textit{et al}.~\cite{PhysRevB.89.014413} provided a detailed quantum mechanical analysis of metallic IFE, relying on the electronic orbital degrees of freedom. Recently, IFE has been predicted in spin-orbit coupled (SOC) Rashba metals~\cite{Tanaka_2020}, semimetals \cite{gao_topological_2020,PhysRevB.101.174429,PhysRevLett.126.247202} and also for superconductors ~\cite{PhysRevLett.126.137002,PhysRevLett.127.087001}. While the realization of IFE using ultrafast control of spin dynamics in rare-earth orthoferrites [ReFeO$_3$, Re $=$ Dy, Ho, Er; (antiferromagnetic insulator)] has been reported in previous works~\cite{Kimel2005,Nemec2018,Paris2021,Kimel2015_1,Kimel2009,PhysRevB.84.104421,PhysRevB.103.014423}, a detailed microscopic analysis of the latter in the Mott insulating regime is still lacking.

\begin{figure}[t]
\centering
\includegraphics[width=1\linewidth]{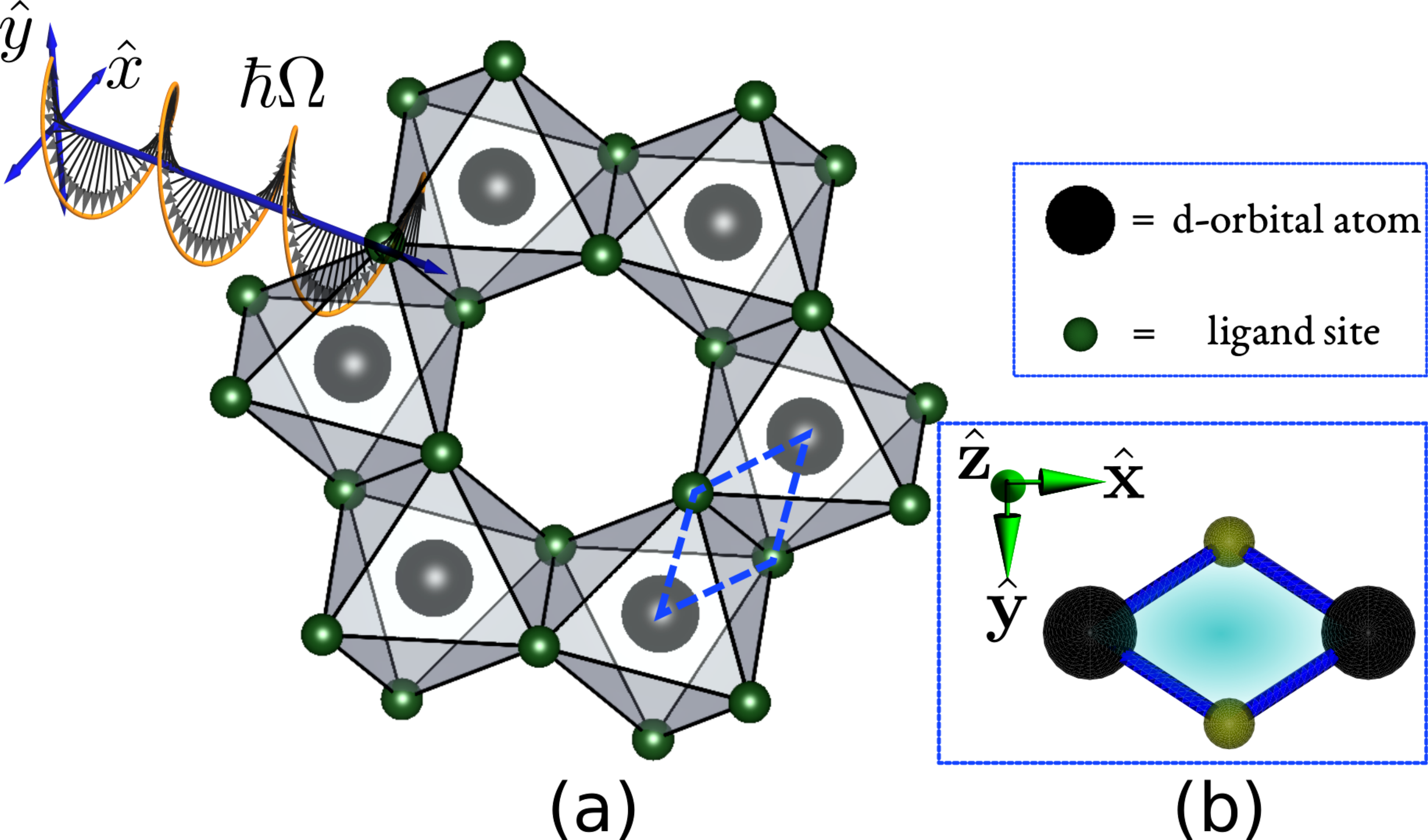} 
\caption{(a) A schematic of the lattice with edge-sharing octahedral geometry where both the direct ($t_{dd}$) and the indirect ($t_{pd}$) hopping amplitudes get modified under the influence of applied circularly polarized light. (b) The four-site cluster [highlighted in panel (a)] including two $d$ orbitals and two ligand atoms, respectively, generates an effective static magnetic field in the presence of circularly polarized light with energy $\hbar \Omega$ (IFE).}\label{fig:Fig1}
\end{figure}

In this work, we consider a periodically driven Mott insulator in the presence of circularly polarized light and analyze the emergent magnetic field in the Floquet regime. We explore both single- and multiorbital models and find that the IFE leads to both antiferromagnetic and ferromagnetic magnetization depending on the inversion symmetry. We employ the time-dependent Schrieffer-Wolff (SW) unitary transformations to derive low-energy spin Hamiltonians. In this case, the transition matrix elements between high-energy (charge excitations) and low-energy states (spin excitations) are removed perturbatively~\cite{PhysRev.149.491,PhysRev.157.295,PhysRevLett.116.125301, PhysRevB.103.064508}. We consider $d$-electron systems with both direct and indirect hopping. The indirect hopping is typically assumed to be mediated through a ligand atom [see Fig.~\ref{fig:Fig1}]. We show that such ligand-mediated hopping in the presence of SOC gives rise to the IFE. In materials with inversion symmetry, such IFE favors a ferromagnetic state; in contrast, the system without inversion symmetry favors antiferromagnetism.

\textit{Symmetry considerations.}\textbf{\---}  Before moving on to the microscopic model calculation, here we investigate the IFE based on symmetry considerations. In Mott insulators, the charge degrees of freedom are gapped, and the system can be described in terms of spin degrees of freedom. The direct Zeeman coupling of the electromagnetic fields to spins is much weaker than the orbital coupling and therefore is neglected here; then, the SOC is an essential ingredient for the IFE. Furthermore, the time-reversal symmetry (TRS) must be broken. We consider a minimal hopping path shown in Fig. \ref{fig:Fig1} (b) for electrons to experience the TRS breaking laser field, where only the in-plane electric-field components couple to the electron hopping. The minimal coupling between the laser electric field and the system's \emph{static} magnetization has the form $\mathcal{L}=\epsilon_{\alpha\beta\gamma} E_\alpha(\Omega) E_\beta(-\Omega) M_\gamma$, where summation over repeated indices is implied, and $\Omega$ is the frequency of the applied laser. Here $\epsilon_{\alpha\beta\gamma}$ is a tensor, and the \emph{static} magnetization is a function of two spin moments in Fig. \ref{fig:Fig1} (b), i.e., $M_\gamma(\mathbf{S_1}, \mathbf{S_2})$, whose form is dictated by the symmetries. The whole system of laser and the Mott insulator has TRS, which enforces $\epsilon_{\alpha\beta\gamma}=-\epsilon_{\beta\alpha\gamma}$. 

In inversion symmetric systems, the atomic SOC is responsible for the IFE. We consider that the system is also symmetric with respect to the mirror plane of the ions (we call it the $xy$-plane with the $z$ axis perpendicular to it). This restricts $\epsilon_{\alpha\beta\gamma}\neq 0$ only when $\gamma=z$. The inversion symmetry requires that $\mathbf{M}=\hat{\mathbf{z}}\cdot (\mathbf{S_1}+\mathbf{S_2})$ with $\hat{z}$ a unit vector normal to the hopping plane. In this case, the IFE can be written as  $\mathcal{L}\propto[\mathbf{E}(\Omega)\times\mathbf{E}^*(\Omega)]\cdot(\mathbf{S_1}+\mathbf{S_2})$ [here $\mathbf{E}^*(\Omega)=\mathbf{E}(-\Omega)$], which is the same as IFE for the isotropic medium~\cite{Pitaevskii_1961}. 

The SOC can also arise due to inversion symmetry breaking, which can be described by vector $\bm{\alpha}$. The direction of $\bm{\alpha}$ is constrained by other symmetries such as rotation and mirror~\cite{PhysRev.120.91}. We consider symmetry transformations, such as inversion and mirror operation, that include the transformation of $\bm{\alpha}$, which leaves the $\mathcal{L}$ invariant. The simplest form that is invariant under these transformations is the scalar $M_z=\bm{\alpha}\cdot(\mathbf{S_1}-\mathbf{S_2})$. Here $M_z$ must be proportional to $\mathbf{S_1}-\mathbf{S_2}$, because $\bm{\alpha}$ is odd under the inversion transformation $1\leftrightarrow2$. The IFE favors the antiferromagnetic arrangement of $\mathbf{S}_1$ and $\mathbf{S}_2$, in contrast with a ferromagnetic arrangement in the inversion symmetric case. This is rather surprising given that the wavelength of light is usually much longer than the atomic lattice parameter. The symmetry analysis is supported by the calculations of the microscopic model below.

\textit{Model.} \textbf{\---} We start with a strongly correlated electronic model for transition metal (TM) compounds forming an edge-sharing octahedral geometry [as shown by black circles in Fig.~\ref{fig:Fig1}(a)]. In this class of materials, the $d$ orbital forms an octahedral geometry with the $p$-block (ligand) elements [chalcogenic or halogenic atoms; see green circles in Fig.~\ref{fig:Fig1}(b)]. Depending on the electronic configuration of the $d$-block elements, such compounds can be modeled by either the single- or the multiorbital Hubbard model~\cite{RevModPhys.70.1039}. A circularly polarized light [see Fig.~\ref{fig:Fig1}(a)] is applied which modifies the hopping between different orbitals. For a typical single-orbital model, the Hamiltonian can be written as $\mathcal{H}(t) = \mathcal{H}_0 + \mathcal{H}_1(t)$, where
\begin{subequations}
\begin{align}
\label{eq.1.1}
\mathcal{H}_0 & = U \sum_{i} n^{d}_{i\uparrow}n^{d}_{i\downarrow} + \Delta \sum_{i,\sigma} p^{\dagger}_{i\sigma}p_{i\sigma}, \\
\label{eq.1.2}
\mathcal{H}_1(t) & = \sum_{\langle ij \rangle} \big[ t^{ij}_{pd}(t)  d^{\dagger}_{i\sigma}p_{j\sigma} + t^{ij}_{\sigma \sigma'}(t) d^{\dagger}_{i\sigma}d_{j\sigma'} \big] + \mathrm{h.c.} , 
\end{align}
\end{subequations}
where $U$ denotes the onsite Coulomb repulsion of the $d$ orbital and $\Delta$ parameterizes the ligand charge transfer energy. Note that we consider only one $d$ orbital along with the ligand $p$ orbital. Here we assume the sum over repeated spin indices $\sigma$, and $t^{ij}_{pd}(t)$ and $t^{ij}_{\sigma \sigma'}(t)$ are the time-dependent hopping amplitudes between $p$ and $d$ and two $d$ orbitals, respectively. In the presence of circularly polarized light with electric-field component $\mathbf{E}(t) = E_0(-\hat{\mathbf{x}} \cos\Omega t  + \hat{\mathbf{y}} \sin\Omega t)$, the hopping depends on Peierls phase as (we work in the unit of $e,\hbar,c = 1$)
\begin{subequations}
\begin{align}
\label{eq.2.1}
t^{ij}_{pd}(t) & = t_{pd}e^{i\theta_{ij}(t)}, \quad \theta_{ij}(t) = -\dr_{pd}\cdot \mathbf{A}(t), \\
\label{eq.2.2}
t^{ij}_{\sigma \sigma'}(t) &= \big[ t_{dd} \mathbb{I}_2 + i\bm{\alpha}_{ij}\cdot\bm{\tau} \big]_{\sigma \sigma'}e^{i\phi_{ij}(t)}, \\
\label{eq.2.3}
\phi_{ij}(t) &= -\dr_{dd}\cdot \mathbf{A}(t),
\end{align}
\end{subequations}
where the vector potential $\mathbf{A}(t) = \frac{E_0}{\Omega} (\hat{\mathbf{x}} \sin\Omega t  + \hat{\mathbf{y}} \cos\Omega t)$, $\bm{\tau}$ denotes the vector of Pauli matrices, $\bm{\alpha}_{ij} = (\alpha_{ij}^1,\alpha_{ij}^2,\alpha_{ij}^3)$ is a real vector corresponding to the strength of the SOC in the $d$-$d$ bond, and $\dr_{pd}$ and $\dr_{dd}$ are the nearest-neighbor vectors between $p$- and $d$- and two $d$- orbitals, respectively. The specific form of the SOC in Eq.~(\ref{eq.2.2}) dictates that the Hamiltonian in Eq.~(\ref{eq.1.2}) is not invariant under inversion, \textit{i.e.}, $\mathcal{I} \mathcal{H}_1(t) \mathcal{I}^{-1} \neq \mathcal{H}_1(t)$. Here $\mathcal{I}$ is the inversion operator which swaps the indices $i$ and $j$. We consider the insulating regime at half filling with $U,\Delta \gg t_{pd}, t_{dd},|\bm{\alpha}|$. Our analysis does not require the energy hierarchy between $U$ and $\Delta$, and therefore is valid both for the Mott and for the charge-transfer type insulator. We broadly term the insulator as the Mott insulator in the following discussions. The presence of $\bm{\alpha}_{ij}$ violates the inversion symmetry but preserves the TRS when the laser is off $\mathbf{A}=0$.

Starting from the Hamiltonian in Eqs.~(\ref{eq.1.1}) and (\ref{eq.1.2}), we go to the rotated frame as $\mathcal{H}_{\mathrm{rot}}(t) = e^{i\mathcal{S}(t)}[\mathcal{H}(t)-i\partial_t]e^{-i\mathcal{S}(t)}$, where $\mathcal{S}(t)$ is a Hermitian operator. Writing $\mathcal{S}(t) = \mathcal{S}^{(1)}(t) + \mathcal{S}^{(2)}(t) + \ldots$ and expanding $\mathcal{H}_{\mathrm{rot}}(t)$ in Taylor series, we obtain order-by-order low-energy effective spin-exchange Hamiltonians. For the subsequent analysis, we consider a simplified four-site cluster model [see Fig.~\ref{fig:Fig1}(b)] containing two $d$-orbitals and two ligand atoms. In the large frequency approximation ($\Omega \gg t_{pd}, t_{dd}, \alpha_{ij}$), we obtain an effective low-energy spin Hamiltonian up to third order in perturbation theory as (see Supplemental Material (SM) \cite{supp})
\begin{align}\label{eq.3}
\mathcal{H}_{\mathrm{eff}} =  \sum_{\langle ij \rangle}  \big[ \mathbf{S}_{i\mu}\Gamma_{\mu , \nu} \mathbf{S}_{j\nu} + \mathbf{h}^{\mathrm{eff}}_{ij} \cdot \left( \mathbf{S}_i -\mathbf{S}_j \right) \big].
\end{align}
The results for the exchange couplings $\Gamma_{\mu,\nu}$ ($\mu,\nu = x,y,z$) are provided in the SM \cite{supp}. In the absence of the SOC and the ligand atoms, we recover the well-known Floquet Hamiltonian $\mathcal{H}_{\mathrm{eff}} = \sum_{\langle ij \rangle} J_{ij} \mathbf{S}_i \cdot \mathbf{S}_j$, where $J_{ij} = 4\sum_{n}\mathcal{J}^2_n(A_0) t^2_{dd}/(U-n\Omega)$, $\mathcal{J}_n(x)$ is the Bessel function of the first kind and $A_0 = r_{dd}E_0/\Omega$~\cite{Mentink2015,Eckstein, PhysRevB.103.064508}. 

The magnetic-field term $\mathbf{h}_{ij}^{\mathrm{eff}}$ is evaluated as
\begin{equation}\label{eq.4}
\mathbf{h}^{\mathrm{eff}}_{ij} = \overline{\sum}\frac{ 8 \mathcal{J}_{n}(\mathrm{A_0}) \mathcal{J}_{m}(\mathrm{A}) \mathcal{J}_{l}(\mathrm{A}) t^2_{pd} \bm{\alpha}_{ij}}{3\big[\Delta + l \Omega\big] \big[U - n\Omega\big]} \sin\psi^{ml}_0, 
\end{equation}
where $\overline{\sum}$ signifies summation over the indices $n,m,l$ with the constraint $n+m+l=0$, $A_0 = r_{dd}E_0/\Omega$, $A = r_{pd}E_0/\Omega$, and $\psi_0^{ml} = (m-l)\psi_0$. Here, $\psi_0$ is the angle between $p$-$d$ and $d$-$d$ orbital bonds [see Fig.~\ref{fig:Fig1}(b)]. Note that the effective magnetic-field $\mathbf{h}^{\mathrm{eff}}_{ij}$ proportional to the SOC $\bm{\alpha}_{ij}$ is a consequence of the \emph{broken time-reversal symmetry} due to the applied circularly polarized light. Since the effective magnetic field couples to ($\mathbf{S}_i-\mathbf{S}_j$), it favors an antiferromagnetic static magnetization, which is consistent with the symmetry analysis.

\begin{figure}[t]
\centering
\includegraphics[width=1\linewidth]{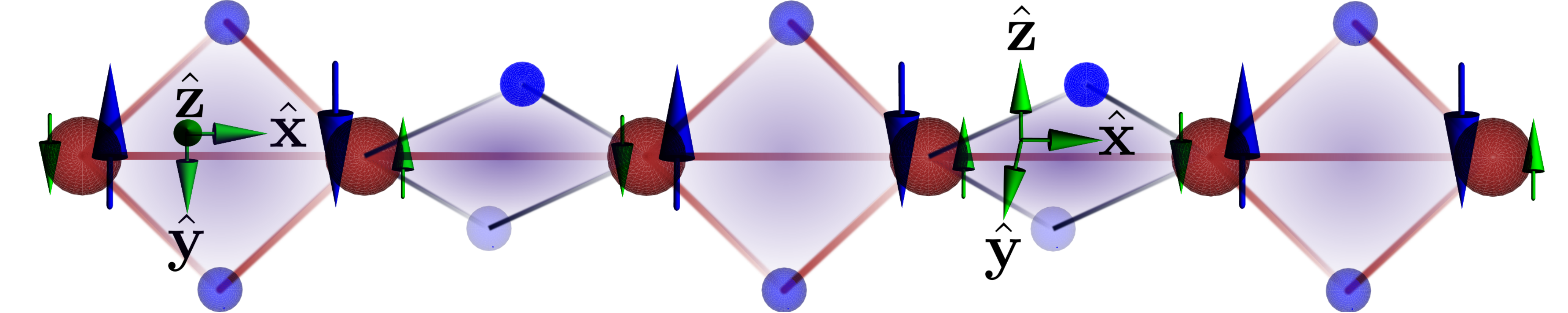} 
\caption{A schematic of the proposed distorted lattice structure where two consecutive (neighboring) four-site clusters of atoms are rotated relatively along the $x$ axis. The circularly polarized light is shined normal to the unrotated $x$-$y$ plane. Big red atoms signify the $d$ orbital and the smaller blue atoms correspond to the ligand site. The spin arrangement (big arrows) on a perfectly aligned (in the $xy$ plane) cluster is oriented along the $z$-axis, whereas the neighboring cluster tilted along the $z$ axis leads to a tilted spin arrangement which has a smaller $z$ component (smaller arrow). The IFE favors an antiferromagnetic order along the chain.}\label{fig:Fig2}
\end{figure}

\begin{figure}[t]
\centering
\includegraphics[width=1\linewidth]{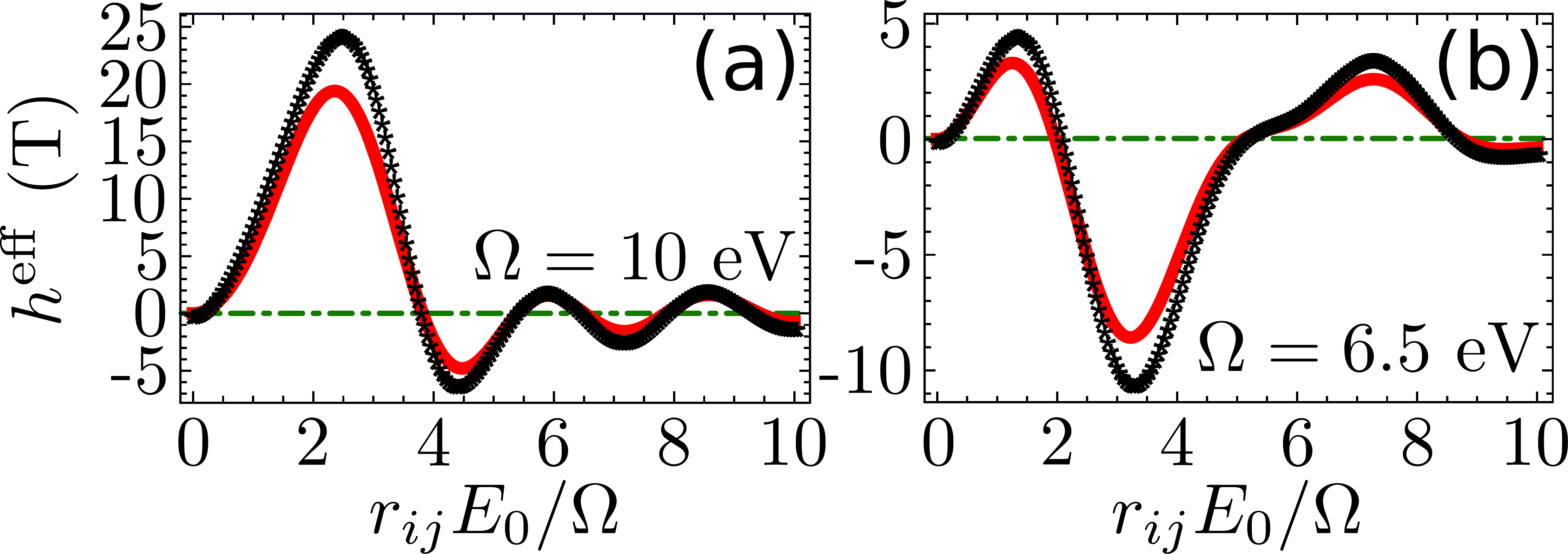} 
\caption{Dependence [\textit{solid line}: perturbation calculations, \textit{symbols}: exact diagonalization results (see SM \cite{supp})] of the IFE magnetic field due to the circularly polarized light for (a) the single-orbital case [Eq.~(\ref{eq.4})] and (b) the mutliorbital case [Eq.~(\ref{eq.9})]. For the single-orbital case, we use generic material parameters as $U = 8$ eV, $\Delta = 16$ eV, $t_{dd} = 1.0$ eV, $t_{pd} = 1.5$ eV, and $\alpha^3_{ij} = 0.05$ eV, whereas we adopt the material parameters for $\alpha$-RuCl$_3$~\cite{PhysRevB.93.155143,Sinn2016} in panel (b). Here we assume $g$-factor $\mathsf{g} = 2$ and plot the amplitude of $\mathbf{h}^{\mathrm{eff}}_{ij}$.}\label{fig:Fig3}
\end{figure}

For a weak laser drive and low frequency, the static magnetic field due to IFE is proportional to square of the electric field and inversely proportional to the frequency. Its [see Eq.~(\ref{eq.4})] asymptotic form is given by
\begin{equation}\label{eq.5}
\mathbf{h}^{\mathrm{eff}}_{ij} \approx \frac{4t_{pd}^2 \bm{\alpha}_{ij}}{3\Omega }\frac{E_0^2 r_{pd} \sin\psi_0}{U\Delta^2} \left( 2r_{pd} \cos \psi_0 + r_{dd}\right),
\end{equation}
which matches qualitatively with our phenomenological ansatz. However, as $\mathbf{h}^{\mathrm{eff}}_{ij}$ couples antiferromagnetically to the localized spins on the $d$-orbital sites, the net magnetization would vanish if all the consecutive four-site clusters are aligned parallel to the $xy$ plane, whereas if the neighboring clusters are tilted along the $z$ axis, the emergent Zeeman magnetic field would point in two different directions as illustrated in Fig.~\ref{fig:Fig2}. In this case, the net magnetization on a particular site ($d$ orbital) would not be zero and this antiferromagnetic order induced by the IFE can be realized in broken inversion symmetric systems. The variation of $h^{\mathrm{eff}}$, at the laser frequency $\Omega = 10$ eV, is illustrated in Fig.~\ref{fig:Fig3}(a) for a set of generic parameters.

\textit{Multi-orbital model.} \textbf{\---} In this case, we consider an inversion symmetric system and necessarily adopt a multiorbital description with atomic SOC. For subsequent analysis, we focus on the Kitaev systems such as $\alpha$-RuCl$_3$, $\beta$-Li$_2$IrO$_3$ where \textit{five} electrons reside in the $t_{2g}$ manifold of the TM $d$ orbital [see Fig.~\ref{fig:Fig1}(a)], which further splits into $j_{\mathrm{eff}}=1/2$ and $j_{\mathrm{eff}} = 3/2$ states due to strong SOC~\cite{PhysRevB.91.144420,PhysRevLett.102.017205,PhysRevLett.110.097204,PhysRevLett.112.077204,PhysRevB.93.214431,Winter_2017,PhysRevB.95.024426}. For $d^5$-electronic configuration, the $j_{\mathrm{eff}} = 3/2$ manifold is completely filled and a lone electron henceforth resides on the $j_{\mathrm{eff}} = 1/2$ manifold. The electronic model to capture the effects of SOC and the charge transfer to the ligand $p$ orbitals is written in terms of the Kanamori Hamiltonian~\cite{Kanamori1957_1,Kanamori1957_2,Georges2013} as 
\begin{align}\label{eq.6}
\mathcal{H}_0  & =  U\sum_{i\alpha} n^d_{i\alpha,\uparrow}n^d_{i\alpha,\downarrow} + \sum_{i\sigma \sigma' \alpha \neq \beta} \left( U' -\delta_{\sigma \sigma'} J_{\mathrm{H}} \right) n^d_{i\alpha \sigma'}n^d_{i\beta \sigma} \nonumber \\
& + J_{\mathrm{H}} \sum_{i\alpha \neq \beta} \left( d^{\dagger}_{i\alpha \uparrow} d^{\dagger}_{i\alpha \downarrow} d_{i\beta \downarrow} d_{i\beta \uparrow} + d^{\dagger}_{i\alpha \uparrow} d^{\dagger}_{i\beta \downarrow} d_{i\alpha \downarrow} d_{i\beta \uparrow} \right) \nonumber \\
& + \frac{\lambda}{2} \sum_{i} d^{\dagger}_i \left( \mathbf{L} \cdot \mathbf{S} \right) d_{i} + \Delta \sum_{i'\sigma} n^{p}_{i'\sigma},
\end{align}
where $U,U'$ denote the intra- and interorbital Coulomb repulsions and $J_{\mathrm{H}}$ stands for the Hund's coupling between the three $t_{2g}$ orbitals: $d_{xy}, d_{yz}$, and $d_{zx}$. Here $\Delta$, as before, denotes the ligand charge-transfer energy and $\lambda$ is the strength of the SOC.

Assuming SOC strength $\lambda$ is much smaller compared with the other parameters as $U,\Delta, \Omega \gg \lambda$, the Kanamori Hamiltonian can be rewritten in terms of the irreducible representation of the doubly occupied states in the $d$-orbital \cite{PhysRevB.65.064442,PhysRevB.94.174416,supp} as 
\begin{equation}\label{eq.7}
\mathcal{H}_0 = \sum_{i} \sum_{\Gamma} \sum_{\mathsf{g}_{\Gamma}} U_{\Gamma} \ket{i;\Gamma, \mathsf{g}_{\Gamma}} \bra{i;\Gamma, \mathsf{g}_{\Gamma}} + \Delta \sum_{i'\sigma} n^{p}_{i'\sigma},
\end{equation}
where $\Gamma$ corresponds to the particular irreducible representation and $\mathsf{g}_{\Gamma}$ characterizes the degeneracy of that state. The total energy of the four different nondegenerate states is given~\cite{PhysRevB.94.174416} as: $U_{A_1}  = U + 2 J_{\mathrm{H}}$, $U_{E}  =  U - J_{\mathrm{H}}$, $U_{T_1}  = U - 3 J_{\mathrm{H}}$, and $U_{T_2} = U - J_{\mathrm{H}}$.

Next, we evaluate the hopping Hamiltonian based on the inherent symmetries of the octahedral geometry. The Hamiltonian in the presence of circularly polarized light is written as
\begin{align}\label{eq.8}
\mathcal{H}_1(t) &= \sum_{ij,\sigma} e^{i\phi_{ij}(t)} \left[ d^{\dagger}_{i xz \sigma}  d^{\dagger}_{i {yz} \sigma}  d^{\dagger}_{i {xy} \sigma} \right] \begin{bmatrix}
t_1 & t_2 & t_4 \\
t_2 & t_1 & t_4 \\
t_4 & t_4 & t_3
\end{bmatrix}
\begin{bmatrix}
d_{j {xz} \sigma} \\
d_{j {yz} \sigma} \\
d_{j {xy} \sigma}
\end{bmatrix}  \nonumber \\
& + t_{pd} \sum_{ij \sigma} \big[ e^{i\theta_{i'j}(t)}p^{\dagger}_{i'\sigma}d_{jyz \sigma} +  e^{i\theta_{j'i}(t)} p^{\dagger}_{j'\sigma}d_{iyz \sigma}  \nonumber \\
& + e^{i\theta_{ji'}(t)} d^{\dagger}_{jxz \sigma} p_{i'\sigma} + e^{i\theta_{jj'}(t)} d^{\dagger}_{jxz \sigma} p_{j'\sigma} \big] + \mathrm{h.c.},
\end{align}
where $p^{\dagger}_{i'\sigma}$ is the creation operator at the ligand sites, surrounding the TM orbitals, and $\phi_{ij}(t)$ and $\theta_{i'j}(t)$ denote the bond-angle-dependent Peierls phases. For the multiorbital analysis, we adopt all the parameters entering Eq.~(\ref{eq.6}) and Eq.~(\ref{eq.7}) from the recent \emph{ab initio}~\cite{PhysRevB.93.155143} and photoemission reports~\cite{Sinn2016} for $\alpha$-RuCl$_3$ as: $U = 3.0$ eV, $J_{\mathrm{H}} = 0.45$ eV, $\Delta = 5$ eV, $t_1 = 0.036$ eV, $t_2 = 0.191$ eV, $t_3 = -0.062$ eV, $t_4 = -0.024$ eV, and $t_{pd} = - 0.9$ eV. 

We employ a similar time-dependent SW transformation and evaluate the low-energy effective spin model up to third order in perturbation. In the high-frequency approximation, the effective Hamiltonian is obtained as $\mathcal{H}_{\mathrm{eff}} = \sum_{\langle ij \rangle} \mathbf{S}_{i\mu} \mathcal{M}_{i\mu,j\nu} \mathbf{S}_{j\nu} + \sum_{\langle ij \rangle} \mathbf{h}^{\mathrm{eff}}_{ij} \cdot (\mathbf{S}_i + \mathbf{S}_j)$, where $\mu,\nu=x, y, z$. The magnetic interactions $\mathcal{M}_{i\mu,j\nu}$ (the expressions are shown in the SM \cite{supp,Kumar2021}) can be controlled by laser, which imply a promising route to stabilize quantum spin liquid by tuning the competing interactions in favor of the quantum spin liquid~\cite{PhysRevLett.102.017205,Rau2016,PhysRevB.103.L100408,Sriram2021}. Here we focus on the photo-induced emergent magnetic-field $\mathbf{h}^{\mathrm{eff}}_{ij}$, which is written in terms of the model parameters as
\begin{align}\label{eq.9}
\mathbf{h}^{\mathrm{eff}}_{ij} = \overline{\sum} & \frac{8\mathcal{J}_n(\mathrm{A}_0) \mathcal{J}_m(\mathrm{A}) \mathcal{J}_l(\mathrm{A})t^2_{pd}}{27} \frac{\sin \psi^{ml}_0}{\Delta + l \Omega} \nonumber \\
& \bigg[ \frac{t_1 - t_3}{U-3J_{\mathrm{H}}-n\Omega} +  \frac{t_1 - t_3}{U-J_{\mathrm{H}}-n\Omega} \bigg]\hat{\mathbf{z}},
\end{align} 
where $\overline{\sum}$ signifies summation over the indices $n,m,l$ with the constraint $n+m+l=0$, $A_0 = r_{dd}E_0/\Omega$, $A = r_{pd}E_0/\Omega$, and $\psi_0^{ml} = (m-l)\psi_0$. In contrast with the single-orbital case, the effective Zeeman magnetic field couples to the symmetric combination of the spins ($\mathbf{S}_i + \mathbf{S}_j$). Consequently, the applied polarized light generates a ferromagnetic magnetization in this case, which was also studied for $\alpha$-RuCl$_3$ in Ref. \cite{Sriram2021} recently, using numerical exact diagonalization. Here we \textit{emphasize} that our analysis is applicable to a wider class of Mott insulators with inversion symmetry.  

For weak laser drive and low frequency, $h^{\mathrm{eff}}_{ij}$ can be expanded asymptotically as
\begin{align}\label{eq.10}
h^{\mathrm{eff}}_{ij} \approx & \frac{4t_{pd}^2(t_1-t_3)}{27\Omega }\frac{E_0^2 r_{pd} \sin\psi_0}{\Delta^2} \left( 2r_{pd} \cos \psi_0 + r_{dd}\right) \nonumber \\
& \left( \frac{1}{U-3J_{\mathrm{H}}} + \frac{1}{U-J_{\mathrm{H}}}\right).
\end{align}
Since the TM atoms in the $\alpha$-RuCl$_3$ unit cell lie in the mirror plane and have additional inversion symmetry, this result is consistent with our phenomenological ansatz. The variation of $h^{\mathrm{eff}}_{ij}$ with the laser drive is shown in Fig.~\ref{fig:Fig3}(b) for $\Omega = 6.5$ eV.

{\textit{Discussion and conclusion.} \textbf{\---}  In this work we use the Floquet theory to study the IFE in Mott insulators. The Floquet formulation allows us to study the strong drive region systematically that goes beyond the weak drive results known before, i.e., the induced IFE Zeeman field $\mathbf{h}_{\mathrm{IFE}} \propto \mathbf{E}(\Omega) \times \mathbf{E}^*(\Omega)$. It also informs the heating associated with IFE due to laser irradiation. Our results are valid in the Floquet prethermal region, which can be exponentially long in time before the system evolves into the infinite temperature state if the laser frequency is tuned away from resonances of the system~\cite{Weidinger2017,PhysRevResearch.1.033202,PhysRevB.99.205111,PhysRevB.97.245122,PhysRevLett.115.256803,PhysRevLett.116.120401,PhysRevB.95.014112,PhysRevX.7.011026,PhysRevLett.120.197601,Hermman_2017,Tatsuhiko2020,PhysRevB.104.134308}. The resonances in our models include the resonances in the Hubbard gap, the charge transfer gap, the crystal field splitting gap and the spin-orbit splitting gap of the $j_{\mathrm{eff}}$ multiplets.} The IFE is resonantly enhanced near resonances in a short time scale, but heating quickly dominates, which invalidates the Floquet description. The IFE magnetic field can be of the order of 10 Tesla even away from the resonances.

We proposed a toy model (see Fig.~\ref{fig:Fig2}) to demonstrate the antiferromagnetic order favored by the IFE in materials with broken inversion symmetry. Certain distorted layered honeycomb compounds, such as Li$_3$Cu$_2$SbO$_6$~\cite{PhysRevB.103.174423}, can also realize our prediction. The single-orbital model can be realized in similar lattice geometries with a $d^9$-electronic configuration. The SOC can be induced by placing the thin films atop a substrate with heavy ions. To clearly distinguish the antiferromagnetic order induced by IFE from the antiferromagnetic Heisenberg exchange interaction, experiments can be performed above the magnetic ordering temperature. Below the ordering temperature a competition between the spin-exchange couplings and the induced magnetic-field $\mathbf{h}^{\mathrm{eff}}_{ij}$ can stabilize complex magnetic orders.

We specifically focused on a $d^5$-electronic configuration in edge-sharing octahedral structure for the IFE in the multiorbital systems. Throughout the analysis, we assumed a perpendicular incidence of light polarization to the TM-ligand-TM atom plane. Apart from the laser amplitude and frequency, the angle between the light polarization and the TM-ligand-TM atom plane, for an oblique incidence, provides yet another tunability to control the spin-exchange couplings and the overall sign of both ferro- and antiferromagnetic IFE Zeeman field \cite{supp}. By choosing the incident angle, we can stabilize an antiferromagnetic order using the IFE by avoiding a complete cancellation of the IFE Zeeman field between neighboring clusters.

To summarize, we studied the IFE in Mott insulators irradiated by a circularly polarized light. Based on both the symmetry consideration and the microscopic model calculations using the Floquet theory, we showed that the IFE in Mott insulators without (with) inversion symmetry favors antiferromagnetic (ferromagnetic) order. Our results suggest a promising route to ultrafast control of magnetic order in Mott insulators by light.

\textit{Acknowledgments.} \textbf{\---} We would like to thank Avadh Saxena, Alexander V. Balatsky, and Dieter Vollhardt for providing important feedback while writing this paper, and thank Nicholas Sirica, Rohit Prasankumar, Sang-Wook Cheong and Jianxin Zhu for the helpful discussions. This work was carried out under the auspices of the U.S. DOE NNSA under Contract No. 89233218CNA000001 through the LDRD Program. S.Z.L. was also supported by the U.S. DOE, Office of Science, Basic Energy Sciences, Materials Sciences and Engineering Division, Condensed Matter Theory Program.

\SB{\textit{Note.} \textbf{\---} We recently became aware of an experiment~\cite{Shan2021} where a similar \textit{anti-ferromagnetic} coupling of response functions (polarization) to neighboring spins is observed in a periodically driven \textit{non-centrosymmetric} Mott insulator (MnPS$_3$)}.

\bibliographystyle{apsrev4-1}
\bibliography{MS_qsl}

\begin{thebibliography}{66}%
\makeatletter
\providecommand \@ifxundefined [1]{%
 \@ifx{#1\undefined}
}%
\providecommand \@ifnum [1]{%
 \ifnum #1\expandafter \@firstoftwo
 \else \expandafter \@secondoftwo
 \fi
}%
\providecommand \@ifx [1]{%
 \ifx #1\expandafter \@firstoftwo
 \else \expandafter \@secondoftwo
 \fi
}%
\providecommand \natexlab [1]{#1}%
\providecommand \enquote  [1]{``#1''}%
\providecommand \bibnamefont  [1]{#1}%
\providecommand \bibfnamefont [1]{#1}%
\providecommand \citenamefont [1]{#1}%
\providecommand \href@noop [0]{\@secondoftwo}%
\providecommand \href [0]{\begingroup \@sanitize@url \@href}%
\providecommand \@href[1]{\@@startlink{#1}\@@href}%
\providecommand \@@href[1]{\endgroup#1\@@endlink}%
\providecommand \@sanitize@url [0]{\catcode `\\12\catcode `\$12\catcode
  `\&12\catcode `\#12\catcode `\^12\catcode `\_12\catcode `\%12\relax}%
\providecommand \@@startlink[1]{}%
\providecommand \@@endlink[0]{}%
\providecommand \url  [0]{\begingroup\@sanitize@url \@url }%
\providecommand \@url [1]{\endgroup\@href {#1}{\urlprefix }}%
\providecommand \urlprefix  [0]{URL }%
\providecommand \Eprint [0]{\href }%
\providecommand \doibase [0]{http://dx.doi.org/}%
\providecommand \selectlanguage [0]{\@gobble}%
\providecommand \bibinfo  [0]{\@secondoftwo}%
\providecommand \bibfield  [0]{\@secondoftwo}%
\providecommand \translation [1]{[#1]}%
\providecommand \BibitemOpen [0]{}%
\providecommand \bibitemStop [0]{}%
\providecommand \bibitemNoStop [0]{.\EOS\space}%
\providecommand \EOS [0]{\spacefactor3000\relax}%
\providecommand \BibitemShut  [1]{\csname bibitem#1\endcsname}%
\let\auto@bib@innerbib\@empty
\bibitem [{\citenamefont {Kirilyuk}\ \emph {et~al.}(2010)\citenamefont
  {Kirilyuk}, \citenamefont {Kimel},\ and\ \citenamefont
  {Rasing}}]{RevModPhys.82.2731}%
  \BibitemOpen
  \bibfield  {author} {\bibinfo {author} {\bibfnamefont {A.}~\bibnamefont
  {Kirilyuk}}, \bibinfo {author} {\bibfnamefont {A.~V.}\ \bibnamefont {Kimel}},
  \ and\ \bibinfo {author} {\bibfnamefont {T.}~\bibnamefont {Rasing}},\ }\href
  {\doibase 10.1103/RevModPhys.82.2731} {\bibfield  {journal} {\bibinfo
  {journal} {Rev. Mod. Phys.}\ }\textbf {\bibinfo {volume} {82}},\ \bibinfo
  {pages} {2731} (\bibinfo {year} {2010})}\BibitemShut {NoStop}%
\bibitem [{\citenamefont {Forn-D\'{\i}az}\ \emph {et~al.}(2019)\citenamefont
  {Forn-D\'{\i}az}, \citenamefont {Lamata}, \citenamefont {Rico}, \citenamefont
  {Kono},\ and\ \citenamefont {Solano}}]{RevModPhys.91.025005}%
  \BibitemOpen
  \bibfield  {author} {\bibinfo {author} {\bibfnamefont {P.}~\bibnamefont
  {Forn-D\'{\i}az}}, \bibinfo {author} {\bibfnamefont {L.}~\bibnamefont
  {Lamata}}, \bibinfo {author} {\bibfnamefont {E.}~\bibnamefont {Rico}},
  \bibinfo {author} {\bibfnamefont {J.}~\bibnamefont {Kono}}, \ and\ \bibinfo
  {author} {\bibfnamefont {E.}~\bibnamefont {Solano}},\ }\href {\doibase
  10.1103/RevModPhys.91.025005} {\bibfield  {journal} {\bibinfo  {journal}
  {Rev. Mod. Phys.}\ }\textbf {\bibinfo {volume} {91}},\ \bibinfo {pages}
  {025005} (\bibinfo {year} {2019})}\BibitemShut {NoStop}%
\bibitem [{\citenamefont {Schatz}\ and\ \citenamefont
  {McCaffery}(1969)}]{Faraday_1845}%
  \BibitemOpen
  \bibfield  {author} {\bibinfo {author} {\bibfnamefont {P.~N.}\ \bibnamefont
  {Schatz}}\ and\ \bibinfo {author} {\bibfnamefont {A.~J.}\ \bibnamefont
  {McCaffery}},\ }\href {\doibase 10.1039/QR9692300552} {\bibfield  {journal}
  {\bibinfo  {journal} {Q. Rev. Chem. Soc.}\ }\textbf {\bibinfo {volume}
  {23}},\ \bibinfo {pages} {552} (\bibinfo {year} {1969})}\BibitemShut
  {NoStop}%
\bibitem [{\citenamefont {Pitaevskii}(1961)}]{Pitaevskii_1961}%
  \BibitemOpen
  \bibfield  {author} {\bibinfo {author} {\bibfnamefont {L.~P.}\ \bibnamefont
  {Pitaevskii}},\ }\href {http://www.jetp.ac.ru/cgi-bin/dn/e_012_05_1008.pdf}
  {\bibfield  {journal} {\bibinfo  {journal} {JETP}\ }\textbf {\bibinfo
  {volume} {12}},\ \bibinfo {pages} {1008} (\bibinfo {year}
  {1961})}\BibitemShut {NoStop}%
\bibitem [{\citenamefont {van~der Ziel}\ \emph {et~al.}(1965)\citenamefont
  {van~der Ziel}, \citenamefont {Pershan},\ and\ \citenamefont
  {Malmstrom}}]{PhysRevLett.15.190}%
  \BibitemOpen
  \bibfield  {author} {\bibinfo {author} {\bibfnamefont {J.~P.}\ \bibnamefont
  {van~der Ziel}}, \bibinfo {author} {\bibfnamefont {P.~S.}\ \bibnamefont
  {Pershan}}, \ and\ \bibinfo {author} {\bibfnamefont {L.~D.}\ \bibnamefont
  {Malmstrom}},\ }\href {\doibase 10.1103/PhysRevLett.15.190} {\bibfield
  {journal} {\bibinfo  {journal} {Phys. Rev. Lett.}\ }\textbf {\bibinfo
  {volume} {15}},\ \bibinfo {pages} {190} (\bibinfo {year} {1965})}\BibitemShut
  {NoStop}%
\bibitem [{\citenamefont {Kimel}\ \emph {et~al.}(2005)\citenamefont {Kimel},
  \citenamefont {Kirilyuk}, \citenamefont {Usachev}, \citenamefont {Pisarev},
  \citenamefont {Balbashov},\ and\ \citenamefont {Rasing}}]{Kimel2005}%
  \BibitemOpen
  \bibfield  {author} {\bibinfo {author} {\bibfnamefont {A.~V.}\ \bibnamefont
  {Kimel}}, \bibinfo {author} {\bibfnamefont {A.}~\bibnamefont {Kirilyuk}},
  \bibinfo {author} {\bibfnamefont {P.~A.}\ \bibnamefont {Usachev}}, \bibinfo
  {author} {\bibfnamefont {R.~V.}\ \bibnamefont {Pisarev}}, \bibinfo {author}
  {\bibfnamefont {A.~M.}\ \bibnamefont {Balbashov}}, \ and\ \bibinfo {author}
  {\bibfnamefont {T.}~\bibnamefont {Rasing}},\ }\href {\doibase
  10.1038/nature03564} {\bibfield  {journal} {\bibinfo  {journal} {Nature}\
  }\textbf {\bibinfo {volume} {435}},\ \bibinfo {pages} {655} (\bibinfo {year}
  {2005})}\BibitemShut {NoStop}%
\bibitem [{\citenamefont {Lottermoser}\ \emph {et~al.}(2004)\citenamefont
  {Lottermoser}, \citenamefont {Lonkai}, \citenamefont {Amann}, \citenamefont
  {Hohlwein}, \citenamefont {Ihringer},\ and\ \citenamefont
  {Fiebig}}]{Lottermoser2004}%
  \BibitemOpen
  \bibfield  {author} {\bibinfo {author} {\bibfnamefont {T.}~\bibnamefont
  {Lottermoser}}, \bibinfo {author} {\bibfnamefont {T.}~\bibnamefont {Lonkai}},
  \bibinfo {author} {\bibfnamefont {U.}~\bibnamefont {Amann}}, \bibinfo
  {author} {\bibfnamefont {D.}~\bibnamefont {Hohlwein}}, \bibinfo {author}
  {\bibfnamefont {J.}~\bibnamefont {Ihringer}}, \ and\ \bibinfo {author}
  {\bibfnamefont {M.}~\bibnamefont {Fiebig}},\ }\href {\doibase
  10.1038/nature02728} {\bibfield  {journal} {\bibinfo  {journal} {Nature}\
  }\textbf {\bibinfo {volume} {430}},\ \bibinfo {pages} {541} (\bibinfo {year}
  {2004})}\BibitemShut {NoStop}%
\bibitem [{\citenamefont {Jungfleisch}\ \emph {et~al.}(2018)\citenamefont
  {Jungfleisch}, \citenamefont {Zhang}, \citenamefont {Zhang}, \citenamefont
  {Pearson}, \citenamefont {Schaller}, \citenamefont {Wen},\ and\ \citenamefont
  {Hoffmann}}]{PhysRevLett.120.207207}%
  \BibitemOpen
  \bibfield  {author} {\bibinfo {author} {\bibfnamefont {M.~B.}\ \bibnamefont
  {Jungfleisch}}, \bibinfo {author} {\bibfnamefont {Q.}~\bibnamefont {Zhang}},
  \bibinfo {author} {\bibfnamefont {W.}~\bibnamefont {Zhang}}, \bibinfo
  {author} {\bibfnamefont {J.~E.}\ \bibnamefont {Pearson}}, \bibinfo {author}
  {\bibfnamefont {R.~D.}\ \bibnamefont {Schaller}}, \bibinfo {author}
  {\bibfnamefont {H.}~\bibnamefont {Wen}}, \ and\ \bibinfo {author}
  {\bibfnamefont {A.}~\bibnamefont {Hoffmann}},\ }\href {\doibase
  10.1103/PhysRevLett.120.207207} {\bibfield  {journal} {\bibinfo  {journal}
  {Phys. Rev. Lett.}\ }\textbf {\bibinfo {volume} {120}},\ \bibinfo {pages}
  {207207} (\bibinfo {year} {2018})}\BibitemShut {NoStop}%
\bibitem [{\citenamefont {Gu}\ and\ \citenamefont {Kornev}(2010)}]{Gu2010}%
  \BibitemOpen
  \bibfield  {author} {\bibinfo {author} {\bibfnamefont {Y.}~\bibnamefont
  {Gu}}\ and\ \bibinfo {author} {\bibfnamefont {K.~G.}\ \bibnamefont
  {Kornev}},\ }\href {\doibase 10.1364/JOSAB.27.002165} {\bibfield  {journal}
  {\bibinfo  {journal} {J. Opt. Soc. Am. B}\ }\textbf {\bibinfo {volume}
  {27}},\ \bibinfo {pages} {2165} (\bibinfo {year} {2010})}\BibitemShut
  {NoStop}%
\bibitem [{\citenamefont {Hertel}(2006)}]{Hertel2006}%
  \BibitemOpen
  \bibfield  {author} {\bibinfo {author} {\bibfnamefont {R.}~\bibnamefont
  {Hertel}},\ }\href {\doibase https://doi.org/10.1016/j.jmmm.2005.10.225}
  {\bibfield  {journal} {\bibinfo  {journal} {J. Magn. Magn. Mater.}\ }\textbf
  {\bibinfo {volume} {303}},\ \bibinfo {pages} {L1} (\bibinfo {year}
  {2006})}\BibitemShut {NoStop}%
\bibitem [{\citenamefont {Woodford}(2009)}]{PhysRevB.79.212412}%
  \BibitemOpen
  \bibfield  {author} {\bibinfo {author} {\bibfnamefont {S.~R.}\ \bibnamefont
  {Woodford}},\ }\href {\doibase 10.1103/PhysRevB.79.212412} {\bibfield
  {journal} {\bibinfo  {journal} {Phys. Rev. B}\ }\textbf {\bibinfo {volume}
  {79}},\ \bibinfo {pages} {212412} (\bibinfo {year} {2009})}\BibitemShut
  {NoStop}%
\bibitem [{\citenamefont {Perroni}\ and\ \citenamefont
  {Liebsch}(2006)}]{PhysRevB.74.134430}%
  \BibitemOpen
  \bibfield  {author} {\bibinfo {author} {\bibfnamefont {C.~A.}\ \bibnamefont
  {Perroni}}\ and\ \bibinfo {author} {\bibfnamefont {A.}~\bibnamefont
  {Liebsch}},\ }\href {\doibase 10.1103/PhysRevB.74.134430} {\bibfield
  {journal} {\bibinfo  {journal} {Phys. Rev. B}\ }\textbf {\bibinfo {volume}
  {74}},\ \bibinfo {pages} {134430} (\bibinfo {year} {2006})}\BibitemShut
  {NoStop}%
\bibitem [{\citenamefont {Battiato}\ \emph {et~al.}(2014)\citenamefont
  {Battiato}, \citenamefont {Barbalinardo},\ and\ \citenamefont
  {Oppeneer}}]{PhysRevB.89.014413}%
  \BibitemOpen
  \bibfield  {author} {\bibinfo {author} {\bibfnamefont {M.}~\bibnamefont
  {Battiato}}, \bibinfo {author} {\bibfnamefont {G.}~\bibnamefont
  {Barbalinardo}}, \ and\ \bibinfo {author} {\bibfnamefont {P.~M.}\
  \bibnamefont {Oppeneer}},\ }\href {\doibase 10.1103/PhysRevB.89.014413}
  {\bibfield  {journal} {\bibinfo  {journal} {Phys. Rev. B}\ }\textbf {\bibinfo
  {volume} {89}},\ \bibinfo {pages} {014413} (\bibinfo {year}
  {2014})}\BibitemShut {NoStop}%
\bibitem [{\citenamefont {Tanaka}\ \emph {et~al.}(2020)\citenamefont {Tanaka},
  \citenamefont {Inoue},\ and\ \citenamefont {Mochizuki}}]{Tanaka_2020}%
  \BibitemOpen
  \bibfield  {author} {\bibinfo {author} {\bibfnamefont {Y.}~\bibnamefont
  {Tanaka}}, \bibinfo {author} {\bibfnamefont {T.}~\bibnamefont {Inoue}}, \
  and\ \bibinfo {author} {\bibfnamefont {M.}~\bibnamefont {Mochizuki}},\ }\href
  {\doibase 10.1088/1367-2630/aba5be} {\bibfield  {journal} {\bibinfo
  {journal} {New J. Phys.}\ }\textbf {\bibinfo {volume} {22}},\ \bibinfo
  {pages} {083054} (\bibinfo {year} {2020})}\BibitemShut {NoStop}%
\bibitem [{\citenamefont {Gao}\ \emph {et~al.}(2020)\citenamefont {Gao},
  \citenamefont {Wang},\ and\ \citenamefont {Xiao}}]{gao_topological_2020}%
  \BibitemOpen
  \bibfield  {author} {\bibinfo {author} {\bibfnamefont {Y.}~\bibnamefont
  {Gao}}, \bibinfo {author} {\bibfnamefont {C.}~\bibnamefont {Wang}}, \ and\
  \bibinfo {author} {\bibfnamefont {D.}~\bibnamefont {Xiao}},\ }\href@noop {}
  {\  (\bibinfo {year} {2020})},\ \Eprint
  {http://arxiv.org/abs/arXiv:2009.13392} {arXiv:2009.13392} \BibitemShut
  {NoStop}%
\bibitem [{\citenamefont {Tokman}\ \emph {et~al.}(2020)\citenamefont {Tokman},
  \citenamefont {Chen}, \citenamefont {Shereshevsky}, \citenamefont
  {Pozdnyakova}, \citenamefont {Oladyshkin}, \citenamefont {Tokman},\ and\
  \citenamefont {Belyanin}}]{PhysRevB.101.174429}%
  \BibitemOpen
  \bibfield  {author} {\bibinfo {author} {\bibfnamefont {I.~D.}\ \bibnamefont
  {Tokman}}, \bibinfo {author} {\bibfnamefont {Q.}~\bibnamefont {Chen}},
  \bibinfo {author} {\bibfnamefont {I.~A.}\ \bibnamefont {Shereshevsky}},
  \bibinfo {author} {\bibfnamefont {V.~I.}\ \bibnamefont {Pozdnyakova}},
  \bibinfo {author} {\bibfnamefont {I.}~\bibnamefont {Oladyshkin}}, \bibinfo
  {author} {\bibfnamefont {M.}~\bibnamefont {Tokman}}, \ and\ \bibinfo {author}
  {\bibfnamefont {A.}~\bibnamefont {Belyanin}},\ }\href {\doibase
  10.1103/PhysRevB.101.174429} {\bibfield  {journal} {\bibinfo  {journal}
  {Phys. Rev. B}\ }\textbf {\bibinfo {volume} {101}},\ \bibinfo {pages}
  {174429} (\bibinfo {year} {2020})}\BibitemShut {NoStop}%
\bibitem [{\citenamefont {Liang}\ \emph {et~al.}(2021)\citenamefont {Liang},
  \citenamefont {Sukhachov},\ and\ \citenamefont
  {Balatsky}}]{PhysRevLett.126.247202}%
  \BibitemOpen
  \bibfield  {author} {\bibinfo {author} {\bibfnamefont {L.}~\bibnamefont
  {Liang}}, \bibinfo {author} {\bibfnamefont {P.~O.}\ \bibnamefont
  {Sukhachov}}, \ and\ \bibinfo {author} {\bibfnamefont {A.~V.}\ \bibnamefont
  {Balatsky}},\ }\href {\doibase 10.1103/PhysRevLett.126.247202} {\bibfield
  {journal} {\bibinfo  {journal} {Phys. Rev. Lett.}\ }\textbf {\bibinfo
  {volume} {126}},\ \bibinfo {pages} {247202} (\bibinfo {year}
  {2021})}\BibitemShut {NoStop}%
\bibitem [{\citenamefont {Mironov}\ \emph {et~al.}(2021)\citenamefont
  {Mironov}, \citenamefont {Mel'nikov}, \citenamefont {Tokman}, \citenamefont
  {Vadimov}, \citenamefont {Lounis},\ and\ \citenamefont
  {Buzdin}}]{PhysRevLett.126.137002}%
  \BibitemOpen
  \bibfield  {author} {\bibinfo {author} {\bibfnamefont {S.~V.}\ \bibnamefont
  {Mironov}}, \bibinfo {author} {\bibfnamefont {A.~S.}\ \bibnamefont
  {Mel'nikov}}, \bibinfo {author} {\bibfnamefont {I.~D.}\ \bibnamefont
  {Tokman}}, \bibinfo {author} {\bibfnamefont {V.}~\bibnamefont {Vadimov}},
  \bibinfo {author} {\bibfnamefont {B.}~\bibnamefont {Lounis}}, \ and\ \bibinfo
  {author} {\bibfnamefont {A.~I.}\ \bibnamefont {Buzdin}},\ }\href {\doibase
  10.1103/PhysRevLett.126.137002} {\bibfield  {journal} {\bibinfo  {journal}
  {Phys. Rev. Lett.}\ }\textbf {\bibinfo {volume} {126}},\ \bibinfo {pages}
  {137002} (\bibinfo {year} {2021})}\BibitemShut {NoStop}%
\bibitem [{\citenamefont {Majedi}(2021)}]{PhysRevLett.127.087001}%
  \BibitemOpen
  \bibfield  {author} {\bibinfo {author} {\bibfnamefont {A.~H.}\ \bibnamefont
  {Majedi}},\ }\href {\doibase 10.1103/PhysRevLett.127.087001} {\bibfield
  {journal} {\bibinfo  {journal} {Phys. Rev. Lett.}\ }\textbf {\bibinfo
  {volume} {127}},\ \bibinfo {pages} {087001} (\bibinfo {year}
  {2021})}\BibitemShut {NoStop}%
\bibitem [{\citenamefont {N{\v{e}}mec}\ \emph {et~al.}(2018)\citenamefont
  {N{\v{e}}mec}, \citenamefont {Fiebig}, \citenamefont {Kampfrath},\ and\
  \citenamefont {Kimel}}]{Nemec2018}%
  \BibitemOpen
  \bibfield  {author} {\bibinfo {author} {\bibfnamefont {P.}~\bibnamefont
  {N{\v{e}}mec}}, \bibinfo {author} {\bibfnamefont {M.}~\bibnamefont {Fiebig}},
  \bibinfo {author} {\bibfnamefont {T.}~\bibnamefont {Kampfrath}}, \ and\
  \bibinfo {author} {\bibfnamefont {A.~V.}\ \bibnamefont {Kimel}},\ }\href
  {\doibase 10.1038/s41567-018-0051-x} {\bibfield  {journal} {\bibinfo
  {journal} {Nat. Phys.}\ }\textbf {\bibinfo {volume} {14}},\ \bibinfo {pages}
  {229} (\bibinfo {year} {2018})}\BibitemShut {NoStop}%
\bibitem [{\citenamefont {Paris}\ \emph {et~al.}(2021)\citenamefont {Paris},
  \citenamefont {Nicholson}, \citenamefont {Johnston}, \citenamefont {Tseng},
  \citenamefont {Rumo}, \citenamefont {Coslovich}, \citenamefont {Zohar},
  \citenamefont {Lin}, \citenamefont {Strocov}, \citenamefont {Saint-Martin},
  \citenamefont {Revcolevschi}, \citenamefont {Kemper}, \citenamefont
  {Schlotter}, \citenamefont {Dakovski}, \citenamefont {Monney},\ and\
  \citenamefont {Schmitt}}]{Paris2021}%
  \BibitemOpen
  \bibfield  {author} {\bibinfo {author} {\bibfnamefont {E.}~\bibnamefont
  {Paris}}, \bibinfo {author} {\bibfnamefont {C.~W.}\ \bibnamefont
  {Nicholson}}, \bibinfo {author} {\bibfnamefont {S.}~\bibnamefont {Johnston}},
  \bibinfo {author} {\bibfnamefont {Y.}~\bibnamefont {Tseng}}, \bibinfo
  {author} {\bibfnamefont {M.}~\bibnamefont {Rumo}}, \bibinfo {author}
  {\bibfnamefont {G.}~\bibnamefont {Coslovich}}, \bibinfo {author}
  {\bibfnamefont {S.}~\bibnamefont {Zohar}}, \bibinfo {author} {\bibfnamefont
  {M.~F.}\ \bibnamefont {Lin}}, \bibinfo {author} {\bibfnamefont {V.~N.}\
  \bibnamefont {Strocov}}, \bibinfo {author} {\bibfnamefont {R.}~\bibnamefont
  {Saint-Martin}}, \bibinfo {author} {\bibfnamefont {A.}~\bibnamefont
  {Revcolevschi}}, \bibinfo {author} {\bibfnamefont {A.}~\bibnamefont
  {Kemper}}, \bibinfo {author} {\bibfnamefont {W.}~\bibnamefont {Schlotter}},
  \bibinfo {author} {\bibfnamefont {G.~L.}\ \bibnamefont {Dakovski}}, \bibinfo
  {author} {\bibfnamefont {C.}~\bibnamefont {Monney}}, \ and\ \bibinfo {author}
  {\bibfnamefont {T.}~\bibnamefont {Schmitt}},\ }\href {\doibase
  10.1038/s41535-021-00350-5} {\bibfield  {journal} {\bibinfo  {journal} {npj
  Quantum Mater.}\ }\textbf {\bibinfo {volume} {6}},\ \bibinfo {pages} {51}
  (\bibinfo {year} {2021})}\BibitemShut {NoStop}%
\bibitem [{\citenamefont {Kimel}\ and\ \citenamefont
  {Zvezdin}(2015)}]{Kimel2015_1}%
  \BibitemOpen
  \bibfield  {author} {\bibinfo {author} {\bibfnamefont {A.~V.}\ \bibnamefont
  {Kimel}}\ and\ \bibinfo {author} {\bibfnamefont {A.~K.}\ \bibnamefont
  {Zvezdin}},\ }\href {\doibase 10.1063/1.4931650} {\bibfield  {journal}
  {\bibinfo  {journal} {Low Temp. Phys.}\ }\textbf {\bibinfo {volume} {41}},\
  \bibinfo {pages} {682} (\bibinfo {year} {2015})}\BibitemShut {NoStop}%
\bibitem [{\citenamefont {Kimel}\ \emph {et~al.}(2009)\citenamefont {Kimel},
  \citenamefont {Ivanov}, \citenamefont {Pisarev}, \citenamefont {Usachev},
  \citenamefont {Kirilyuk},\ and\ \citenamefont {Rasing}}]{Kimel2009}%
  \BibitemOpen
  \bibfield  {author} {\bibinfo {author} {\bibfnamefont {A.~V.}\ \bibnamefont
  {Kimel}}, \bibinfo {author} {\bibfnamefont {B.~A.}\ \bibnamefont {Ivanov}},
  \bibinfo {author} {\bibfnamefont {R.~V.}\ \bibnamefont {Pisarev}}, \bibinfo
  {author} {\bibfnamefont {P.~A.}\ \bibnamefont {Usachev}}, \bibinfo {author}
  {\bibfnamefont {A.}~\bibnamefont {Kirilyuk}}, \ and\ \bibinfo {author}
  {\bibfnamefont {T.}~\bibnamefont {Rasing}},\ }\href {\doibase
  10.1038/nphys1369} {\bibfield  {journal} {\bibinfo  {journal} {Nat. Phys.}\
  }\textbf {\bibinfo {volume} {5}},\ \bibinfo {pages} {727} (\bibinfo {year}
  {2009})}\BibitemShut {NoStop}%
\bibitem [{\citenamefont {de~Jong}\ \emph {et~al.}(2011)\citenamefont
  {de~Jong}, \citenamefont {Kimel}, \citenamefont {Pisarev}, \citenamefont
  {Kirilyuk},\ and\ \citenamefont {Rasing}}]{PhysRevB.84.104421}%
  \BibitemOpen
  \bibfield  {author} {\bibinfo {author} {\bibfnamefont {J.~A.}\ \bibnamefont
  {de~Jong}}, \bibinfo {author} {\bibfnamefont {A.~V.}\ \bibnamefont {Kimel}},
  \bibinfo {author} {\bibfnamefont {R.~V.}\ \bibnamefont {Pisarev}}, \bibinfo
  {author} {\bibfnamefont {A.}~\bibnamefont {Kirilyuk}}, \ and\ \bibinfo
  {author} {\bibfnamefont {T.}~\bibnamefont {Rasing}},\ }\href {\doibase
  10.1103/PhysRevB.84.104421} {\bibfield  {journal} {\bibinfo  {journal} {Phys.
  Rev. B}\ }\textbf {\bibinfo {volume} {84}},\ \bibinfo {pages} {104421}
  (\bibinfo {year} {2011})}\BibitemShut {NoStop}%
\bibitem [{\citenamefont {Popov}\ \emph {et~al.}(2021)\citenamefont {Popov},
  \citenamefont {Zvezdin}, \citenamefont {Gareeva}, \citenamefont {Kimel},\
  and\ \citenamefont {Zvezdin}}]{PhysRevB.103.014423}%
  \BibitemOpen
  \bibfield  {author} {\bibinfo {author} {\bibfnamefont {A.~I.}\ \bibnamefont
  {Popov}}, \bibinfo {author} {\bibfnamefont {K.~A.}\ \bibnamefont {Zvezdin}},
  \bibinfo {author} {\bibfnamefont {Z.~V.}\ \bibnamefont {Gareeva}}, \bibinfo
  {author} {\bibfnamefont {A.~V.}\ \bibnamefont {Kimel}}, \ and\ \bibinfo
  {author} {\bibfnamefont {A.~K.}\ \bibnamefont {Zvezdin}},\ }\href {\doibase
  10.1103/PhysRevB.103.014423} {\bibfield  {journal} {\bibinfo  {journal}
  {Phys. Rev. B}\ }\textbf {\bibinfo {volume} {103}},\ \bibinfo {pages}
  {014423} (\bibinfo {year} {2021})}\BibitemShut {NoStop}%
\bibitem [{\citenamefont {Schrieffer}\ and\ \citenamefont
  {Wolff}(1966)}]{PhysRev.149.491}%
  \BibitemOpen
  \bibfield  {author} {\bibinfo {author} {\bibfnamefont {J.~R.}\ \bibnamefont
  {Schrieffer}}\ and\ \bibinfo {author} {\bibfnamefont {P.~A.}\ \bibnamefont
  {Wolff}},\ }\href {\doibase 10.1103/PhysRev.149.491} {\bibfield  {journal}
  {\bibinfo  {journal} {Phys. Rev.}\ }\textbf {\bibinfo {volume} {149}},\
  \bibinfo {pages} {491} (\bibinfo {year} {1966})}\BibitemShut {NoStop}%
\bibitem [{\citenamefont {Harris}\ and\ \citenamefont
  {Lange}(1967)}]{PhysRev.157.295}%
  \BibitemOpen
  \bibfield  {author} {\bibinfo {author} {\bibfnamefont {A.~B.}\ \bibnamefont
  {Harris}}\ and\ \bibinfo {author} {\bibfnamefont {R.~V.}\ \bibnamefont
  {Lange}},\ }\href {\doibase 10.1103/PhysRev.157.295} {\bibfield  {journal}
  {\bibinfo  {journal} {Phys. Rev.}\ }\textbf {\bibinfo {volume} {157}},\
  \bibinfo {pages} {295} (\bibinfo {year} {1967})}\BibitemShut {NoStop}%
\bibitem [{\citenamefont {Bukov}\ \emph {et~al.}(2016)\citenamefont {Bukov},
  \citenamefont {Kolodrubetz},\ and\ \citenamefont
  {Polkovnikov}}]{PhysRevLett.116.125301}%
  \BibitemOpen
  \bibfield  {author} {\bibinfo {author} {\bibfnamefont {M.}~\bibnamefont
  {Bukov}}, \bibinfo {author} {\bibfnamefont {M.}~\bibnamefont {Kolodrubetz}},
  \ and\ \bibinfo {author} {\bibfnamefont {A.}~\bibnamefont {Polkovnikov}},\
  }\href {\doibase 10.1103/PhysRevLett.116.125301} {\bibfield  {journal}
  {\bibinfo  {journal} {Phys. Rev. Lett.}\ }\textbf {\bibinfo {volume} {116}},\
  \bibinfo {pages} {125301} (\bibinfo {year} {2016})}\BibitemShut {NoStop}%
\bibitem [{\citenamefont {Kumar}\ and\ \citenamefont
  {Lin}(2021)}]{PhysRevB.103.064508}%
  \BibitemOpen
  \bibfield  {author} {\bibinfo {author} {\bibfnamefont {U.}~\bibnamefont
  {Kumar}}\ and\ \bibinfo {author} {\bibfnamefont {S.-Z.}\ \bibnamefont
  {Lin}},\ }\href {\doibase 10.1103/PhysRevB.103.064508} {\bibfield  {journal}
  {\bibinfo  {journal} {Phys. Rev. B}\ }\textbf {\bibinfo {volume} {103}},\
  \bibinfo {pages} {064508} (\bibinfo {year} {2021})}\BibitemShut {NoStop}%
\bibitem [{\citenamefont {Moriya}(1960)}]{PhysRev.120.91}%
  \BibitemOpen
  \bibfield  {author} {\bibinfo {author} {\bibfnamefont {T.}~\bibnamefont
  {Moriya}},\ }\href {\doibase 10.1103/PhysRev.120.91} {\bibfield  {journal}
  {\bibinfo  {journal} {Phys. Rev.}\ }\textbf {\bibinfo {volume} {120}},\
  \bibinfo {pages} {91} (\bibinfo {year} {1960})}\BibitemShut {NoStop}%
\bibitem [{\citenamefont {Imada}\ \emph {et~al.}(1998)\citenamefont {Imada},
  \citenamefont {Fujimori},\ and\ \citenamefont {Tokura}}]{RevModPhys.70.1039}%
  \BibitemOpen
  \bibfield  {author} {\bibinfo {author} {\bibfnamefont {M.}~\bibnamefont
  {Imada}}, \bibinfo {author} {\bibfnamefont {A.}~\bibnamefont {Fujimori}}, \
  and\ \bibinfo {author} {\bibfnamefont {Y.}~\bibnamefont {Tokura}},\ }\href
  {\doibase 10.1103/RevModPhys.70.1039} {\bibfield  {journal} {\bibinfo
  {journal} {Rev. Mod. Phys.}\ }\textbf {\bibinfo {volume} {70}},\ \bibinfo
  {pages} {1039} (\bibinfo {year} {1998})}\BibitemShut {NoStop}%
\bibitem [{sup()}]{supp}%
  \BibitemOpen
  \href@noop {} {\bibinfo  {journal} {See the supplementary for more details.}\
  }\BibitemShut {NoStop}%
\bibitem [{\citenamefont {Mentink}\ \emph {et~al.}(2015)\citenamefont
  {Mentink}, \citenamefont {Balzer},\ and\ \citenamefont
  {Eckstein}}]{Mentink2015}%
  \BibitemOpen
\bibfield  {journal} {  }\bibfield  {author} {\bibinfo {author} {\bibfnamefont
  {J.~H.}\ \bibnamefont {Mentink}}, \bibinfo {author} {\bibfnamefont
  {K.}~\bibnamefont {Balzer}}, \ and\ \bibinfo {author} {\bibfnamefont
  {M.}~\bibnamefont {Eckstein}},\ }\href {\doibase 10.1038/ncomms7708}
  {\bibfield  {journal} {\bibinfo  {journal} {Nat. Commun.}\ }\textbf {\bibinfo
  {volume} {6}},\ \bibinfo {pages} {6708} (\bibinfo {year} {2015})}\BibitemShut
  {NoStop}%
\bibitem [{\citenamefont {{Eckstein}}\ \emph {et~al.}()\citenamefont
  {{Eckstein}}, \citenamefont {{Mentink}},\ and\ \citenamefont
  {{Werner}}}]{Eckstein}%
  \BibitemOpen
  \bibfield  {author} {\bibinfo {author} {\bibfnamefont {M.}~\bibnamefont
  {{Eckstein}}}, \bibinfo {author} {\bibfnamefont {J.~H.}\ \bibnamefont
  {{Mentink}}}, \ and\ \bibinfo {author} {\bibfnamefont {P.}~\bibnamefont
  {{Werner}}},\ }\href@noop {} {\ }\Eprint {http://arxiv.org/abs/1703.03269}
  {arXiv:1703.03269} \BibitemShut {NoStop}%
\bibitem [{\citenamefont {Sears}\ \emph {et~al.}(2015)\citenamefont {Sears},
  \citenamefont {Songvilay}, \citenamefont {Plumb}, \citenamefont {Clancy},
  \citenamefont {Qiu}, \citenamefont {Zhao}, \citenamefont {Parshall},\ and\
  \citenamefont {Kim}}]{PhysRevB.91.144420}%
  \BibitemOpen
  \bibfield  {author} {\bibinfo {author} {\bibfnamefont {J.~A.}\ \bibnamefont
  {Sears}}, \bibinfo {author} {\bibfnamefont {M.}~\bibnamefont {Songvilay}},
  \bibinfo {author} {\bibfnamefont {K.~W.}\ \bibnamefont {Plumb}}, \bibinfo
  {author} {\bibfnamefont {J.~P.}\ \bibnamefont {Clancy}}, \bibinfo {author}
  {\bibfnamefont {Y.}~\bibnamefont {Qiu}}, \bibinfo {author} {\bibfnamefont
  {Y.}~\bibnamefont {Zhao}}, \bibinfo {author} {\bibfnamefont {D.}~\bibnamefont
  {Parshall}}, \ and\ \bibinfo {author} {\bibfnamefont {Y.-J.}\ \bibnamefont
  {Kim}},\ }\href {\doibase 10.1103/PhysRevB.91.144420} {\bibfield  {journal}
  {\bibinfo  {journal} {Phys. Rev. B}\ }\textbf {\bibinfo {volume} {91}},\
  \bibinfo {pages} {144420} (\bibinfo {year} {2015})}\BibitemShut {NoStop}%
\bibitem [{\citenamefont {Jackeli}\ and\ \citenamefont
  {Khaliullin}(2009)}]{PhysRevLett.102.017205}%
  \BibitemOpen
  \bibfield  {author} {\bibinfo {author} {\bibfnamefont {G.}~\bibnamefont
  {Jackeli}}\ and\ \bibinfo {author} {\bibfnamefont {G.}~\bibnamefont
  {Khaliullin}},\ }\href {\doibase 10.1103/PhysRevLett.102.017205} {\bibfield
  {journal} {\bibinfo  {journal} {Phys. Rev. Lett.}\ }\textbf {\bibinfo
  {volume} {102}},\ \bibinfo {pages} {017205} (\bibinfo {year}
  {2009})}\BibitemShut {NoStop}%
\bibitem [{\citenamefont {Chaloupka}\ \emph {et~al.}(2013)\citenamefont
  {Chaloupka}, \citenamefont {Jackeli},\ and\ \citenamefont
  {Khaliullin}}]{PhysRevLett.110.097204}%
  \BibitemOpen
  \bibfield  {author} {\bibinfo {author} {\bibfnamefont {J.}~\bibnamefont
  {Chaloupka}}, \bibinfo {author} {\bibfnamefont {G.}~\bibnamefont {Jackeli}},
  \ and\ \bibinfo {author} {\bibfnamefont {G.}~\bibnamefont {Khaliullin}},\
  }\href {\doibase 10.1103/PhysRevLett.110.097204} {\bibfield  {journal}
  {\bibinfo  {journal} {Phys. Rev. Lett.}\ }\textbf {\bibinfo {volume} {110}},\
  \bibinfo {pages} {097204} (\bibinfo {year} {2013})}\BibitemShut {NoStop}%
\bibitem [{\citenamefont {Rau}\ \emph {et~al.}(2014)\citenamefont {Rau},
  \citenamefont {Lee},\ and\ \citenamefont {Kee}}]{PhysRevLett.112.077204}%
  \BibitemOpen
  \bibfield  {author} {\bibinfo {author} {\bibfnamefont {J.~G.}\ \bibnamefont
  {Rau}}, \bibinfo {author} {\bibfnamefont {E.~K.-H.}\ \bibnamefont {Lee}}, \
  and\ \bibinfo {author} {\bibfnamefont {H.-Y.}\ \bibnamefont {Kee}},\ }\href
  {\doibase 10.1103/PhysRevLett.112.077204} {\bibfield  {journal} {\bibinfo
  {journal} {Phys. Rev. Lett.}\ }\textbf {\bibinfo {volume} {112}},\ \bibinfo
  {pages} {077204} (\bibinfo {year} {2014})}\BibitemShut {NoStop}%
\bibitem [{\citenamefont {Winter}\ \emph {et~al.}(2016)\citenamefont {Winter},
  \citenamefont {Li}, \citenamefont {Jeschke},\ and\ \citenamefont
  {Valent\'{\i}}}]{PhysRevB.93.214431}%
  \BibitemOpen
  \bibfield  {author} {\bibinfo {author} {\bibfnamefont {S.~M.}\ \bibnamefont
  {Winter}}, \bibinfo {author} {\bibfnamefont {Y.}~\bibnamefont {Li}}, \bibinfo
  {author} {\bibfnamefont {H.~O.}\ \bibnamefont {Jeschke}}, \ and\ \bibinfo
  {author} {\bibfnamefont {R.}~\bibnamefont {Valent\'{\i}}},\ }\href {\doibase
  10.1103/PhysRevB.93.214431} {\bibfield  {journal} {\bibinfo  {journal} {Phys.
  Rev. B}\ }\textbf {\bibinfo {volume} {93}},\ \bibinfo {pages} {214431}
  (\bibinfo {year} {2016})}\BibitemShut {NoStop}%
\bibitem [{\citenamefont {Winter}\ \emph {et~al.}(2017)\citenamefont {Winter},
  \citenamefont {Tsirlin}, \citenamefont {Daghofer}, \citenamefont {van~den
  Brink}, \citenamefont {Singh}, \citenamefont {Gegenwart},\ and\ \citenamefont
  {Valent{\'{\i}}}}]{Winter_2017}%
  \BibitemOpen
  \bibfield  {author} {\bibinfo {author} {\bibfnamefont {S.~M.}\ \bibnamefont
  {Winter}}, \bibinfo {author} {\bibfnamefont {A.~A.}\ \bibnamefont {Tsirlin}},
  \bibinfo {author} {\bibfnamefont {M.}~\bibnamefont {Daghofer}}, \bibinfo
  {author} {\bibfnamefont {J.}~\bibnamefont {van~den Brink}}, \bibinfo {author}
  {\bibfnamefont {Y.}~\bibnamefont {Singh}}, \bibinfo {author} {\bibfnamefont
  {P.}~\bibnamefont {Gegenwart}}, \ and\ \bibinfo {author} {\bibfnamefont
  {R.}~\bibnamefont {Valent{\'{\i}}}},\ }\href {\doibase
  10.1088/1361-648x/aa8cf5} {\bibfield  {journal} {\bibinfo  {journal} {J.
  Phys. Condens. Matter.}\ }\textbf {\bibinfo {volume} {29}},\ \bibinfo {pages}
  {493002} (\bibinfo {year} {2017})}\BibitemShut {NoStop}%
\bibitem [{\citenamefont {Gotfryd}\ \emph {et~al.}(2017)\citenamefont
  {Gotfryd}, \citenamefont {Rusna\ifmmode~\check{c}\else \v{c}\fi{}ko},
  \citenamefont {Wohlfeld}, \citenamefont {Jackeli}, \citenamefont
  {Chaloupka},\ and\ \citenamefont {Ole\ifmmode~\acute{s}\else
  \'{s}\fi{}}}]{PhysRevB.95.024426}%
  \BibitemOpen
  \bibfield  {author} {\bibinfo {author} {\bibfnamefont {D.}~\bibnamefont
  {Gotfryd}}, \bibinfo {author} {\bibfnamefont {J.}~\bibnamefont
  {Rusna\ifmmode~\check{c}\else \v{c}\fi{}ko}}, \bibinfo {author}
  {\bibfnamefont {K.}~\bibnamefont {Wohlfeld}}, \bibinfo {author}
  {\bibfnamefont {G.}~\bibnamefont {Jackeli}}, \bibinfo {author} {\bibfnamefont
  {J.~c.~v.}\ \bibnamefont {Chaloupka}}, \ and\ \bibinfo {author}
  {\bibfnamefont {A.~M.}\ \bibnamefont {Ole\ifmmode~\acute{s}\else
  \'{s}\fi{}}},\ }\href {\doibase 10.1103/PhysRevB.95.024426} {\bibfield
  {journal} {\bibinfo  {journal} {Phys. Rev. B}\ }\textbf {\bibinfo {volume}
  {95}},\ \bibinfo {pages} {024426} (\bibinfo {year} {2017})}\BibitemShut
  {NoStop}%
\bibitem [{\citenamefont {Kanamori}(1957{\natexlab{a}})}]{Kanamori1957_1}%
  \BibitemOpen
  \bibfield  {author} {\bibinfo {author} {\bibfnamefont {J.}~\bibnamefont
  {Kanamori}},\ }\href {\doibase 10.1143/PTP.17.177} {\bibfield  {journal}
  {\bibinfo  {journal} {Prog. Theor. Phys.}\ }\textbf {\bibinfo {volume}
  {17}},\ \bibinfo {pages} {177} (\bibinfo {year}
  {1957}{\natexlab{a}})}\BibitemShut {NoStop}%
\bibitem [{\citenamefont {Kanamori}(1957{\natexlab{b}})}]{Kanamori1957_2}%
  \BibitemOpen
  \bibfield  {author} {\bibinfo {author} {\bibfnamefont {J.}~\bibnamefont
  {Kanamori}},\ }\href {\doibase 10.1143/PTP.17.197} {\bibfield  {journal}
  {\bibinfo  {journal} {Prog. Theor. Phys.}\ }\textbf {\bibinfo {volume}
  {17}},\ \bibinfo {pages} {197} (\bibinfo {year}
  {1957}{\natexlab{b}})}\BibitemShut {NoStop}%
\bibitem [{\citenamefont {Georges}\ \emph {et~al.}(2013)\citenamefont
  {Georges}, \citenamefont {Medici},\ and\ \citenamefont
  {Mravlje}}]{Georges2013}%
  \BibitemOpen
  \bibfield  {author} {\bibinfo {author} {\bibfnamefont {A.}~\bibnamefont
  {Georges}}, \bibinfo {author} {\bibfnamefont {L.~d.}\ \bibnamefont {Medici}},
  \ and\ \bibinfo {author} {\bibfnamefont {J.}~\bibnamefont {Mravlje}},\ }\href
  {\doibase 10.1146/annurev-conmatphys-020911-125045} {\bibfield  {journal}
  {\bibinfo  {journal} {Annu. Rev. Condens. Matter Phys.}\ }\textbf {\bibinfo
  {volume} {4}},\ \bibinfo {pages} {137} (\bibinfo {year} {2013})}\BibitemShut
  {NoStop}%
\bibitem [{\citenamefont {Ishihara}\ \emph {et~al.}(2002)\citenamefont
  {Ishihara}, \citenamefont {Hatakeyama},\ and\ \citenamefont
  {Maekawa}}]{PhysRevB.65.064442}%
  \BibitemOpen
  \bibfield  {author} {\bibinfo {author} {\bibfnamefont {S.}~\bibnamefont
  {Ishihara}}, \bibinfo {author} {\bibfnamefont {T.}~\bibnamefont
  {Hatakeyama}}, \ and\ \bibinfo {author} {\bibfnamefont {S.}~\bibnamefont
  {Maekawa}},\ }\href {\doibase 10.1103/PhysRevB.65.064442} {\bibfield
  {journal} {\bibinfo  {journal} {Phys. Rev. B}\ }\textbf {\bibinfo {volume}
  {65}},\ \bibinfo {pages} {064442} (\bibinfo {year} {2002})}\BibitemShut
  {NoStop}%
\bibitem [{\citenamefont {Arakawa}(2016)}]{PhysRevB.94.174416}%
  \BibitemOpen
  \bibfield  {author} {\bibinfo {author} {\bibfnamefont {N.}~\bibnamefont
  {Arakawa}},\ }\href {\doibase 10.1103/PhysRevB.94.174416} {\bibfield
  {journal} {\bibinfo  {journal} {Phys. Rev. B}\ }\textbf {\bibinfo {volume}
  {94}},\ \bibinfo {pages} {174416} (\bibinfo {year} {2016})}\BibitemShut
  {NoStop}%
\bibitem [{\citenamefont {Kim}\ and\ \citenamefont
  {Kee}(2016)}]{PhysRevB.93.155143}%
  \BibitemOpen
  \bibfield  {author} {\bibinfo {author} {\bibfnamefont {H.-S.}\ \bibnamefont
  {Kim}}\ and\ \bibinfo {author} {\bibfnamefont {H.-Y.}\ \bibnamefont {Kee}},\
  }\href {\doibase 10.1103/PhysRevB.93.155143} {\bibfield  {journal} {\bibinfo
  {journal} {Phys. Rev. B}\ }\textbf {\bibinfo {volume} {93}},\ \bibinfo
  {pages} {155143} (\bibinfo {year} {2016})}\BibitemShut {NoStop}%
\bibitem [{\citenamefont {Sinn}\ \emph {et~al.}(2016)\citenamefont {Sinn},
  \citenamefont {Kim}, \citenamefont {Kim}, \citenamefont {Lee}, \citenamefont
  {Won}, \citenamefont {Oh}, \citenamefont {Han}, \citenamefont {Chang},
  \citenamefont {Hur}, \citenamefont {Sato}, \citenamefont {Park},
  \citenamefont {Kim}, \citenamefont {Kim},\ and\ \citenamefont
  {Noh}}]{Sinn2016}%
  \BibitemOpen
  \bibfield  {author} {\bibinfo {author} {\bibfnamefont {S.}~\bibnamefont
  {Sinn}}, \bibinfo {author} {\bibfnamefont {C.~H.}\ \bibnamefont {Kim}},
  \bibinfo {author} {\bibfnamefont {B.~H.}\ \bibnamefont {Kim}}, \bibinfo
  {author} {\bibfnamefont {K.~D.}\ \bibnamefont {Lee}}, \bibinfo {author}
  {\bibfnamefont {C.~J.}\ \bibnamefont {Won}}, \bibinfo {author} {\bibfnamefont
  {J.~S.}\ \bibnamefont {Oh}}, \bibinfo {author} {\bibfnamefont
  {M.}~\bibnamefont {Han}}, \bibinfo {author} {\bibfnamefont {Y.~J.}\
  \bibnamefont {Chang}}, \bibinfo {author} {\bibfnamefont {N.}~\bibnamefont
  {Hur}}, \bibinfo {author} {\bibfnamefont {H.}~\bibnamefont {Sato}}, \bibinfo
  {author} {\bibfnamefont {B.-G.}\ \bibnamefont {Park}}, \bibinfo {author}
  {\bibfnamefont {C.}~\bibnamefont {Kim}}, \bibinfo {author} {\bibfnamefont
  {H.-D.}\ \bibnamefont {Kim}}, \ and\ \bibinfo {author} {\bibfnamefont
  {T.~W.}\ \bibnamefont {Noh}},\ }\href {\doibase 10.1038/srep39544} {\bibfield
   {journal} {\bibinfo  {journal} {Sci. Rep.}\ }\textbf {\bibinfo {volume}
  {6}},\ \bibinfo {pages} {39544} (\bibinfo {year} {2016})}\BibitemShut
  {NoStop}%
\bibitem [{\citenamefont {Kumar}\ \emph {et~al.}(2021)\citenamefont {Kumar},
  \citenamefont {Banerjee},\ and\ \citenamefont {Lin}}]{Kumar2021}%
  \BibitemOpen
  \bibfield  {author} {\bibinfo {author} {\bibfnamefont {U.}~\bibnamefont
  {Kumar}}, \bibinfo {author} {\bibfnamefont {S.}~\bibnamefont {Banerjee}}, \
  and\ \bibinfo {author} {\bibfnamefont {S.-Z.}\ \bibnamefont {Lin}},\
  }\href@noop {} {\  (\bibinfo {year} {2021})},\ \Eprint
  {http://arxiv.org/abs/arXiv:2111.01316} {arXiv:2111.01316} \BibitemShut
  {NoStop}%
\bibitem [{\citenamefont {Rau}\ \emph {et~al.}(2016)\citenamefont {Rau},
  \citenamefont {Lee},\ and\ \citenamefont {Kee}}]{Rau2016}%
  \BibitemOpen
  \bibfield  {author} {\bibinfo {author} {\bibfnamefont {J.~G.}\ \bibnamefont
  {Rau}}, \bibinfo {author} {\bibfnamefont {E.~K.-H.}\ \bibnamefont {Lee}}, \
  and\ \bibinfo {author} {\bibfnamefont {H.-Y.}\ \bibnamefont {Kee}},\ }\href
  {\doibase 10.1146/annurev-conmatphys-031115-011319} {\bibfield  {journal}
  {\bibinfo  {journal} {Annu. Rev. Condens. Matter Phys.}\ }\textbf {\bibinfo
  {volume} {7}},\ \bibinfo {pages} {195} (\bibinfo {year} {2016})}\BibitemShut
  {NoStop}%
\bibitem [{\citenamefont {Arakawa}\ and\ \citenamefont
  {Yonemitsu}(2021)}]{PhysRevB.103.L100408}%
  \BibitemOpen
  \bibfield  {author} {\bibinfo {author} {\bibfnamefont {N.}~\bibnamefont
  {Arakawa}}\ and\ \bibinfo {author} {\bibfnamefont {K.}~\bibnamefont
  {Yonemitsu}},\ }\href {\doibase 10.1103/PhysRevB.103.L100408} {\bibfield
  {journal} {\bibinfo  {journal} {Phys. Rev. B}\ }\textbf {\bibinfo {volume}
  {103}},\ \bibinfo {pages} {L100408} (\bibinfo {year} {2021})}\BibitemShut
  {NoStop}%
\bibitem [{\citenamefont {Sriram}\ and\ \citenamefont
  {Claassen}(2021)}]{Sriram2021}%
  \BibitemOpen
  \bibfield  {author} {\bibinfo {author} {\bibfnamefont {A.}~\bibnamefont
  {Sriram}}\ and\ \bibinfo {author} {\bibfnamefont {M.}~\bibnamefont
  {Claassen}},\ }\href@noop {} {\  (\bibinfo {year} {2021})},\ \Eprint
  {http://arxiv.org/abs/arXiv:2105.01062} {arXiv:2105.01062} \BibitemShut
  {NoStop}%
\bibitem [{\citenamefont {Weidinger}\ and\ \citenamefont
  {Knap}(2017)}]{Weidinger2017}%
  \BibitemOpen
  \bibfield  {author} {\bibinfo {author} {\bibfnamefont {S.~A.}\ \bibnamefont
  {Weidinger}}\ and\ \bibinfo {author} {\bibfnamefont {M.}~\bibnamefont
  {Knap}},\ }\href {\doibase 10.1038/srep45382} {\bibfield  {journal} {\bibinfo
   {journal} {Sci. Rep.}\ }\textbf {\bibinfo {volume} {7}},\ \bibinfo {pages}
  {45382} (\bibinfo {year} {2017})}\BibitemShut {NoStop}%
\bibitem [{\citenamefont {Machado}\ \emph {et~al.}(2019)\citenamefont
  {Machado}, \citenamefont {Kahanamoku-Meyer}, \citenamefont {Else},
  \citenamefont {Nayak},\ and\ \citenamefont {Yao}}]{PhysRevResearch.1.033202}%
  \BibitemOpen
  \bibfield  {author} {\bibinfo {author} {\bibfnamefont {F.}~\bibnamefont
  {Machado}}, \bibinfo {author} {\bibfnamefont {G.~D.}\ \bibnamefont
  {Kahanamoku-Meyer}}, \bibinfo {author} {\bibfnamefont {D.~V.}\ \bibnamefont
  {Else}}, \bibinfo {author} {\bibfnamefont {C.}~\bibnamefont {Nayak}}, \ and\
  \bibinfo {author} {\bibfnamefont {N.~Y.}\ \bibnamefont {Yao}},\ }\href
  {\doibase 10.1103/PhysRevResearch.1.033202} {\bibfield  {journal} {\bibinfo
  {journal} {Phys. Rev. Research}\ }\textbf {\bibinfo {volume} {1}},\ \bibinfo
  {pages} {033202} (\bibinfo {year} {2019})}\BibitemShut {NoStop}%
\bibitem [{\citenamefont {Hejazi}\ \emph {et~al.}(2019)\citenamefont {Hejazi},
  \citenamefont {Liu},\ and\ \citenamefont {Balents}}]{PhysRevB.99.205111}%
  \BibitemOpen
  \bibfield  {author} {\bibinfo {author} {\bibfnamefont {K.}~\bibnamefont
  {Hejazi}}, \bibinfo {author} {\bibfnamefont {J.}~\bibnamefont {Liu}}, \ and\
  \bibinfo {author} {\bibfnamefont {L.}~\bibnamefont {Balents}},\ }\href
  {\doibase 10.1103/PhysRevB.99.205111} {\bibfield  {journal} {\bibinfo
  {journal} {Phys. Rev. B}\ }\textbf {\bibinfo {volume} {99}},\ \bibinfo
  {pages} {205111} (\bibinfo {year} {2019})}\BibitemShut {NoStop}%
\bibitem [{\citenamefont {Haldar}\ \emph {et~al.}(2018)\citenamefont {Haldar},
  \citenamefont {Moessner},\ and\ \citenamefont {Das}}]{PhysRevB.97.245122}%
  \BibitemOpen
  \bibfield  {author} {\bibinfo {author} {\bibfnamefont {A.}~\bibnamefont
  {Haldar}}, \bibinfo {author} {\bibfnamefont {R.}~\bibnamefont {Moessner}}, \
  and\ \bibinfo {author} {\bibfnamefont {A.}~\bibnamefont {Das}},\ }\href
  {\doibase 10.1103/PhysRevB.97.245122} {\bibfield  {journal} {\bibinfo
  {journal} {Phys. Rev. B}\ }\textbf {\bibinfo {volume} {97}},\ \bibinfo
  {pages} {245122} (\bibinfo {year} {2018})}\BibitemShut {NoStop}%
\bibitem [{\citenamefont {Abanin}\ \emph {et~al.}(2015)\citenamefont {Abanin},
  \citenamefont {De~Roeck},\ and\ \citenamefont
  {Huveneers}}]{PhysRevLett.115.256803}%
  \BibitemOpen
  \bibfield  {author} {\bibinfo {author} {\bibfnamefont {D.~A.}\ \bibnamefont
  {Abanin}}, \bibinfo {author} {\bibfnamefont {W.}~\bibnamefont {De~Roeck}}, \
  and\ \bibinfo {author} {\bibfnamefont {F.}~\bibnamefont {Huveneers}},\ }\href
  {\doibase 10.1103/PhysRevLett.115.256803} {\bibfield  {journal} {\bibinfo
  {journal} {Phys. Rev. Lett.}\ }\textbf {\bibinfo {volume} {115}},\ \bibinfo
  {pages} {256803} (\bibinfo {year} {2015})}\BibitemShut {NoStop}%
\bibitem [{\citenamefont {Mori}\ \emph {et~al.}(2016)\citenamefont {Mori},
  \citenamefont {Kuwahara},\ and\ \citenamefont
  {Saito}}]{PhysRevLett.116.120401}%
  \BibitemOpen
  \bibfield  {author} {\bibinfo {author} {\bibfnamefont {T.}~\bibnamefont
  {Mori}}, \bibinfo {author} {\bibfnamefont {T.}~\bibnamefont {Kuwahara}}, \
  and\ \bibinfo {author} {\bibfnamefont {K.}~\bibnamefont {Saito}},\ }\href
  {\doibase 10.1103/PhysRevLett.116.120401} {\bibfield  {journal} {\bibinfo
  {journal} {Phys. Rev. Lett.}\ }\textbf {\bibinfo {volume} {116}},\ \bibinfo
  {pages} {120401} (\bibinfo {year} {2016})}\BibitemShut {NoStop}%
\bibitem [{\citenamefont {Abanin}\ \emph {et~al.}(2017)\citenamefont {Abanin},
  \citenamefont {De~Roeck}, \citenamefont {Ho},\ and\ \citenamefont
  {Huveneers}}]{PhysRevB.95.014112}%
  \BibitemOpen
  \bibfield  {author} {\bibinfo {author} {\bibfnamefont {D.~A.}\ \bibnamefont
  {Abanin}}, \bibinfo {author} {\bibfnamefont {W.}~\bibnamefont {De~Roeck}},
  \bibinfo {author} {\bibfnamefont {W.~W.}\ \bibnamefont {Ho}}, \ and\ \bibinfo
  {author} {\bibfnamefont {F.}~\bibnamefont {Huveneers}},\ }\href {\doibase
  10.1103/PhysRevB.95.014112} {\bibfield  {journal} {\bibinfo  {journal} {Phys.
  Rev. B}\ }\textbf {\bibinfo {volume} {95}},\ \bibinfo {pages} {014112}
  (\bibinfo {year} {2017})}\BibitemShut {NoStop}%
\bibitem [{\citenamefont {Else}\ \emph {et~al.}(2017)\citenamefont {Else},
  \citenamefont {Bauer},\ and\ \citenamefont {Nayak}}]{PhysRevX.7.011026}%
  \BibitemOpen
  \bibfield  {author} {\bibinfo {author} {\bibfnamefont {D.~V.}\ \bibnamefont
  {Else}}, \bibinfo {author} {\bibfnamefont {B.}~\bibnamefont {Bauer}}, \ and\
  \bibinfo {author} {\bibfnamefont {C.}~\bibnamefont {Nayak}},\ }\href
  {\doibase 10.1103/PhysRevX.7.011026} {\bibfield  {journal} {\bibinfo
  {journal} {Phys. Rev. X}\ }\textbf {\bibinfo {volume} {7}},\ \bibinfo {pages}
  {011026} (\bibinfo {year} {2017})}\BibitemShut {NoStop}%
\bibitem [{\citenamefont {Peronaci}\ \emph {et~al.}(2018)\citenamefont
  {Peronaci}, \citenamefont {Schir\'o},\ and\ \citenamefont
  {Parcollet}}]{PhysRevLett.120.197601}%
  \BibitemOpen
  \bibfield  {author} {\bibinfo {author} {\bibfnamefont {F.}~\bibnamefont
  {Peronaci}}, \bibinfo {author} {\bibfnamefont {M.}~\bibnamefont {Schir\'o}},
  \ and\ \bibinfo {author} {\bibfnamefont {O.}~\bibnamefont {Parcollet}},\
  }\href {\doibase 10.1103/PhysRevLett.120.197601} {\bibfield  {journal}
  {\bibinfo  {journal} {Phys. Rev. Lett.}\ }\textbf {\bibinfo {volume} {120}},\
  \bibinfo {pages} {197601} (\bibinfo {year} {2018})}\BibitemShut {NoStop}%
\bibitem [{\citenamefont {Herrmann}\ \emph {et~al.}(2017)\citenamefont
  {Herrmann}, \citenamefont {Murakami}, \citenamefont {Eckstein},\ and\
  \citenamefont {Werner}}]{Hermman_2017}%
  \BibitemOpen
  \bibfield  {author} {\bibinfo {author} {\bibfnamefont {A.}~\bibnamefont
  {Herrmann}}, \bibinfo {author} {\bibfnamefont {Y.}~\bibnamefont {Murakami}},
  \bibinfo {author} {\bibfnamefont {M.}~\bibnamefont {Eckstein}}, \ and\
  \bibinfo {author} {\bibfnamefont {P.}~\bibnamefont {Werner}},\ }\href
  {\doibase 10.1209/0295-5075/120/57001} {\bibfield  {journal} {\bibinfo
  {journal} {EPL(Europhys. Lett.)}\ }\textbf {\bibinfo {volume} {120}},\
  \bibinfo {pages} {57001} (\bibinfo {year} {2017})}\BibitemShut {NoStop}%
\bibitem [{\citenamefont {Ikeda}\ and\ \citenamefont
  {Sato}(2020)}]{Tatsuhiko2020}%
  \BibitemOpen
  \bibfield  {author} {\bibinfo {author} {\bibfnamefont {T.~N.}\ \bibnamefont
  {Ikeda}}\ and\ \bibinfo {author} {\bibfnamefont {M.}~\bibnamefont {Sato}},\
  }\href {\doibase 10.1126/sciadv.abb4019} {\bibfield  {journal} {\bibinfo
  {journal} {Sci. Adv.}\ }\textbf {\bibinfo {volume} {6}},\ \bibinfo {pages}
  {eabb4019} (\bibinfo {year} {2020})}\BibitemShut {NoStop}%
\bibitem [{\citenamefont {Ikeda}\ and\ \citenamefont
  {Polkovnikov}(2021)}]{PhysRevB.104.134308}%
  \BibitemOpen
  \bibfield  {author} {\bibinfo {author} {\bibfnamefont {T.~N.}\ \bibnamefont
  {Ikeda}}\ and\ \bibinfo {author} {\bibfnamefont {A.}~\bibnamefont
  {Polkovnikov}},\ }\href {\doibase 10.1103/PhysRevB.104.134308} {\bibfield
  {journal} {\bibinfo  {journal} {Phys. Rev. B}\ }\textbf {\bibinfo {volume}
  {104}},\ \bibinfo {pages} {134308} (\bibinfo {year} {2021})}\BibitemShut
  {NoStop}%
\bibitem [{\citenamefont {Bhattacharyya}\ \emph {et~al.}(2021)\citenamefont
  {Bhattacharyya}, \citenamefont {Bhowmik}, \citenamefont {Adroja},
  \citenamefont {Rahaman}, \citenamefont {Kar}, \citenamefont {Das},
  \citenamefont {Saha-Dasgupta}, \citenamefont {Biswas}, \citenamefont {Sinha},
  \citenamefont {Ewings}, \citenamefont {Khalyavin},\ and\ \citenamefont
  {Strydom}}]{PhysRevB.103.174423}%
  \BibitemOpen
  \bibfield  {author} {\bibinfo {author} {\bibfnamefont {A.}~\bibnamefont
  {Bhattacharyya}}, \bibinfo {author} {\bibfnamefont {T.~K.}\ \bibnamefont
  {Bhowmik}}, \bibinfo {author} {\bibfnamefont {D.~T.}\ \bibnamefont {Adroja}},
  \bibinfo {author} {\bibfnamefont {B.}~\bibnamefont {Rahaman}}, \bibinfo
  {author} {\bibfnamefont {S.}~\bibnamefont {Kar}}, \bibinfo {author}
  {\bibfnamefont {S.}~\bibnamefont {Das}}, \bibinfo {author} {\bibfnamefont
  {T.}~\bibnamefont {Saha-Dasgupta}}, \bibinfo {author} {\bibfnamefont {P.~K.}\
  \bibnamefont {Biswas}}, \bibinfo {author} {\bibfnamefont {T.~P.}\
  \bibnamefont {Sinha}}, \bibinfo {author} {\bibfnamefont {R.~A.}\ \bibnamefont
  {Ewings}}, \bibinfo {author} {\bibfnamefont {D.~D.}\ \bibnamefont
  {Khalyavin}}, \ and\ \bibinfo {author} {\bibfnamefont {A.~M.}\ \bibnamefont
  {Strydom}},\ }\href {\doibase 10.1103/PhysRevB.103.174423} {\bibfield
  {journal} {\bibinfo  {journal} {Phys. Rev. B}\ }\textbf {\bibinfo {volume}
  {103}},\ \bibinfo {pages} {174423} (\bibinfo {year} {2021})}\BibitemShut
  {NoStop}%
\bibitem [{\citenamefont {Shan}\ \emph {et~al.}(2021)\citenamefont {Shan},
  \citenamefont {Ye}, \citenamefont {Chu}, \citenamefont {Lee}, \citenamefont
  {Park}, \citenamefont {Balents},\ and\ \citenamefont {Hsieh}}]{Shan2021}%
  \BibitemOpen
  \bibfield  {author} {\bibinfo {author} {\bibfnamefont {J.-Y.}\ \bibnamefont
  {Shan}}, \bibinfo {author} {\bibfnamefont {M.}~\bibnamefont {Ye}}, \bibinfo
  {author} {\bibfnamefont {H.}~\bibnamefont {Chu}}, \bibinfo {author}
  {\bibfnamefont {S.}~\bibnamefont {Lee}}, \bibinfo {author} {\bibfnamefont
  {J.-G.}\ \bibnamefont {Park}}, \bibinfo {author} {\bibfnamefont
  {L.}~\bibnamefont {Balents}}, \ and\ \bibinfo {author} {\bibfnamefont
  {D.}~\bibnamefont {Hsieh}},\ }\href {\doibase 10.1038/s41586-021-04051-8}
  {\bibfield  {journal} {\bibinfo  {journal} {Nature}\ }\textbf {\bibinfo
  {volume} {600}},\ \bibinfo {pages} {235} (\bibinfo {year}
  {2021})}\BibitemShut {NoStop}%
\end{thebibliography}%


\begin{thebibliography}{12}%
\makeatletter
\providecommand \@ifxundefined [1]{%
 \@ifx{#1\undefined}
}%
\providecommand \@ifnum [1]{%
 \ifnum #1\expandafter \@firstoftwo
 \else \expandafter \@secondoftwo
 \fi
}%
\providecommand \@ifx [1]{%
 \ifx #1\expandafter \@firstoftwo
 \else \expandafter \@secondoftwo
 \fi
}%
\providecommand \natexlab [1]{#1}%
\providecommand \enquote  [1]{``#1''}%
\providecommand \bibnamefont  [1]{#1}%
\providecommand \bibfnamefont [1]{#1}%
\providecommand \citenamefont [1]{#1}%
\providecommand \href@noop [0]{\@secondoftwo}%
\providecommand \href [0]{\begingroup \@sanitize@url \@href}%
\providecommand \@href[1]{\@@startlink{#1}\@@href}%
\providecommand \@@href[1]{\endgroup#1\@@endlink}%
\providecommand \@sanitize@url [0]{\catcode `\\12\catcode `\$12\catcode
  `\&12\catcode `\#12\catcode `\^12\catcode `\_12\catcode `\%12\relax}%
\providecommand \@@startlink[1]{}%
\providecommand \@@endlink[0]{}%
\providecommand \url  [0]{\begingroup\@sanitize@url \@url }%
\providecommand \@url [1]{\endgroup\@href {#1}{\urlprefix }}%
\providecommand \urlprefix  [0]{URL }%
\providecommand \Eprint [0]{\href }%
\providecommand \doibase [0]{http://dx.doi.org/}%
\providecommand \selectlanguage [0]{\@gobble}%
\providecommand \bibinfo  [0]{\@secondoftwo}%
\providecommand \bibfield  [0]{\@secondoftwo}%
\providecommand \translation [1]{[#1]}%
\providecommand \BibitemOpen [0]{}%
\providecommand \bibitemStop [0]{}%
\providecommand \bibitemNoStop [0]{.\EOS\space}%
\providecommand \EOS [0]{\spacefactor3000\relax}%
\providecommand \BibitemShut  [1]{\csname bibitem#1\endcsname}%
\let\auto@bib@innerbib\@empty
\bibitem [{\citenamefont {Schrieffer}\ and\ \citenamefont
  {Wolff}(1966)}]{PhysRev.149.491}%
  \BibitemOpen
  \bibfield  {author} {\bibinfo {author} {\bibfnamefont {J.~R.}\ \bibnamefont
  {Schrieffer}}\ and\ \bibinfo {author} {\bibfnamefont {P.~A.}\ \bibnamefont
  {Wolff}},\ }\href {\doibase 10.1103/PhysRev.149.491} {\bibfield  {journal}
  {\bibinfo  {journal} {Phys. Rev.}\ }\textbf {\bibinfo {volume} {149}},\
  \bibinfo {pages} {491} (\bibinfo {year} {1966})}\BibitemShut {NoStop}%
\bibitem [{\citenamefont {Kumar}\ and\ \citenamefont
  {Lin}(2021)}]{PhysRevB.103.064508}%
  \BibitemOpen
  \bibfield  {author} {\bibinfo {author} {\bibfnamefont {U.}~\bibnamefont
  {Kumar}}\ and\ \bibinfo {author} {\bibfnamefont {S.-Z.}\ \bibnamefont
  {Lin}},\ }\href {\doibase 10.1103/PhysRevB.103.064508} {\bibfield  {journal}
  {\bibinfo  {journal} {Phys. Rev. B}\ }\textbf {\bibinfo {volume} {103}},\
  \bibinfo {pages} {064508} (\bibinfo {year} {2021})}\BibitemShut {NoStop}%
\bibitem [{\citenamefont {Boström}\ \emph {et~al.}(2020)\citenamefont
  {Boström}, \citenamefont {Rubio},\ and\ \citenamefont
  {Verdozzi}}]{Rubio2021}%
  \BibitemOpen
  \bibfield  {author} {\bibinfo {author} {\bibfnamefont {E.~V.}\ \bibnamefont
  {Boström}}, \bibinfo {author} {\bibfnamefont {A.}~\bibnamefont {Rubio}}, \
  and\ \bibinfo {author} {\bibfnamefont {C.}~\bibnamefont {Verdozzi}},\
  }\href@noop {} {\  (\bibinfo {year} {2020})},\ \Eprint
  {http://arxiv.org/abs/arXiv:2010.16125} {arXiv:2010.16125} \BibitemShut
  {NoStop}%
\bibitem [{\citenamefont {Mentink}\ \emph {et~al.}(2015)\citenamefont
  {Mentink}, \citenamefont {Balzer},\ and\ \citenamefont
  {Eckstein}}]{Mentink2015}%
  \BibitemOpen
  \bibfield  {author} {\bibinfo {author} {\bibfnamefont {J.~H.}\ \bibnamefont
  {Mentink}}, \bibinfo {author} {\bibfnamefont {K.}~\bibnamefont {Balzer}}, \
  and\ \bibinfo {author} {\bibfnamefont {M.}~\bibnamefont {Eckstein}},\ }\href
  {\doibase 10.1038/ncomms7708} {\bibfield  {journal} {\bibinfo  {journal}
  {Nat. Commun.}\ }\textbf {\bibinfo {volume} {6}},\ \bibinfo {pages} {6708}
  (\bibinfo {year} {2015})}\BibitemShut {NoStop}%
\bibitem [{\citenamefont {Arakawa}(2016)}]{PhysRevB.94.174416}%
  \BibitemOpen
  \bibfield  {author} {\bibinfo {author} {\bibfnamefont {N.}~\bibnamefont
  {Arakawa}},\ }\href {\doibase 10.1103/PhysRevB.94.174416} {\bibfield
  {journal} {\bibinfo  {journal} {Phys. Rev. B}\ }\textbf {\bibinfo {volume}
  {94}},\ \bibinfo {pages} {174416} (\bibinfo {year} {2016})}\BibitemShut
  {NoStop}%
\bibitem [{\citenamefont {Ishihara}\ \emph {et~al.}(2002)\citenamefont
  {Ishihara}, \citenamefont {Hatakeyama},\ and\ \citenamefont
  {Maekawa}}]{PhysRevB.65.064442}%
  \BibitemOpen
  \bibfield  {author} {\bibinfo {author} {\bibfnamefont {S.}~\bibnamefont
  {Ishihara}}, \bibinfo {author} {\bibfnamefont {T.}~\bibnamefont
  {Hatakeyama}}, \ and\ \bibinfo {author} {\bibfnamefont {S.}~\bibnamefont
  {Maekawa}},\ }\href {\doibase 10.1103/PhysRevB.65.064442} {\bibfield
  {journal} {\bibinfo  {journal} {Phys. Rev. B}\ }\textbf {\bibinfo {volume}
  {65}},\ \bibinfo {pages} {064442} (\bibinfo {year} {2002})}\BibitemShut
  {NoStop}%
\bibitem [{\citenamefont {Arakawa}\ and\ \citenamefont
  {Yonemitsu}(2021)}]{PhysRevB.103.L100408}%
  \BibitemOpen
  \bibfield  {author} {\bibinfo {author} {\bibfnamefont {N.}~\bibnamefont
  {Arakawa}}\ and\ \bibinfo {author} {\bibfnamefont {K.}~\bibnamefont
  {Yonemitsu}},\ }\href {\doibase 10.1103/PhysRevB.103.L100408} {\bibfield
  {journal} {\bibinfo  {journal} {Phys. Rev. B}\ }\textbf {\bibinfo {volume}
  {103}},\ \bibinfo {pages} {L100408} (\bibinfo {year} {2021})}\BibitemShut
  {NoStop}%
\bibitem [{\citenamefont {Wright}\ and\ \citenamefont
  {Shastry}(2013)}]{Shastry2013}%
  \BibitemOpen
  \bibfield  {author} {\bibinfo {author} {\bibfnamefont {J.~G.}\ \bibnamefont
  {Wright}}\ and\ \bibinfo {author} {\bibfnamefont {B.~S.}\ \bibnamefont
  {Shastry}},\ }\href@noop {} {\  (\bibinfo {year} {2013})},\ \Eprint
  {http://arxiv.org/abs/arXiv:1301.4494} {arXiv:1301.4494} \BibitemShut
  {NoStop}%
\bibitem [{\citenamefont {Rau}\ \emph {et~al.}(2014)\citenamefont {Rau},
  \citenamefont {Lee},\ and\ \citenamefont {Kee}}]{PhysRevLett.112.077204}%
  \BibitemOpen
  \bibfield  {author} {\bibinfo {author} {\bibfnamefont {J.~G.}\ \bibnamefont
  {Rau}}, \bibinfo {author} {\bibfnamefont {E.~K.-H.}\ \bibnamefont {Lee}}, \
  and\ \bibinfo {author} {\bibfnamefont {H.-Y.}\ \bibnamefont {Kee}},\ }\href
  {\doibase 10.1103/PhysRevLett.112.077204} {\bibfield  {journal} {\bibinfo
  {journal} {Phys. Rev. Lett.}\ }\textbf {\bibinfo {volume} {112}},\ \bibinfo
  {pages} {077204} (\bibinfo {year} {2014})}\BibitemShut {NoStop}%
\bibitem [{\citenamefont {Kumar}\ \emph {et~al.}(2021)\citenamefont {Kumar},
  \citenamefont {Banerjee},\ and\ \citenamefont {Lin}}]{Kumar2021}%
  \BibitemOpen
  \bibfield  {author} {\bibinfo {author} {\bibfnamefont {U.}~\bibnamefont
  {Kumar}}, \bibinfo {author} {\bibfnamefont {S.}~\bibnamefont {Banerjee}}, \
  and\ \bibinfo {author} {\bibfnamefont {S.-Z.}\ \bibnamefont {Lin}},\
  }\href@noop {} {\  (\bibinfo {year} {2021})},\ \Eprint
  {http://arxiv.org/abs/arXiv:2111.01316} {arXiv:2111.01316} \BibitemShut
  {NoStop}%
\bibitem [{\citenamefont {Jackeli}\ and\ \citenamefont
  {Khaliullin}(2009)}]{PhysRevLett.102.017205}%
  \BibitemOpen
  \bibfield  {author} {\bibinfo {author} {\bibfnamefont {G.}~\bibnamefont
  {Jackeli}}\ and\ \bibinfo {author} {\bibfnamefont {G.}~\bibnamefont
  {Khaliullin}},\ }\href {\doibase 10.1103/PhysRevLett.102.017205} {\bibfield
  {journal} {\bibinfo  {journal} {Phys. Rev. Lett.}\ }\textbf {\bibinfo
  {volume} {102}},\ \bibinfo {pages} {017205} (\bibinfo {year}
  {2009})}\BibitemShut {NoStop}%
\bibitem [{\citenamefont {Eckardt}\ and\ \citenamefont
  {Anisimovas}(2015)}]{Eckardt_2015}%
  \BibitemOpen
  \bibfield  {author} {\bibinfo {author} {\bibfnamefont {A.}~\bibnamefont
  {Eckardt}}\ and\ \bibinfo {author} {\bibfnamefont {E.}~\bibnamefont
  {Anisimovas}},\ }\href {\doibase 10.1088/1367-2630/17/9/093039} {\bibfield
  {journal} {\bibinfo  {journal} {New J. Phys.}\ }\textbf {\bibinfo {volume}
  {17}},\ \bibinfo {pages} {093039} (\bibinfo {year} {2015})}\BibitemShut
  {NoStop}%
\end{thebibliography}%
\end{document}


\title{Supplementary Material: \--- \\ Inverse Faraday effect in Mott insulators}

\author{Saikat Banerjee}
\email{saikatb@lanl.gov}
\affiliation{Theoretical Division, T-4, Los Alamos National Laboratory, Los Alamos, New Mexico 87545, USA}
\author{Umesh Kumar}
\affiliation{Theoretical Division, T-4, Los Alamos National Laboratory, Los Alamos, New Mexico 87545, USA}
\author{Shi-Zeng Lin} 
\email{szl@lanl.gov}
\affiliation{Theoretical Division, T-4 and CNLS, Los Alamos National Laboratory, Los Alamos, New Mexico 87545, USA}

\maketitle
\tableofcontents 
\clearpage

\section{Schrieffer-Wolff Transformation \label{sec:1}}

In this section, we provide the details of the perturbative Schrieffer-Wolff transformation (SWT)~\cite{PhysRev.149.491}  for time-dependent Hamiltonians. Let us consider a model Hamiltonian for a strongly correlated system as 
\begin{equation}\label{eq.1}
\mathcal{H} = \mathcal{H}_0 + \mathcal{H}_1,
\end{equation}
where \SZL{$\mathcal{H}_0$ contains strong Coulomb interaction and is diagonal in the occupation basis, and  $\mathcal{H}_1$ describes the weak hopping between different sites.}  Often, we are interested in the physics within a particular low-energy window, and consequently the full Hamiltonian $\mathcal{H}$ is not needed. SWT is therefore used to derive the low-energy effective model from the full Hamiltonian. We consider a unitary operator $U = e^{-i\vS}$ and rewrite the Hamiltonian in the rotated frame as
\begin{align}\label{eq.2}
\mathcal{H}' & = U^{\dagger} \mathcal{H} U = e^{i\vS} \mathcal{H}e^{-i\vS} \nonumber \\
& = \mathcal{H} +  i\big[\vS, \mathcal{H} \big] - \frac{1}{2!} \big[\vS, \big[ \vS,\mathcal{H} \big]\big] - \frac{i}{3!} \big[\vS, \big[\vS, \big[\vS, \mathcal{H} \big] \big] \big] + \cdots \nonumber \\
& = \mathcal{H}_0 + \mathcal{H}_1 + i\big[\vS, \mathcal{H}_0] + i\big[\vS, \mathcal{H}_1] - \frac{1}{2}\big[\vS, \big[ \vS,\mathcal{H}_0 \big]\big] - \frac{1}{2} \big[\vS, \big[ \vS,\mathcal{H}_1 \big]\big] + \cdots.
\end{align}
The trick is now to choose a suitable hermitian operator $\vS$ such that $i\big[ \vS,\mathcal{H}_0 \big] = -\mathcal{H}_1$. Eventually, the respective cancellations lead to an effective Hamiltonian upto second order as
\begin{equation}\label{eq.3}
\mathcal{H}_{\mathrm{eff}} = \mathcal{H}_0 + \frac{i}{2} \big[\vS,\mathcal{H}_1 \big].
\end{equation}
In case of time-dependent Hamiltonian [$\mathcal{H}(t) = \mathcal{H}_0 + \mathcal{H}_1(t)$], the analogous unitary transformation becomes time-dependent. We first consider the transformation of the wave-function as $\ket{\psi'} = e^{i\vS(t)} \ket{\psi}$ . We argue that the form of the Schr\"odinger equation remains invariant in the rotated coordinate frame, i.e. (in the unit of $\hbar = 1$)
\begin{align}\label{eq.4}
\mathcal{H} \ket{\psi} = i \partial_t \ket{\psi} \Leftrightarrow  \mathcal{H}' \ket{\psi'} & = i \partial_t \ket{\psi'} = i \partial_t \left(e^{i\vS(t)} \ket{\psi} \right)\nonumber \\
& = i\partial_t  e^{i\vS(t)} \ket{\psi} + i e^{i\vS(t)} \partial_t \ket{\psi}  \nonumber \\
& = e^{i\vS(t)} \mathcal{H}\ket{\psi} + i \partial_t e^{i\vS(t)} \ket{\psi},
\end{align}
hence, the Hamiltonian in the rotated frame transforms as
\begin{equation}\label{eq.5}
\mathcal{H}' = e^{i\vS(t)} \mathcal{H} e^{-i\vS(t)} - i \partial_t \vS(t) e^{i\vS(t)}.
\end{equation}

\subsection{Low-energy effective Hamiltonian \label{subsec:1}}

We expand Eq.~(\ref{eq.5}) in Taylor series as
\begin{align}\label{eq.6}
\mathcal{H}'(t) & = e^{i\vS(t)} \mathcal{H}(t)e^{-i\vS(t)} - e^{i\vS(t)} i\partial_t e^{-i\vS(t)}  = \mathcal{H}(t) + \big[i\vS(t), \mathcal{H}(t) - i\partial_t \big] + \frac{1}{2}\big[ i\vS(t), \big[i\vS(t), \mathcal{H}(t)-i\partial_t \big] \big] + \cdots 
\end{align}
The generator $S(t)$ is thereafter written as a perturbative expansion in the small parameters of the non-diagonal Hamiltonian $\mathcal{H}_1(t)$ as $\vS(t) = \vS^{(1)}(t) + \vS^{(2)}(t) + \vS^{(3)}(t) + \cdots$. Rewriting Eq.~(\ref{eq.6}) in each order of the generator $\vS(t)$ we obtain the effective Hamiltonian perturbatively. Right hand side of Eq.~(\ref{eq.6}) is written (upto third order) as follows 
\begin{subequations}
\begin{align}
\label{eq.7.1}
\mathrm{\text{First order}:} \quad & \mathcal{H}_1(t) + i \big[ \vS^{(1)}(t), \mathcal{H}_0 \big] - \partial_t \vS^{(1)}(t), \\
\label{eq.7.2}
\mathrm{\text{Second order}:} \quad & i \big[ \vS^{(1)}(t), \mathcal{H}_1(t) \big] + i\big[ \vS^{(2)}(t), \mathcal{H}_0 \big] - \frac{1}{2} \big[\vS^{(1)}(t), i\partial_t \vS^{(1)}(t) + \big[ \vS^{(1)}(t),\mathcal{H}_0 \big]\big] - \partial_t \vS^{(2)}(t),  \\
\label{eq.7.3}
\mathrm{\text{Third order}:} \quad & i \big[ \vS^{(3)}(t), \mathcal{H}_0 \big] + i\big[ \vS^{(2)}(t), \mathcal{H}_1(t)\big] - \frac{1}{2} \big[\vS^{(1)}(t),  i\partial_t \vS^{(2)}(t) + \big[ \vS^{(1)}(t),\mathcal{H}_1(t) \big] + \big[ \vS^{(2)}(t),\mathcal{H}_0 \big]\big] -  \frac{1}{2} \big[\vS^{(2)}(t), \nonumber \\
& i\partial_t \vS^{(1)}(t) + \big[ \vS^{(1)}(t),\mathcal{H}_0 \big]\big] - \frac{i}{3!} \big[ \vS^{(1)}(t), \big[ \vS^{(1)}(t), i\partial_t \vS^{(1)}(t) +\big[ \vS^{(1)}(t), \mathcal{H}_0 \big]\big]\big] -\partial_t \vS^{(3)}(t).
\end{align}
\end{subequations}
We now diagonalize Eqs.~(\ref{eq.7.1})-(\ref{eq.7.3}) in each order by introducing two projection operators $\mathcal{P}$ and $\mathcal{Q}$ = $(1 - \mathcal{P})$, which project a generic operator $A$ into the low- and the high-energy Hilbert spaces, respectively. The transition matrix elements $A_{pq}$ are defined as $\mathcal{A}_{pq} = \mathcal{P}\mathcal{A}\mathcal{Q}$. The Hamiltonian in Eq.~(\ref{eq.6}) can be recasted as
\begin{equation}\label{eq.8}
\mathcal{H}'(t) = \sum_{m=0}^n \mathcal{H}_{\mathrm{eff}}^{(m)}(t) + \mathcal{O}(n+1),
\end{equation}
where $m$ corresponds to the each order in the perturbative expansion. The final goal is to find out the generating functions $\vS^{(m)}(t)$, so that the effective Hamiltonian of a certain order $m$, $\mathcal{H}_{\mathrm{eff}}^{(m)}(t)$ will not have the mixing terms, i.e. $\mathcal{P} \mathcal{H}^{(m)}_{\mathrm{eff}}(t) \mathcal{Q} = \mathcal{Q} \mathcal{H}^{(m)}_{\mathrm{eff}}(t) \mathcal{P} = 0$.

\subsection{Generating functions: First and second order \label{subsec:2}}

To find the generating functions in each order which satisfies the conditions as mentioned above, we read-off the dynamical equations from Eqs.~(\ref{eq.7.1})-(\ref{eq.7.2}) as [\SB{note that the second last term in Eq.~(\ref{eq.7.2}) is already recasted as low-energy effective Hamiltonian when we utilize the equation of motion for the generating function $\vS^{(1)}(t)$}]
\begin{subequations}
\begin{align}
\label{eq.9.1}
\partial_t \vS^{(1)}(t) & =  i \big[ \vS^{(1)}(t), \mathcal{H}_0 \big] + \mathcal{H}_1(t),\\
\label{eq.9.2}
\partial_t \vS^{(2)}(t) & = i\big[ \vS^{(2)}(t), \mathcal{H}_0 \big] + i \big[ \vS^{(1)}(t), \mathcal{H}_1(t) \big].
\end{align}
\end{subequations}
To solve the Liouville-like equations in Eq.~(\ref{eq.9.1})-(\ref{eq.9.2}), we introduce the retarded and the advanced Green's function as 
\begin{subequations}
\begin{align}
\label{eq.10.1}
\mathcal{G}^{\mathrm{R}}(t,t') & = -ie^{-i(\mathcal{H}_0-i\eta)(t-t')} \theta(t-t') \\ 
\label{eq.10.2}
\mathcal{G}^{\mathrm{A}}(t,t') & = ie^{i(\mathcal{H}_0+i\eta)(t'-t)} \theta(t'-t),
\end{align}
\end{subequations}
where $\theta(t-t')$ is a Heaviside step function. It is important to notice that in the basis of $\mathcal{P}$ and $\mathcal{Q}$, $S^{(1)}(t)$ is a $2 \times 2$ matrix written as
\begin{equation}\label{eq.11}
\vS^{(1)}(t) = \begin{pmatrix}
\mathcal{P} \vS^{(1)}(t) \mathcal{P} & \mathcal{P} \vS^{(1)}(t) \mathcal{Q} \\
\mathcal{Q} \vS^{(1)}(t) \mathcal{P} & \mathcal{Q} \vS^{(1)}(t) \mathcal{Q}
\end{pmatrix}.
\end{equation}
The diagonal terms of the matrix in Eq.~(\ref{eq.11}) can be chosen to be zero. Furthermore since $\vS^{(1)}(t)$ is hermitian, we have $\big[ \mathcal{P} \vS^{(1)}(t) \mathcal{Q} \big]^{\dagger} = \mathcal{Q} \vS^{(1)}(t) \mathcal{P}$. Writing down the dynamical equations of $\vS^{(1)}(t)$ projected into the individual energy subspaces we obtain (assuming $\mathcal{H}_0$ is diagonal in $\mathcal{Q}$ space)
\begin{subequations}
\begin{align}
\partial_t \mathcal{P} \vS^{(1)}(t) \mathcal{Q} & = i \mathcal{P} \vS^{(1)}(t) \mathcal{H}_0\mathcal{Q} -i \mathcal{P} \mathcal{H}_0 \vS^{(1)}(t)\mathcal{Q} + \mathcal{P}\mathcal{H}_1(t) \mathcal{Q} \nonumber \\
\label{eq.12.1}
& = i \mathcal{P} \vS^{(1)}(t) \mathcal{Q} \mathcal{Q} \mathcal{H}_0\mathcal{Q} - \cancel{i \mathcal{P} \mathcal{H}_0 \mathcal{Q} \mathcal{Q} \vS^{(1)}(t)\mathcal{Q} } + \mathcal{P}\mathcal{H}_1(t) \mathcal{Q}  = i \mathcal{P} \vS^{(1)}(t) \mathcal{Q} \mathcal{Q} \mathcal{H}_0\mathcal{Q} + \mathcal{P}\mathcal{H}_1(t) \mathcal{Q}, \\
\partial_t \mathcal{Q} \vS^{(1)}(t) \mathcal{P} & = i \mathcal{Q} \vS^{(1)}(t) \mathcal{H}_0\mathcal{P} -i \mathcal{Q} \mathcal{H}_0 \vS^{(1)}(t)\mathcal{P} + \mathcal{Q}\mathcal{H}_1(t) \mathcal{P} \nonumber \\
\label{eq.12.2}
& = \cancel{i \mathcal{Q} \vS^{(1)}(t) \mathcal{Q} \mathcal{Q} \mathcal{H}_0\mathcal{P}} - i \mathcal{Q} \mathcal{H}_0 \mathcal{Q} \mathcal{Q} \vS^{(1)}(t)\mathcal{P} + \mathcal{Q}\mathcal{H}_1(t) \mathcal{P} = - i \mathcal{Q} \mathcal{H}_0 \mathcal{Q} \mathcal{Q} \vS^{(1)}(t)\mathcal{P} + \mathcal{Q}\mathcal{H}_1(t) \mathcal{P}.
\end{align}
\end{subequations}
From Eqs.~(\ref{eq.12.1})-(\ref{eq.12.1}) and using the Green's functions, we obtain the solution of the projected operators as 
\begin{subequations}
\begin{align}
\label{eq.13.1}
\mathcal{P} \vS^{(1)}(t)\mathcal{Q} & = -i\int dt' \mathcal{P}\mathcal{H}_1(t')\mathcal{Q}\mathcal{G}^{\mathrm{A}}(t',t) = \int dt' \theta(t-t')  \mathcal{P}\mathcal{H}_1(t')\mathcal{Q}e^{i(\mathcal{H}_0+i\eta)(t-t')}, \\ 
\label{eq.13.2}
\mathcal{Q} \vS^{(1)}(t)\mathcal{P} & = i\int dt' \mathcal{G}^{\mathrm{R}}(t,t')\mathcal{Q}\mathcal{H}_1(t')\mathcal{P} = \int dt' \theta(t-t')  e^{-i(\mathcal{H}_0-i\eta)(t-t')}\mathcal{Q}\mathcal{H}_1(t')\mathcal{P}.
\end{align}
\end{subequations}
Utilizing Eqs.~(\ref{eq.13.1})-(\ref{eq.13.2}) in Eq.~(\ref{eq.7.2}), the effective low-energy Hamiltonian is evaluated upto the second-order as 
\begin{equation}\label{eq.14}
\mathcal{H}^{(2)}_{\mathrm{eff}}(t) = \frac{i}{2} \mathcal{P}\big[ \vS^{(1)}(t),\mathcal{H}_1(t) \big]\mathcal{P}  = \frac{i}{2} \Big[  \mathcal{P} \vS^{(1)}(t)\mathcal{Q}\mathcal{Q}\mathcal{H}_1(t) \mathcal{P} -\mathcal{P} \mathcal{H}_1(t) \mathcal{Q}\mathcal{Q} \vS^{(1)}(t) \mathcal{P}\Big].
\end{equation} 
In the similar manner, we obtain the solution of the generating function $\vS^{(2)}(t)$ from Eq.~(\ref{eq.9.2}) as 
\begin{equation}\label{eq.15}
\mathcal{P}\vS^{(2)}(t)\mathcal{Q}  = \int dt' \mathcal{P} \left(\big[ \vS^{(1)}(t'), \mathcal{H}_1(t')\big] \right)\mathcal{Q}\mathcal{G}^{\mathrm{A}}(t',t), \;\;  \mathcal{Q}\vS^{(2)}(t)\mathcal{P}  = -\int dt' \mathcal{G}^{\mathrm{R}}(t,t')\mathcal{Q}\left(\big[ \vS^{(1)}(t'), \mathcal{H}_1(t')\big] \right)\mathcal{P}.
\end{equation}
With the above Eq.~(\ref{eq.15}), we finally obtain the effective low-energy Hamiltonian upto third-order as
\begin{equation}\label{eq.16}
\mathcal{H}^{(3)}_{\mathrm{eff}}(t) = \frac{i}{2} \big[ \vS^{(2)}(t),\mathcal{H}_1(t) \big] + \frac{1}{6} \big[ \vS^{(1)}(t),\big[ \vS^{(1)}(t), \mathcal{H}_1(t) \big] \big].
\end{equation}

\section{Effective Hamiltonian: Single-orbital case \label{sec:2}}	

In this section, we provide the details of the derivation of Eq.~(3) in the main text. Then, we evaluate the low-energy (third-order) effective spin-exchange Hamiltonian in the presence of spin-orbit coupling (SOC). In the second-order perturbation only $d$-orbital hopping contributes. We identify the Hamiltonians $\mathcal{H}_0$ and $\mathcal{H}_1(t)$ as 
\begin{equation}\label{eq.17}
\mathcal{H}_0 = U \sum_{i}n^d_{i\uparrow} n^d_{i\downarrow}, \quad \mathcal{H}_1(t) = t_{dd}\sum_{\langle ij \rangle, \sigma} e^{i\phi_{ij}(t)} d^{\dagger}_{i\sigma}d_{j\sigma}.
\end{equation}
Using Eqs.~(\ref{eq.13.1})-(\ref{eq.13.2}) we obtain the low-energy effective Hamiltonian in the second-order as
\begin{align}\label{eq.18}
\mathcal{H}^{(2)}_{\mathrm{eff}}(t) & = \frac{i}{2} \int_{-\infty}^{t} dt' \left[ \mathcal{P} \mathcal{H}_1(t') \mathcal{Q} e^{i(\mathcal{H}_0+i\eta)(t-t')} \mathcal{Q} \mathcal{H}_1(t) \mathcal{P} - \mathcal{P} \mathcal{H}_1(t) \mathcal{Q} e^{-i(\mathcal{H}_0-i\eta)(t-t')} \mathcal{Q} \mathcal{H}_1(t') \mathcal{P} \right] \nonumber \\
& = \frac{i}{2}\int_{-\infty}^{t} dt' \left( e^{i(U+i\eta)(t-t')} \mathcal{P} \mathcal{H}_1(t') \mathcal{Q} \mathcal{Q} \mathcal{H}_1(t) \mathcal{P} - e^{-i(U-i\eta)(t-t')} \mathcal{P} \mathcal{H}_1(t) \mathcal{Q} \mathcal{Q} \mathcal{H}_1(t') \mathcal{P} \right) \nonumber \\
& = \frac{it^2_{dd}}{2} \sum_{\langle ij \rangle}\int_{-\infty}^{t} dt' \left( e^{i(U+i\eta)(t-t')} e^{i\phi_{ij}(t')} d^{\dagger}_{i\sigma}d_{j\sigma} e^{i\phi_{ji}(t)} d^{\dagger}_{j\sigma'}d_{i\sigma'} - e^{-i(U-i\eta)(t-t')} e^{i\phi_{ij}(t)} d^{\dagger}_{i\sigma}d_{j\sigma} e^{i\phi_{ji}(t')} d^{\dagger}_{j\sigma'}d_{i\sigma'} \right) \nonumber \\
& = \frac{it^2_{dd}}{2} \sum_{\langle ij \rangle}\int_{-\infty}^{t} dt' \left( e^{i(U+i\eta)(t-t')} e^{i\phi_{ij}(t') - i\phi_{ij}(t)} - e^{-i(U-i\eta)(t-t')} e^{i\phi_{ij}(t) - i\phi_{ij}(t')} \right) d^{\dagger}_{i\sigma}d_{j\sigma} d^{\dagger}_{j\sigma'}d_{i\sigma'},
\end{align}
where we used the fact that $\phi_{ij}(t) = - \phi_{ji}(t)$. The operator product $d^{\dagger}_{i\sigma}d_{j\sigma} d^{\dagger}_{j\sigma'}d_{i\sigma'}$ can be simplified in terms of the spin operators as~\cite{PhysRevB.103.064508,Rubio2021} 
\begin{align}\label{eq.19}
d^{\dagger}_{i\sigma}d_{j\sigma} d^{\dagger}_{j\sigma'}d_{i\sigma'} = - 2 \left( \mathbf{S}_i \cdot \mathbf{S}_j  - \frac{1}{4} \right),
\end{align}
where we used the fact that $\mathrm{Tr} \, \mathbb{I}_2 = 2$ and $\mathrm{Tr}\, \bm{\tau} = 0$, $\bm{\tau}$ being the vector of the Pauli matrices. The electronic bilnears $d^{\dagger}_{i\sigma}d_{i\sigma'}$ are rewritten in terms of the spin variables as $d^{\dagger}_{i\sigma}d_{i\sigma'} = \left(\mathbb{I}_2/2 + \mathbf{S}_i\cdot \bm{\tau} \right)_{\sigma'\sigma}$. Ignoring the constant term in Eq.~(\ref{eq.19}), the effective Hamiltonian in Eq.~(\ref{eq.18}) becomes
\begin{align}\label{eq.20}
\mathcal{H}^{(2)}_{\mathrm{eff}}(t) & = 2t^2_{dd} \sum_{\langle ij \rangle} \mathrm{Im}\;\left( \int_{-\infty}^{t} dt' e^{i(U+i\eta)(t-t')}  e^{i(\phi_{ij}(t')- \phi_{ij}(t))} \right)  \mathbf{S}_i \cdot \mathbf{S}_j = \sum_{\langle ij \rangle} J_{ij}^{(2)}(t) \mathbf{S}_i \cdot \mathbf{S}_j,
\end{align}
where, we defined a new variable $J^{(2)}_{ij}(t)$ as $2t^2_{dd}\mathrm{Im}\;\left( \int_{-\infty}^{t} dt' e^{i(U+i\eta)(t-t')}  e^{i(\phi_{ij}(t')- \phi_{ij}(t))} \right)$. Assuming the nearest-neighbor vector $\dr_{dd}$ oriented along $\hat{x}$-axis [see Fig.~1(b) in the main text], the Peierls phase $\phi_{ij}(t)$ is simplified as $\phi_{ij}(t) = -r_{dd}E_0\sin\Omega t/\Omega$. Rewriting $e^{i\phi_{ij}(t)} = \sum_{n=-\infty}^{\infty} \mathcal{J}_n(r_{dd} E_0/\Omega) e^{in\Omega t}$ and evaluating the integral in Eq.~(\ref{eq.20}), we finally obtain
\begin{equation}\label{eq.21}
J^{(2)}_{ij}(t) = 4 t^2_{dd} \sum_{n,m = -\infty}^{\infty} \mathcal{J}_n(r_{dd}E_0/\Omega) \mathcal{J}_m(r_{dd}E_0/\Omega) \mathrm{Im} \Bigg[ i  \frac{e^{i(m-n)\Omega t}}{U- m \Omega} \Bigg],
\end{equation}
where $\mathcal{J}_{n}(x)$  is the Bessel function of the first kind. In the large frequency limit $\Omega \gg t_{dd}$, we obtain
\begin{equation}\label{eq.22}
J^{(2)}_{ij} =  \sum_{n = -\infty}^{\infty} \frac{4 t^2_{dd}\mathcal{J}^2_n(r_{dd}E_0/\Omega)}{U- n \Omega}.
\end{equation}
This result matches with those as shown in Ref.~\cite{Mentink2015}. We now move on to the analysis for the third-order SW transformation in the presence of spin-orbit coupling.

\subsection{Spin-orbit coupling: Second-order SWT}

In this section, we provide the details of the effective spin-exchange Hamitlonian as described in the main text for the single-orbital case. The diagonal and off-diagonal part of the Hamiltonian for second-order perturbation theory is identified as 
\begin{equation}\label{seq.1}
\mathcal{H}_0 = U \sum_{i}n_{i\uparrow} n_{i\downarrow} , \quad \mathcal{H}_1(t) = - \sum_{\langle ij \rangle,\sigma\sigma'} t^{ij}_{\sigma \sigma'}e^{i\phi_{ij}(t)} d^{\dagger}_{i\sigma}d_{j\sigma'},
\end{equation}
where $ t^{ij}_{\sigma \sigma'} =(t_{dd} \mathbb{I}_2 + i\bm{\alpha}_{ij}\cdot \bm{\tau})_{\sigma \sigma'}$. As explained in Eq.~(\ref{eq.14}), the effective Hamiltonian is written as
\begin{equation}\label{seq.2}
\mathcal{H}^{(2)}_{\mathrm{eff}}(t)  = \frac{i}{2} \int_{-\infty}^{t} dt' \left[ \mathcal{P} \mathcal{H}_1(t') \mathcal{Q} e^{i(\mathcal{H}_0+i\eta)(t-t')} \mathcal{Q} \mathcal{H}_1(t) \mathcal{P} - \mathcal{P} \mathcal{H}_1(t) \mathcal{Q} e^{-i(\mathcal{H}_0-i\eta)(t-t')} \mathcal{Q} \mathcal{H}_1(t') \mathcal{P} \right].
\end{equation}
The first term in Eq.~(\ref{seq.2}) can be simplified as 
\begin{align}\label{seq.3}
& \int_{-\infty}^{t} dt' e^{i(U+i\eta)(t-t')} \mathcal{P} \mathcal{H}_1(t') \mathcal{Q} \mathcal{Q} \mathcal{H}_1(t) \mathcal{P} \nonumber \\
& = \int_{-\infty}^{t} dt' e^{i(U+i\eta)(t-t')} \sum_{\langle ij \rangle, \sigma \sigma'} e^{i\phi_{ij}(t')} \left( t_{dd} \mathbb{I}_2 + i\bm{\alpha}_{ij} \cdot \bm{\tau}\right)_{\sigma''' \sigma''} d^{\dagger}_{i\sigma'''}d_{j\sigma''} e^{i\phi_{ji}(t)} \left( t_{dd} \mathbb{I}_2 -i \bm{\alpha}_{ij} \cdot \bm{\tau}\right)_{\sigma' \sigma} d^{\dagger}_{j\sigma'}d_{i\sigma} \nonumber \\
& = \int_{-\infty}^{t} dt' e^{i(U+i\eta)(t-t')} \sum_{\langle ij \rangle, \sigma,\sigma'} e^{i[\phi_{ij}(t')-\phi_{ij}(t)]} \underbrace{\left( t_{dd} \mathbb{I}_2 + i\bm{\alpha}_{ij} \cdot \bm{\tau}\right)_{\sigma''' \sigma''} d^{\dagger}_{i\sigma'''}d_{i\sigma} \left( t_{dd} \mathbb{I}_2 -i \bm{\alpha}_{ij} \cdot \bm{\tau}\right)_{\sigma' \sigma} d_{j\sigma''} d^{\dagger}_{j\sigma'}}_{\mathbf{M}}.
\end{align}
The operator product $\mathbf{M}$ can be simplified as
\begin{align}\label{seq.4}
\mathbf{M} & = \left( t_{dd} \mathbb{I}_2 + i\bm{\alpha}_{ij} \cdot \bm{\tau}\right)_{\sigma''' \sigma''} d^{\dagger}_{i\sigma'''}d_{i\sigma} \left( t_{dd} \mathbb{I}_2 -i \bm{\alpha}_{ij} \cdot \bm{\tau}\right)_{\sigma' \sigma} d_{j\sigma''} d^{\dagger}_{j\sigma'} \nonumber \\
& = \left( t_{dd} \mathbb{I}_2 + i\bm{\alpha}_{ij} \cdot \bm{\tau}\right)_{\sigma''' \sigma''} \left( \frac{n_i}{2} + \mathbf{S}_i \cdot \tau \right)_{\sigma \sigma'''} \left( t_{dd} \mathbb{I}_2 -i \bm{\alpha}_{ij} \cdot \bm{\tau}\right)_{\sigma' \sigma} \Bigg[ \mathbb{I}_2 - \left( \frac{n_j}{2} +  \mathbf{S}_j \cdot \tau \right) \Bigg]_{\sigma'' \sigma'} \nonumber \\
& = \frac{t_{dd}^2}{2} + \frac{\bm{\alpha}_{ij}^2}{2}  - \Big[ 2t_{dd}^2 \mathbf{S}_i \cdot \mathbf{S}_j + 4t_{dd}\bm{\alpha}_{ij} \cdot \mathbf{S}_i \times \mathbf{S}_j - 2 (\bm{\alpha}_{ij} \times \mathbf{S}_i). (\bm{\alpha}_{ij} \times \mathbf{S}_j) + 2 (\bm{\alpha}_{ij} \cdot \mathbf{S}_i)\cdot ( \bm{\alpha}_{ij} \cdot \mathbf{S}_j) \Big],
\end{align}
Ignoring the constant terms (which provides an energy shift to the overall Hamiltonian), Eq.~\ref{seq.4} is simplified as
\begin{equation}\label{seq.5}
\mathbf{M} \equiv  - 2t_{dd}^2 \mathbf{S}_i \cdot \mathbf{S}_j - 4t_{dd}  \bm{\alpha}_{ij}  \cdot \left( \mathbf{S}_i \times \mathbf{S}_j \right) + 2 (\bm{\alpha}_{ij} \times \mathbf{S}_i). (\bm{\alpha}_{ij} \times \mathbf{S}_j) - 2 (\bm{\alpha}_{ij} \cdot \mathbf{S}_i)\cdot ( \bm{\alpha}_{ij} \cdot \mathbf{S}_j). 
\end{equation}
This expression matches with the previously known theoretical work in Ref.~\onlinecite{Rubio2021}. Hence, we obtain the effective exchange Hamiltonian as 
\begin{equation}\label{seq.6}
\mathcal{H}^{(2)}_{\mathrm{eff}} = \sum_{ij} \bigg[ J^{(2)}_{ij} \mathbf{S}_i \cdot\mathbf{S}_j + \mathbf{D}^{(2)}_{ij} \cdot ( \mathbf{S}_i \times \mathbf{S}_j ) + S_{i\mu} \Gamma^{(2)}_{i\mu,j\nu} S_{j\nu}\bigg],
\end{equation}
where $\mu,\nu = x,y,z$. The exchange couplings are evaluated in the Floquet approximation as (where $A_0 = r_{dd}E_0/\Omega$)
\begin{subequations}
\begin{align}
\label{seq.7.1}
J^{(2)}_{ij} & = 4\sum_{n = -\infty}^{\infty} \mathcal{J}^2_n(\mathrm{A}_0) \frac{t^2_{dd}}{U- n \Omega}, \\
\label{seq.7.2}
\mathbf{D}^{(2)}_{ij} & = -8\sum_{n = -\infty}^{\infty} \mathcal{J}^2_n(\mathrm{A}_0) \frac{t_{dd} \bm{\alpha}_{ij}}{U- n \Omega}, \\
\label{seq.7.3}
\Gamma^{(2)}_{i\mu,j\nu} & = -\sum_{n = -\infty}^{\infty} \mathcal{J}^2_n(\mathrm{A}_0) \frac{ 8 \alpha_{ij,\mu} \alpha_{ij,\nu} + 4 \delta_{\mu\nu}\bm{\alpha}_{ij}\cdot \bm{\alpha}_{ij}}{U- n \Omega}.
\end{align}
\end{subequations}

\subsection{Spin-orbit coupling: Third-order SWT}

\begin{figure}[t]
\centering
\includegraphics[width=0.3\linewidth]{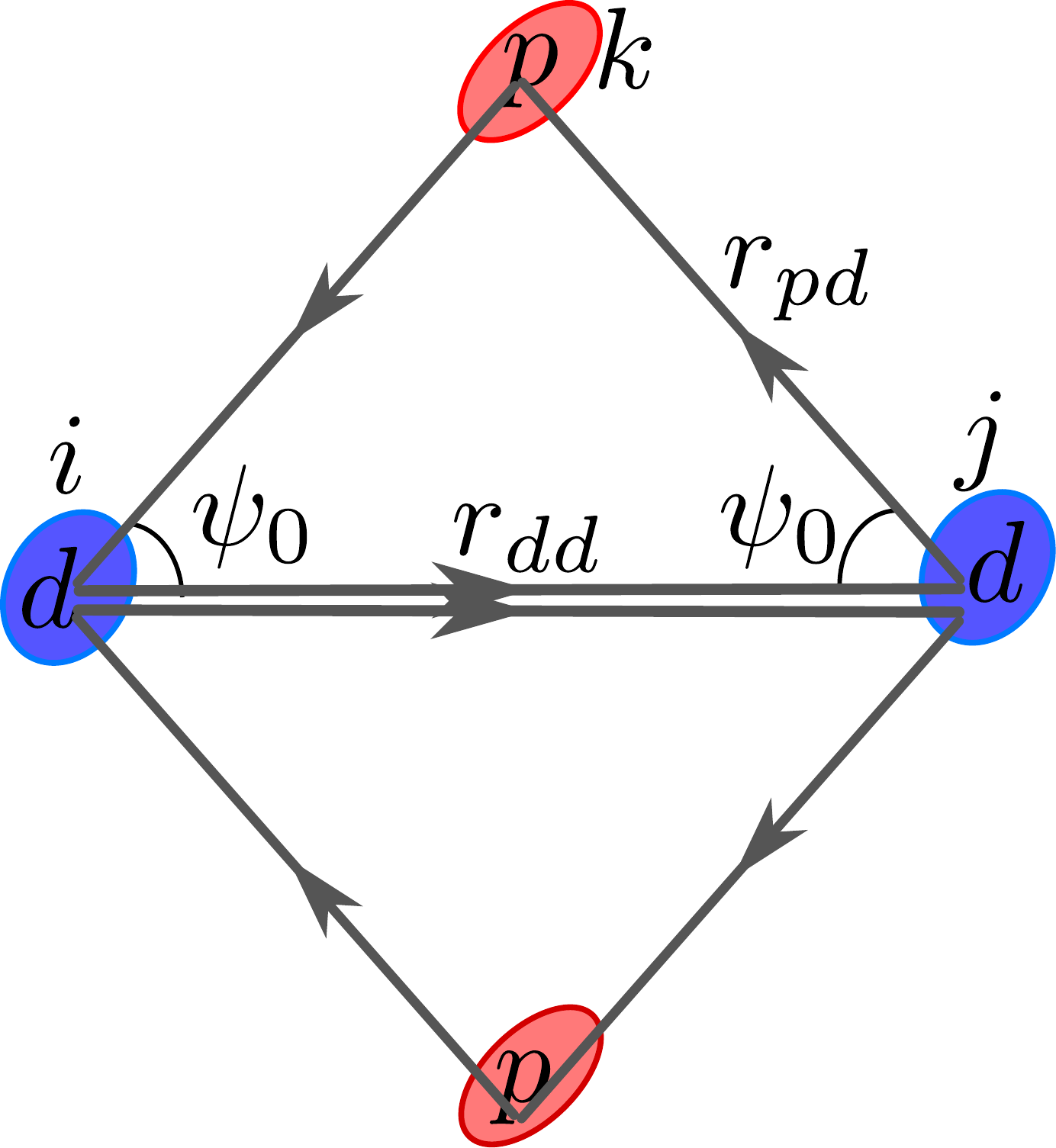} 
\caption{A schematic diagram of the hopping of electrons in the third-order Schrieffer-Wolff process. The angle between the transition metal-transition metal bond and the transition metal-ligand atom bond is labeled by $\psi_0$. The site indices are placed as $i,j,k$ on the respective sites.}
\label{fig:Fig1}
\end{figure} 

In this case, we identify the diagonal and non-diagonal parts of the Hamiltonian as 
\begin{equation}\label{eq.23}
\mathcal{H}_0 = U \sum_{i}n_{i\uparrow} n_{i\downarrow} + \Delta \sum_{i,\sigma} p^{\dagger}_{i\sigma}p_{i'\sigma}, \quad \mathcal{H}_1(t) = -t_{pd} \sum_{\langle ij \rangle, \sigma}e^{i\theta_{ij}(t)} d^{\dagger}_{i\sigma}p_{j\sigma} - \sum_{\langle ij \rangle,\sigma\sigma'} t^{ij}_{\sigma\sigma'} e^{i\phi_{ij}(t)} d^{\dagger}_{i\sigma}d_{j\sigma'},
\end{equation}
$t^{ij}_{\sigma \sigma'}$ is again written as  $(t_{dd} \mathbb{I}_2 + {i} \bm{\alpha}_{ij}\cdot \bm{\tau})_{\sigma \sigma'}$ where $\bm{\tau}$ is the vector of the Pauli matrices. From Eq.~(\ref{eq.16}) the low-energy effective Hamiltonian in third-order is written as
\begin{align}\label{eq.24}
\mathcal{H}^{(3)}_{\mathrm{eff}}(t)  & = \frac{i}{2} \mathcal{P}\big[ \vS^{(2)}(t),\mathcal{H}_1(t) \big]\mathcal{P} + \frac{1}{6} \mathcal{P}\big[ \vS^{(1)}(t),\big[ \vS^{(1)}(t), \mathcal{H}_1(t) \big] \big]\mathcal{P}  \nonumber \\
 = & \frac{i}{2} \left( \mathcal{P} \vS^{(2)}(t) \mathcal{H}_1(t) \mathcal{P} - \mathcal{P}\mathcal{H}_1(t) \vS^{(2)}(t)\mathcal{P} \right) + \frac{1}{6} \mathcal{P} \left( \vS^{(1)}(t) \big[ \vS^{(1)}(t), \mathcal{H}_1(t) \big] - \big[ \vS^{(1)}(t), \mathcal{H}_1(t) \big] \vS^{(1)}(t) \right) \mathcal{P} \nonumber \\
 = & - \frac{1}{2} \int_{-\infty}^{t} dt' \big[ \mathcal{P}\mathcal{H}_1(t) \mathcal{Q}_i e^{-i(\mathcal{H}_0-i\eta)(t-t')}\mathcal{Q}_i  \mathcal{H}_1(t') \mathcal{Q}_j \mathcal{Q}_j \vS^{(1)}(t') \mathcal{P} + \mathcal{P}  \vS^{(1)}(t') \mathcal{Q}_j \mathcal{Q}_j \mathcal{H}_1(t') \mathcal{Q}_ie^{i(\mathcal{H}_0+i\eta)(t-t')} \mathcal{Q}_i \mathcal{H}_1(t) \mathcal{P} \big] \nonumber \\
 & - \frac{1}{6} \mathcal{P} \vS^{(1)}(t) \mathcal{Q}_i \mathcal{Q}_i \mathcal{H}_1(t) \mathcal{Q}_j \mathcal{Q}_j \vS^{(1)}(t)\mathcal{P} - \frac{1}{6} \mathcal{P} \vS^{(1)}(t) \mathcal{Q}_j \mathcal{Q}_j \mathcal{H}_1(t) \mathcal{Q}_i \mathcal{Q}_i \vS^{(1)}(t)\mathcal{P}, \quad [i\neq j \wedge i,j =1,2]
\end{align}
where we have introduced two high-energy projection operators $\mathcal{Q}_1$ and $\mathcal{Q}_2$ corresponding to one electron at the ligand site with energy $\Delta$, and the double occupancy at the $d$-orbital sites with energy $U$, respectively. To derive Eq.~(\ref{eq.24}), we assumed $\mathcal{Q}_1 \vS^{(1)}(t) \mathcal{Q}_2 = \mathcal{Q}_2 \vS^{(1)}(t) \mathcal{Q}_1 = 0$ using the gauge freedom of the generating function $\vS^{(1)}(t)$. Considering the geometry as shown in Fig.~\ref{fig:Fig1}, we evaluate the effective Hamiltonian in Eq.~(\ref{eq.24}). The first term in Eq.~(\ref{eq.24}) can be simplified as
\begin{align}\label{eq.25}
& \int_{-\infty}^{t} dt' \mathcal{P}\mathcal{H}_1(t) \mathcal{Q}_i e^{-i(\mathcal{H}_0-i\eta)(t-t')}\mathcal{Q}_i  \mathcal{H}_1(t') \mathcal{Q}_j \mathcal{Q}_j \vS^{(1)}(t') \mathcal{P} \nonumber \\
& = \int_{-\infty}^{t} dt'\int_{-\infty}^{t'} dt'' e^{-i(U-i\eta)(t-t')} e^{-i(\Delta-i\eta)(t'-t'')} \mathcal{P}\mathcal{H}_1(t) \mathcal{Q}_2 \mathcal{Q}_2\mathcal{H}_1(t')\mathcal{Q}_1\mathcal{Q}_1 \mathcal{H}_1(t'') \mathcal{P} \nonumber \\
& = t^2_{pd}\sum_{\langle ij \rangle} \int_{-\infty}^{t} dt'\int_{-\infty}^{t'} dt'' e^{-i(U-i\eta)(t-t')} e^{-i(\Delta-i\eta)(t'-t'')} e^{i[\phi_{ij}(t) + \theta_{jk}(t') + \theta_{ki}(t'')]} \left( t_{dd} \mathbb{I}_2 + i\bm{\alpha}_{ij} \cdot \bm{\tau} \right)_{\sigma''' \sigma''} d^{\dagger}_{i\sigma'''}d_{j\sigma''} d^{\dagger}_{j\sigma'}\underbrace{p_{k\sigma'}p^{\dagger}_{k\sigma}}_{\delta_{\sigma \sigma'}}d_{i\sigma} \nonumber \\
& = \sum_{\langle ij \rangle}\sum_{n,m,l = -\infty}^{\infty} \mathcal{J}_n(A_0) \mathcal{J}_m(A) \mathcal{J}_l(A)  \frac{e^{i(n+m+l)\Omega t}e^{i(m-l)\psi_0}t_{pd}^2}{\big[\Delta + l \Omega\big] \big[U + (l+m)\Omega\big]} \left[i\bm{\alpha}_{ij}\cdot (\mathbf{S}_i - \mathbf{S}_j) -2 t_{dd} \mathbf{S}_i \cdot \mathbf{S}_j - 2\bm{\alpha}_{ij} \cdot (\mathbf{S}_i \times \mathbf{S}_j) \right],
\end{align}
where $A_0 = r_{dd}E_0/\Omega$ and $A = r_{pd} E_0/\Omega$. Here, we have rewritten the Peierl's phases in terms of the Bessel functions. Adding the other three terms in Eq.~(\ref{eq.24}) and taking large frequency average, we obtain the the Eqs.~(5a)-(5c) in the main text. Note that the averaging over time for large $\Omega$ introduces the constraint $n+m+l = 0$. Evaluating the rest of the terms in Eq.~(\ref{eq.24}) the final spin-exchange Hamiltonian is obtained as (after performing the high-frequency time-average)
\begin{equation}\label{eq.26}
\mathcal{H}_{\mathrm{eff}}^{(3)} = \sum_{\langle ij \rangle} J_{ij}^{(3)} \mathbf{S}_i \cdot \mathbf{S}_j + \sum_{\langle ij \rangle} \mathbf{h}^{\mathrm{eff}}_{ij} \cdot (\mathbf{S}_i - \mathbf{S}_j) + \sum_{\langle ij \rangle}\mathbf{D}^{(3)}_{ij} \cdot (\mathbf{S}_i \times \mathbf{S}_j),
\end{equation}
where $J_{ij}^{(3)}$ and $\mathbf{D}_{ij}^{(3)}$ are the Heisenberg and Dzyaloshinskii-Moriya interactions, respectively and are given by
\begin{subequations}
\begin{align}
\label{seq.8.1}
J^{(3)}_{ij} & = \sum_{n,m,l = -\infty}^{\infty}\frac{8 \mathcal{J}_{n}(\mathrm{A_0}) \mathcal{J}_{m}(\mathrm{A}) \mathcal{J}_{l}(\mathrm{A}) t^2_{pd} t_{dd}}{3}  \frac{\cos \psi^{ml}_0}{\big[\Delta + l \Omega\big] \big[U - n\Omega\big]}, \\
\label{seq.8.2}
\mathbf{D}^{(3)}_{ij} & = \sum_{n,m,l = -\infty}^{\infty}\frac{8 \mathcal{J}_{n}(\mathrm{A_0}) \mathcal{J}_{m}(\mathrm{A}) \mathcal{J}_{l}(\mathrm{A}) t^2_{pd} \bm{\alpha}_{ij}}{3}  \frac{\cos \psi^{ml}_0}{\big[\Delta + l \Omega\big] \big[U - n\Omega\big]},
\end{align}
\end{subequations}
where $\psi^{ml}_0 = (m-l) \psi_0$ and $\psi_0$ is the angle between the TM-TM bond and TM-ligand atom bond (TM = transition metal atom). Since, $\mathbf{h}^{\mathrm{eff}}_{ij}$ is proportional to $\sin (m-l)\psi_0$ [see main text in Eq.~(5)], the  emergent effective magnetic field vanishes for a linearly polarized light. 

Finally, we evaluate the spin-exchange couplings $\Gamma_{\mu,\nu}$ in Eq.~(3) of the main text. The $3 \times 3$ matrix can be written as
\begin{equation}\label{meq.1}
\Gamma_{\mu,\nu} = \begin{pmatrix}
\Gamma_{xx} & \Gamma_{xy} & \Gamma_{xz} \\
\Gamma_{yx} & \Gamma_{yy} & \Gamma_{yz} \\
\Gamma_{zx} & \Gamma_{zy} & \Gamma_{zz}
\end{pmatrix},
\end{equation}
where the diagonal (symmetric) components $\Gamma_{xx}, \Gamma_{yy}, \Gamma_{zz}$ are given by 
\begin{equation}\label{meq.2}
\Gamma_{xx} = \Gamma_{yy} = \Gamma_{zz} = 4\sum_{n = -\infty}^{\infty} \mathcal{J}^2_n(\mathrm{A}_0) \frac{t^2_{dd}}{U- n \Omega} + \sum_{n,m,l = -\infty}^{\infty}\frac{8 \mathcal{J}_{n}(\mathrm{A_0}) \mathcal{J}_{m}(\mathrm{A}) \mathcal{J}_{l}(\mathrm{A}) t^2_{pd} t_{dd}}{3}  \frac{\cos \psi^{ml}_0}{\big[\Delta + l \Omega\big] \big[U - n\Omega\big]}.
\end{equation}
The off-diagonal components $\Gamma_{\mu,\nu}$ ($\mu \neq \nu$) can be written as sum of a symmetric and an anti-symmetric part as $\Gamma_{\mu,\nu} = \Gamma^{\mathrm{sym}}_{\mu,\nu} + \Gamma^{\mathrm{asym}}_{\mu,\nu}$. The symmetric part is given by
\begin{equation}\label{meq.3}
\Gamma^{\mathrm{sym}}_{\mu,\nu} = -\sum_{n = -\infty}^{\infty} \mathcal{J}^2_n(\mathrm{A}_0) \frac{ 8 \alpha_{ij,\mu} \alpha_{ij,\nu} + 4 \delta_{\mu\nu}\bm{\alpha}_{ij}\cdot \bm{\alpha_{ij}} }{U- n \Omega},
\end{equation}
whereas the anti-symmetric part corresponds to the Dzyaloshinskii-Moriya interaction as 
\begin{equation}\label{meq.4}
\Gamma^{\mathrm{asym}}_{\mu,\nu} = \bigg[-8\sum_{n = -\infty}^{\infty} \mathcal{J}^2_n(\mathrm{A}_0) \frac{t_{dd}}{U- n \Omega} + \sum_{n,m,l = -\infty}^{\infty}\frac{8 \mathcal{J}_{n}(\mathrm{A_0}) \mathcal{J}_{m}(\mathrm{A}) \mathcal{J}_{l}(\mathrm{A}) t^2_{pd}}{3}  \frac{\cos \psi^{ml}_0}{\big[\Delta + l \Omega\big] \big[U - n\Omega\big]}\bigg] \epsilon_{\kappa\mu\nu}\alpha^{\kappa}_{ij},
\end{equation}
where $ \epsilon_{\kappa\mu\nu}$ ($\kappa,\mu,\nu = x,y,z$) is the Levi-Civita symbol and $\alpha^{\kappa}_{ij}$ is the $\kappa$-th component of the SOC vector $\bm{\alpha}_{ij}$.

\section{Multi-orbital Kanamori model: Schrieffer-Wolff transformation}

The mutli-orbital Hamiltonian [see Eq.~(6) and Eq.~(8) in the main text] is written in terms of the irreducible representation of the doubly occupied states in the $d$-orbitals. In terms of three $d$-orbitals: $d_{xy}$, $d_{yz}$ and $d_{zx}$ the Kanamori Hamiltonian is 
\begin{align}\label{eq.27}
\mathcal{H}_0  = & U\sum_{i\alpha} n_{i\alpha,\uparrow}n_{i\alpha,\downarrow} + \sum_{i\sigma \sigma' \alpha \neq \beta} \left( U' -\delta_{\sigma \sigma'} J_{\mathrm{H}} \right) n_{i\alpha \sigma'}n_{i\beta \sigma} + J_{\mathrm{H}} \sum_{i\alpha \neq \beta} \left( d^{\dagger}_{i\alpha \uparrow} d^{\dagger}_{i\alpha \downarrow} d_{i\beta \downarrow} d_{i\beta \uparrow} + d^{\dagger}_{i\alpha \uparrow} d^{\dagger}_{i\beta \downarrow} d_{i\alpha \downarrow} d_{i\beta \uparrow} \right) \nonumber \\
& + \frac{\lambda}{2} \sum_{i} d^{\dagger}_i \left( \mathbf{L} \cdot \mathbf{S} \right) d_{i} + \Delta \sum_{i\sigma} n^{p}_{i\sigma}.
\end{align}
Eq.~(\ref{eq.27}) is now written in the irreducible representation doubly occupied states as $\mathcal{H}_0 = \sum_{i} \sum_{\Gamma} \sum_{\mathsf{g}_{\Gamma}} U_{\Gamma} \ket{i;\Gamma, \mathsf{g}_{\Gamma}} \bra{i;\Gamma, \mathsf{g}_{\Gamma}}$, where the fifteen states (two particles are to placed in six possible states corresponding to three orbitals and two spins i.e. $^6C_2$ combinations) are given as follows~\cite{PhysRevB.94.174416,PhysRevB.65.064442,PhysRevB.103.L100408}
\begin{subequations}
\begin{align}
\label{eq.28.1}
& \ket{i;A_1} = \frac{1}{\sqrt{3}} (d^{\dagger}_{ixz \uparrow}d^{\dagger}_{ixz \downarrow} + d^{\dagger}_{iyz \uparrow}d^{\dagger}_{iyz \downarrow} + d^{\dagger}_{ixy \uparrow}d^{\dagger}_{ixy \downarrow}) \ket{0}, \\ 
\label{eq.28.2}
& \ket{i;E,u}  = \frac{1}{\sqrt{6}} (d^{\dagger}_{iyz \uparrow}d^{\dagger}_{iyz \downarrow} + d^{\dagger}_{ixz \uparrow}d^{\dagger}_{ixz \downarrow} -2 d^{\dagger}_{ixy \uparrow}d^{\dagger}_{ixy \downarrow}) \ket{0}, \\
\label{eq.28.3}
& \ket{i;E,v}  = \frac{1}{\sqrt{2}} (d^{\dagger}_{iyz \uparrow}d^{\dagger}_{iyz \downarrow} - d^{\dagger}_{ixz \uparrow}d^{\dagger}_{ixz \downarrow}) \ket{0}, \\
\label{eq.28.4}
& \ket{i;T_1,\alpha_+} = d^{\dagger}_{iyz \uparrow} d^{\dagger}_{i zx \uparrow} \ket{0}, \\ 
\label{eq.28.5}
& \ket{i;T_1,\alpha_-} = d^{\dagger}_{iyz \downarrow} d^{\dagger}_{i zx \downarrow} \ket{0}, \\ 
\label{eq.28.6}
& \ket{i;T_1,\alpha} = \frac{1}{\sqrt{2}} ( d^{\dagger}_{iyz \uparrow} d^{\dagger}_{i zx \downarrow} + d^{\dagger}_{iyz \downarrow} d^{\dagger}_{i zx \uparrow} ) \ket{0}, \\ 
\label{eq.28.7}
& \ket{i;T_1,\beta_+} = d^{\dagger}_{izx \uparrow} d^{\dagger}_{i xy \uparrow} \ket{0}, \\ 
\label{eq.28.8}
& \ket{i;T_1,\beta_-} = d^{\dagger}_{izx \downarrow} d^{\dagger}_{i xy \downarrow} \ket{0}, \\ 
\label{eq.28.9}
& \ket{i;T_1,\beta} = \frac{1}{\sqrt{2}} ( d^{\dagger}_{izx \uparrow} d^{\dagger}_{i xy \downarrow} + d^{\dagger}_{izx \downarrow} d^{\dagger}_{i xy \uparrow} ) \ket{0}, \\
\label{eq.28.10}
& \ket{i;T_1,\gamma_+} = d^{\dagger}_{ixy \uparrow} d^{\dagger}_{i yz \uparrow} \ket{0}, \\ 
\label{eq.28.11}
& \ket{i;T_1,\gamma_-} = d^{\dagger}_{ixy \downarrow} d^{\dagger}_{i yz \downarrow} \ket{0}, \\ 
\label{eq.28.12}
& \ket{i;T_1,\gamma} = \frac{1}{\sqrt{2}} ( d^{\dagger}_{ixy \uparrow} d^{\dagger}_{i yz \downarrow} + d^{\dagger}_{ixy \downarrow} d^{\dagger}_{i yz \uparrow} ) \ket{0}, \\
\label{eq.28.13}
& \ket{i;T_2,\alpha} = \frac{1}{\sqrt{2}} ( d^{\dagger}_{iyz \uparrow} d^{\dagger}_{i zx \downarrow} - d^{\dagger}_{iyz \downarrow} d^{\dagger}_{i zx \uparrow} ) \ket{0}, \\ 
\label{eq.28.14}
& \ket{i;T_2,\beta} = \frac{1}{\sqrt{2}} ( d^{\dagger}_{izx \uparrow} d^{\dagger}_{i xy \downarrow} - d^{\dagger}_{izx \downarrow} d^{\dagger}_{i xy \uparrow} ) \ket{0}, \\
\label{eq.28.15}
& \ket{i;T_2,\gamma} = \frac{1}{\sqrt{2}} ( d^{\dagger}_{ixy \uparrow} d^{\dagger}_{i yz \downarrow} - d^{\dagger}_{ixy \downarrow} d^{\dagger}_{i yz \uparrow} ) \ket{0}. 
\end{align}
\end{subequations}
where we have ignored the SOC term assuming $U, \Delta \gg \lambda$. In the similar manner to the single-orbital case, we defined the high- and low-energy projection operators as $\mathcal{Q}$ and $\mathcal{P}$, respectively. Assuming the $j_{\mathrm{eff}} = 1/2$ basis, we identify the following four-states spanning the low-energy Hilbert space as~\cite{PhysRevB.94.174416,PhysRevB.65.064442,PhysRevB.103.L100408}
\begin{subequations}
\begin{align}
\label{eq.29.1}
\ket{i,+;j,+} & = \frac{1}{3} \left( id^{\dagger}_{ixz\downarrow} + d^{\dagger}_{iyz\downarrow} + d^{\dagger}_{ixy\uparrow} \right) \left( id^{\dagger}_{jxz\downarrow} + d^{\dagger}_{jyz\downarrow} + d^{\dagger}_{jxy\uparrow} \right) \ket{0}, \\
\label{eq.29.2}
\ket{i,+;j,-} & = \frac{1}{3} \left( id^{\dagger}_{ixz\downarrow} + d^{\dagger}_{iyz\downarrow} + d^{\dagger}_{ixy\uparrow} \right) \left( id^{\dagger}_{jxz\uparrow} - d^{\dagger}_{jyz\uparrow} + d^{\dagger}_{jxy\downarrow} \right)\ket{0},  \\
\label{eq.29.3}
\ket{i,-;j,+} & = \frac{1}{3} \left( id^{\dagger}_{ixz\uparrow} - d^{\dagger}_{iyz\uparrow} + d^{\dagger}_{ixy\downarrow} \right) \left( id^{\dagger}_{jxz\downarrow} + d^{\dagger}_{jyz\downarrow} + d^{\dagger}_{jxy\uparrow} \right) \ket{0},  \\
\label{eq.29.4}
\ket{i,-;j,-} & = \frac{1}{3} \left( id^{\dagger}_{ixz\uparrow} + d^{\dagger}_{iyz\uparrow} + d^{\dagger}_{ixy\downarrow} \right) \left( id^{\dagger}_{jxz\uparrow} + d^{\dagger}_{jyz\uparrow} + d^{\dagger}_{jxy\downarrow} \right) \ket{0},
\end{align}
\end{subequations}
where the states $\ket{i,\pm}$ are defined as~\cite{PhysRevB.94.174416,PhysRevB.65.064442,PhysRevB.103.L100408}
\begin{subequations}
\begin{align}
\label{eq.30.1}
\ket{i,+} & = \frac{1}{\sqrt{3}} \left( id^{\dagger}_{ixz\downarrow} + d^{\dagger}_{iyz\downarrow} + d^{\dagger}_{ixy\uparrow}\right) \ket{0}, \\
\label{eq.30.2}
\ket{i,-} & = \frac{1}{\sqrt{3}} \left( id^{\dagger}_{ixz\uparrow} - d^{\dagger}_{iyz\uparrow} + d^{\dagger}_{ixy\downarrow}\right) \ket{0}.
\end{align}
\end{subequations}
In defining the states in Eqs.~(\ref{eq.29.1})-(\ref{eq.29.4}) we again constrained our system to the four-site cluster with two $d$-orbitals and two $p$-orbitals, respectively [see Fig.~\ref{fig:Fig1}]. The high-energy subspace of projection operator $\mathcal{Q}$ are spanned by the fifteen states as defined in Eqs.~(\ref{eq.28.1})-(\ref{eq.28.15}). We now write the hopping Hamiltonian in the multi-orbital setup as
\begin{equation}\label{eq.31}
\mathcal{H}_1(t) = -\sum_{ij,\sigma} e^{i\phi_{ij}(t)}\begin{pmatrix}
d^{\dagger}_{i xz \sigma} & d^{\dagger}_{i {yz} \sigma} & d^{\dagger}_{i {xy} \sigma} 
\end{pmatrix} \begin{pmatrix}
t_1 & t_2 & t_4 \\
t_2 & t_1 & t_4 \\
t_4 & t_4 & t_3
\end{pmatrix}
\begin{pmatrix}
d_{j {xz} \sigma} \\
d_{j {yz} \sigma} \\
d_{j {xy} \sigma}
\end{pmatrix} = -\sum_{ij \alpha \beta \sigma} e^{i \phi_{ij}(t)} t_{ij\alpha \beta}d^{\dagger}_{i\alpha \sigma} d_{j \beta \sigma},
\end{equation}
where $t_1$, $t_2$, $t_3$ and $t_4$ are the hopping amplitudes between the respective orbitals. For subsequent analysis, we use the compact notation as $-\sum_{ij \alpha \beta \sigma} e^{i \phi_{ij}(t)} t_{ij\alpha \beta}d^{\dagger}_{i\alpha \sigma} d_{j \beta \sigma}$. Finally, the hopping Hamiltonian between the ligand atom and the $d$-orbital is written as
\begin{equation}\label{eq.32}
\mathcal{H}_2(t) = -t_{pd} \sum_{ij \sigma} \left( e^{i\theta_{i'j}(t)}p^{\dagger}_{i'\sigma}d_{jyz \sigma} +  e^{i\theta_{j'i}(t)} p^{\dagger}_{j'\sigma}d_{iyz \sigma} + e^{i\theta_{ji'}(t)} d^{\dagger}_{jxz \sigma} p_{i'\sigma} + e^{i\theta_{jj'}(t)} d^{\dagger}_{jxz \sigma} p_{j'\sigma} \right) + \mathrm{h.c.}
\end{equation}
The Peierls phases are denied as in the main text. Note that the orientation of the various $d$-orbitals ($d_{xy}$, $d_{yz}$ and $d_{zx}$) and the ligand $p$-orbital ($p_x$ and $p_y$) restricts the hopping between only $d_{yz}$ and $d_{zx}$ orbitals and $d_{xy}$-orbital does not take part in the Hamiltonian $\mathcal{H}_2(t)$. The total Hamiltonian is written as
\begin{equation}\label{eq.33}
\mathcal{H}(t) = \mathcal{H}_0 + \mathcal{H}_1(t) + \mathcal{H}_2(t),
\end{equation}
where $\mathcal{H}_0$, $\mathcal{H}_1(t)$ and $\mathcal{H}_2(t)$ are defined in Eq.~(\ref{eq.27}), Eq.~(\ref{eq.31}) and Eq.~(\ref{eq.32}), respectively.

\subsection{Second-order Schrieffer-Wolff transformation}

In this section, we provide the details of the derivation of the low-energy effective spin-exchange Hamiltonian using SW transformation upto second-order in perturbation. The effective Hamiltonian is written from in Eq.~(\ref{eq.14}) as
\begin{align}\label{eq.34}
\mathcal{H}^{(2)}_{\mathrm{eff}} & = \frac{i}{2} \mathcal{P}\big[ \vS^{(1)}(t), \mathcal{H}_1(t) \big]\mathcal{P} \nonumber \\
& = \frac{i}{2} \left( \underbrace{\mathcal{P} \vS^{(1)}(t) \mathcal{Q} \mathcal{Q} \mathcal{H}_1(t) \mathcal{P}}_{\mathcal{A}_1} - \underbrace{\mathcal{P} \mathcal{H}_1(t) \mathcal{Q} \mathcal{Q} \vS^{(1)}(t) \mathcal{P}}_{\mathcal{A}_2} \right),
\end{align}
where we only consider the process through the Hamiltonian $\mathcal{H}_1(t)$, as the ligand mediated hopping does not contribute in the second-order process. We start by analyzing the first term $\mathbf{A}_1$ as follows 
\begin{align}\label{eq.35}
\mathcal{A}_1 & = \mathcal{P} S^{(1)}(t) \mathcal{Q} \mathcal{Q} \mathcal{H}_1(t) \mathcal{P} \nonumber \\
& = \int_{-\infty}^{t} dt' \mathcal{P} \mathcal{H}_1(t') \mathcal{Q} e^{i(\mathcal{H}_0 + i \eta)(t-t')} \mathcal{Q}\mathcal{H}_1(t) \mathcal{P}, \, [\text{see Eq.~(\ref{eq.13.1})}] \nonumber \\
& = \sum_{\{\Gamma \}}\int_{-\infty}^{t} dt' e^{i(U_{\Gamma}+ i \eta)(t-t')} \mathcal{P} \mathcal{H}_1(t') \mathcal{Q} \mathcal{Q} \mathcal{H}_1(t) \mathcal{P} \nonumber \\
& = \sum_{\{\Gamma \}} \int_{-\infty}^{t}dt' e^{i\phi_{ij}(t') - i \phi_{ij}(t) + i(U_{\Gamma}+ i\eta)(t-t')} \mathcal{P} t_{ij\alpha \beta} d^{\dagger}_{i\alpha \sigma} d_{j\beta \sigma} \ket{i;\Gamma,\mathsf{g}_{\Gamma}}\bra{i;\Gamma,\mathsf{g}_{\Gamma}} t_{ji\gamma \delta} d^{\dagger}_{j\gamma \sigma'}d_{i\delta \sigma'} \mathcal{P} \nonumber \\ 
& = \sum_{\{\Gamma \}} \sum_{\mathcal{F},\mathcal{I}} \int_{-\infty}^{t}dt' e^{i\phi_{ij}(t') - i \phi_{ij}(t) + i(U_{\Gamma}+ i\eta)(t-t')} \ket{\mathcal{F}} \bra{\mathcal{F}} t_{ij\alpha \beta} d^{\dagger}_{i\alpha \sigma} d_{j\beta \sigma} \ket{i;\Gamma,\mathsf{g}_{\Gamma}}\bra{i;\Gamma,\mathsf{g}_{\Gamma}} t_{ji\gamma \delta} d^{\dagger}_{j\gamma \sigma'}d_{i\delta \sigma'} \ket{\mathcal{I}}\bra{\mathcal{I}} \nonumber \\
& = \sum_{\{\Gamma \}} \sum_{\mathcal{F},\mathcal{I}} \int_{-\infty}^{t}dt' e^{i\phi_{ij}(t') - i \phi_{ij}(t) + i(U_{\Gamma}+ i\eta)(t-t')}  \bra{\mathcal{F}} t_{ij\alpha \beta}d^{\dagger}_{i\alpha \sigma} d_{j\beta \sigma} \ket{i;\Gamma,\mathsf{g}_{\Gamma}} \bra{i;\Gamma,\mathsf{g}_{\Gamma}}  t_{ji\gamma \delta}d^{\dagger}_{j\gamma \sigma'}d_{i\delta \sigma'} \ket{\mathcal{I}}  \ket{\mathcal{F}} \bra{\mathcal{I}},
\end{align}
where $\{\Gamma \}$ corresponds to the summation over all the fifteen high-energy states [see Eqs.~(\ref{eq.28.1})-(\ref{eq.28.15})], and $\ket{\mathcal{F},\mathcal{I}}$ corresponds to one of low-energy states as written in Eqs.~(\ref{eq.29.1})-(\ref{eq.29.4}). The operators $\ket{\mathcal{F}}\bra{\mathcal{I}}$ can be further written in terms of the spin operators as
\begin{subequations}
\begin{align}
\label{eq.36.1}
\ket{i,+;j,+} \bra{i,+;j,+} & = \left( \frac{1}{2} + S_i^z \right) \left( \frac{1}{2} + S_j^z \right), \\
\label{eq.36.2}
\ket{i,-;j,-} \bra{i,-;j,-} & = \left( \frac{1}{2} - S_i^z \right) \left( \frac{1}{2} - S_j^z \right), \\
\label{eq.36.3}
\ket{i,+;j,-} \bra{i,+;j,-} & = \left( \frac{1}{2} + S_i^z \right) \left( \frac{1}{2} - S_j^z \right), \\
\label{eq.36.4}
\ket{i,+j,-} \bra{i,-;j,+} & = S_i^+ S_j^-, \\
\label{eq.36.5}
\ket{i,-,j,+} \bra{i,+;j,-} & = S_i^- S_j^+, \quad \ldots\ldots 
\end{align}
\end{subequations}
The generic spin exchange Hamiltonian can be written in terms of the spin operators as
\begin{equation}\label{eq.37}
\mathcal{H}_{\mathrm{eff}} = \sum_{ij} \begin{pmatrix}
S_i^x & S_i^y & S_i^z
\end{pmatrix}
\begin{pmatrix}
J & \Gamma & \Gamma' \\
\Gamma & J & \Gamma' \\
\Gamma' & \Gamma' & J + K_{\gamma}
\end{pmatrix}
\begin{pmatrix}
S_j^x \\
S_j^y \\
S_j^z
\end{pmatrix}.
\end{equation}
Now we can easily identify the spin-exchange couplings in terms of the parameters of the original microscopic multi-orbital Hamiltonian in Eq.~(\ref{eq.33}). To do so, we need to evaluate the matrix elements $\bra{\mathcal{F}} t_{ij\alpha \beta}d^{\dagger}_{i\alpha \sigma} d_{j\beta \sigma} \ket{i;\Gamma,g_{\Gamma}}$. We use the mathematical package DiracQ~\cite{Shastry2013} to evaluate these matrix elements as follows 
\begin{subequations}
\begin{align}
\label{eq.38.1}
&\Braket{i;A_1|t_{ij\alpha \beta}d^{\dagger}_{i\alpha \sigma} d_{j\beta \sigma}|+,+} = \Braket{i;A_1|t_{ij\alpha \beta}d^{\dagger}_{i\alpha \sigma} d_{j\beta \sigma}|-,-} = 0, \\
\label{eq.38.2}
&\Braket{i;A_1|t_{ij\alpha \beta}d^{\dagger}_{i\alpha \sigma} d_{j\beta \sigma}|+,-} = -\Braket{i;A_1|t_{ij\alpha \beta}d^{\dagger}_{i\alpha \sigma} d_{j\beta \sigma}|-,+} = \frac{2t_1+t_3}{3\sqrt{3}}, \\
\label{eq.38.3}
&\Braket{i;E,u|t_{ij\alpha \beta}d^{\dagger}_{i\alpha \sigma} d_{j\beta \sigma}|+,+} = -\frac{1+i}{\sqrt{6}}t_4, \Braket{i;E,u|t_{ij\alpha \beta}d^{\dagger}_{i\alpha \sigma} d_{j\beta \sigma}|-,-} = -\frac{1-i}{\sqrt{6}}t_4, \\
\label{eq.38.4}
&\Braket{i;E,u|t_{ij\alpha \beta}d^{\dagger}_{i\alpha \sigma} d_{j\beta \sigma}|+,-} = -\Braket{i;E,u|t_{ij\alpha \beta}d^{\dagger}_{i\alpha \sigma} d_{j\beta \sigma}|-,+} = -\frac{1}{3}\sqrt{\frac{2}{3}}(t_1-t_3), \\
\label{eq.38.5}
&\Braket{i;E,v|t_{ij\alpha \beta}d^{\dagger}_{i\alpha \sigma} d_{j\beta \sigma}|+,+} = -\frac{1}{3\sqrt{2}}(1-i)t_4, \Braket{i;E,v|t_{ij\alpha \beta}d^{\dagger}_{i\alpha \sigma} d_{j\beta \sigma}|-,-} = -\frac{1}{3\sqrt{2}}(1+i)t_4, \\
\label{eq.38.6}
&\Braket{i;E,v|t_{ij\alpha \beta}d^{\dagger}_{i\alpha \sigma} d_{j\beta \sigma}|+,-} = \Braket{i;E,v|t_{ij\alpha \beta}d^{\dagger}_{i\alpha \sigma} d_{j\beta \sigma}|-,+} = -\frac{\sqrt{2}}{3} i t_2, \quad  \ldots \ldots
\end{align}
\end{subequations}
where we have written a few of the all possible combinations. Following the similar procedure of rewriting the Peierls phases in terms of the Bessel functions, the Hamiltonian in Eq.~(\ref{eq.34}) can be simplified as (in the Floquet regime)
\begin{equation}\label{eq.39}
\mathcal{H}^{(2)}_{\mathrm{eff}} = - \sum_{n=-\infty}^{\infty} \sum_{i,\Gamma,\mathsf{g}_{\Gamma}} \sum_{\mathcal{F},\mathcal{I}}  \frac{\mathcal{J}_n(\mathrm{A}_0)^2\Braket{\mathcal{F}|\mathcal{H}^{(1)}_{ij}|i,\Gamma,\mathsf{g}_{\Gamma}} \Braket{i,\Gamma,\mathsf{g}_{\Gamma}|\mathcal{H}^{(1)}_{ji}|\mathcal{I}}}{U_{\Gamma} - n\Omega} \ket{\mathcal{F}} \bra{\mathcal{I}},
\end{equation}
where we used a short-hand notation for $t_{ij\alpha \beta} \equiv \mathcal{H}^{(1)}_{ij}$, and $t_{ji\gamma \delta} \equiv \mathcal{H}^{(1)}_{ji}$. The time-dependent part has already been integrated out. Plugging back the matrix elements, the effective couplings, up to the second order, are evaluated as
\begin{subequations}
\begin{align}
\label{eq.39.1}
J^{(2)}_{ij} & = \sum_{n=-\infty}^{\infty} \frac{4}{27} \mathcal{J}^2_{n}(\mathrm{A}_0)\bigg[  \frac{(2t_1+t_3)^2}{U+2J_{\mathrm{H}}-n\Omega} + \frac{2(t_1-t_3)^2 + 9t_4^2}{U-J_{\mathrm{H}}-n\Omega} + \frac{6t_1(t_1+2t_3) - 9 t^2_4}{U-3J_{\mathrm{H}}-n\Omega} \bigg], \\
\label{eq.39.2}
K^{(2)}_{\gamma,ij} & = \frac{8}{9}\sum_{n=-\infty}^{\infty} \mathcal{J}^2_{n}(\mathrm{A}_0)  \frac{J_{\mathrm{H}}[(t_1-t_3)^2 -3t_2^2+ 3t_4^2]}{(U-3J_{\mathrm{H}}-n\Omega)(U-J_{\mathrm{H}}-n\Omega)}, \\
\label{eq.39.3}
\Gamma^{(2)}_{ij} & = \frac{8}{9}\sum_{n=-\infty}^{\infty} \mathcal{J}^2_{n}(\mathrm{A}_0)  \frac{J_{\mathrm{H}}[2t_2(t_1-t_3) + 3t_4^2]}{(U-3J_{\mathrm{H}}-n\Omega)(U-J_{\mathrm{H}}-n\Omega)}, \\
\label{eq.39.4}
\Gamma'^{(2)}_{ij} & = -\frac{8}{9}\sum_{n=-\infty}^{\infty} \mathcal{J}^2_{n}(\mathrm{A}_0) \frac{J_{\mathrm{H}}[t_4(t_1-t_3-3t_2)]}{(U-3J_{\mathrm{H}}-n\Omega)(U-J_{\mathrm{H}}-n\Omega)}.
\end{align}
\end{subequations}
The static limit of the above results matches exactly with Refs.~\cite{PhysRevB.103.L100408,PhysRevLett.112.077204}. In the second order process, the ligand induced hoppings do not contribute and hence, we also do not have any induced Zeeman magnetic field. The inverse Faraday effect is absent in this case.

\subsection{Third-order Schrieffer-Wolff transformation}

In this section, we finally analyze the details of the derivation of the effective emergent magnetic field $\mathbf{h}^{\mathrm{eff}}$ [Eq.~(9) in the main text]. The third-order effective Hamiltonian is written from Eq.~(\ref{eq.16}) as
\begin{equation}\label{eq.40}
\mathcal{H}^{(3)}_{\mathrm{eff}}(t) = \frac{i}{2} \underbrace{\big[ \vS^{(2)}(t),\mathcal{H}_1(t) + \mathcal{H}_2(t)\big]}_{\mathcal{T}_1} + \frac{1}{6} \underbrace{\big[ \vS^{(1)}(t),\big[ \vS^{(1)}(t), \mathcal{H}_1(t) + \mathcal{H}_2(t) \big] \big]}_{\mathcal{T}_2},
\end{equation}
where each of the two terms $\mathcal{T}_1$ and $\mathcal{T}_2$ are comprised of two expressions as evaluated in Eq.~(\ref{eq.24}). The first term $\mathcal{T}_1$ can be simplified as
\begin{align}\label{eq.41}
\mathcal{T}_1 = & -\frac{1}{2}\sum_{\Gamma, \mathcal{F}, \mathcal{I}} \int_{-\infty}^t dt' \int_{-\infty}^{t'} dt''\bra{\mathcal{F}} \mathcal{H}_1(t)\ket{i;\Gamma,\mathsf{g}_{\Gamma}}\bra{i;\Gamma,\mathsf{g}_{\Gamma}} e^{-i(\mathcal{H}_0 - i \eta)(t-t')}\ket{i;\Gamma,\mathsf{g}_{\Gamma}}\bra{i;\Gamma,\mathsf{g}_{\Gamma}} \mathcal{H}_2(t') \ket{\mathcal{L}} \nonumber  \\
& \bra{\mathcal{L}} e^{-i(\mathcal{H}_0 -i\eta)(t'-t'')} \ket{\mathcal{L}}\bra{\mathcal{L}} \mathcal{H}_2(t'')\ket{\mathcal{I}} \ket{\mathcal{F}} \bra{\mathcal{I}}  + \mathrm{c.c},
\end{align}
where $\mathrm{c.c}$  stands for the complex conjugate and $\mathcal{L}$ corresponds to the ligand states. The respective Hamiltonians $\mathcal{H}_1(t)$ and $\mathcal{H}_2(t)$ are defined before. Again introducing the Bessel functions for
the three different Peierl’s phases and taking the high-frequency average we obtain the Floquet regime expression for $\mathcal{T}_1$ as 
\begin{align}\label{eq.42}
\overline{\mathbf{T}}_1 =  & \overline{\sum} \mathcal{J}_n(\mathrm{A}_0) \mathcal{J}_m(\mathrm{A}) \mathcal{J}_l(\mathrm{A})t_{pd}^2 \mathrm{Re} \big[ \frac{e^{-i(m-l)\psi_0}}{[\Delta + l \Omega][U_{\Gamma} -n\Omega]} \bra{\mathcal{F}} d^{\dagger}_{j\delta \sigma'}d_{i \gamma \sigma'}\ket{i;\Gamma,g_{\Gamma}}  \bra{i;\Gamma,g_{\Gamma}} t_{ij\alpha \beta} d^{\dagger}_{i\alpha\sigma} d_{j\beta \sigma} \ket{\mathcal{I}}\big] \ket{\mathcal{F}} \bra{\mathcal{I}},
\end{align} 
where $\overline{\sum}$ corresponds to the summation over the fifteen high-energy states and also over the indices $n,m,l$ with the constraint $n+m+l = 0$. The latter condition emerges due to the high-frequency average. We again evaluate the matrix elements $\bra{\mathcal{F}} d^{\dagger}_{j\delta \sigma'}d_{i \gamma \sigma'}\ket{i;\Gamma,g_{\Gamma}}$ using the DiracQ Mathematica package~\cite{Shastry2013}. A few of the matrix elements are shown here for example as 
\begin{subequations}
\begin{align}
\label{eq.43.1}
&\Braket{i;A_1|d^{\dagger}_{j\delta \sigma'}d_{i \gamma \sigma'}|+,+} = \Braket{i;A_1|d^{\dagger}_{j\delta \sigma'}d_{i \gamma \sigma'}|-,-} = 0, \\
\label{eq.43.2}
&\Braket{i;A_1|d^{\dagger}_{j\delta \sigma'}d_{i \gamma \sigma'}|+,-} = \Braket{i;A_1|d^{\dagger}_{j\delta \sigma'}d_{i \gamma \sigma'}|-,+} = \frac{i}{3\sqrt{3}}, \\
\label{eq.43.3}
&\Braket{i;E,u|d^{\dagger}_{j\delta \sigma'}d_{i \gamma \sigma'}|+,+} = \Braket{i;E,u|d^{\dagger}_{j\delta \sigma'}d_{i \gamma \sigma'}|-,-} = 0, \quad \ldots \ldots.
\end{align}
\end{subequations}
Utilizing the above matrix elements we evaluate the analytical structure of the effective Hamiltonian as 
\begin{equation}\label{dq.44}
\mathcal{H}_{\mathrm{eff}}^{(3)} = \sum_{\langle ij \rangle} \left[ J^{(3)}_{ij} (\mathbf{S}_i \cdot \mathbf{S}_j + K^{(3)}_{ij} S_i^zS_j^z + \Gamma^{(3)}_{ij} (S_i^xS_j^y + S_i^yS_j^x) \right] + \mathbf{h}^{\mathrm{eff}}_{ij} \cdot \left( \mathbf{S}_i + \mathbf{S}_j \right),
\end{equation}
where the emergent magnetic field $\mathbf{h}^{\mathrm{eff}}$ is a result of the incident circularly polarized light within the triangular plaquet formed by the two transition metal and a ligand atom. The exchange couplings $J^{(3)}_{ij}$, $K^{(3)}_{\gamma,ij}$, $\Gamma^{(3)}_{ij}$, evaluated using third-order SWT and perturbation results, are provided in our recent work~\cite{Kumar2021}. Here, we focus on the effective emergent magnetic field $h^{\mathrm{eff}}_{ij}$. However, for completeness, we also provide the expressions for the other couplings as follows
\begin{subequations}
\begin{align}
\label{eq.137.1}
J^{(3)}_{ij} & =  \sum_{\{n\}} \frac{16 \mathcal{J}_n(\mathrm{A}_0) \mathcal{J}_m(\mathrm{A}) \mathcal{J}_l(\mathrm{A})t^2_{pd}}{81} \Bigg[ \frac{\sin \psi^{ml}_0}{\Delta + l\Omega} \left( \frac{2t_1+t_3}{U + 2 J_{\mathrm{H}}-n\Omega} +  \frac{t_1-t_3}{U-J_{\mathrm{H}}-n\Omega} +  \frac{3t_1+3t_3}{U-3J_{\mathrm{H}}-n\Omega} \right) \Bigg],  \\
\label{eq.137.2}
K^{(3)}_{ij} & = \sum_{\{n\}} \frac{16 \mathcal{J}_n(\mathrm{A}_0) \mathcal{J}_m(\mathrm{A}) \mathcal{J}_l(\mathrm{A})t^2_{pd}}{27} \bigg[ \frac{J_{\mathrm{H}}}{(\Delta + l \Omega)} \frac{(t_1-t_3) \sin \psi^{ml}_0 - 3t_2 \cos \psi^{ml}_0 }{(U-3J_{\mathrm{H}}-n\Omega)(U-J_{\mathrm{H}}-n\Omega)}\bigg],  \\
\label{eq.137.3}
\Gamma^{(3)}_{ij} & = \sum_{\{n\}} \frac{16 \mathcal{J}_n(\mathrm{A}_0) \mathcal{J}_m(\mathrm{A}) \mathcal{J}_l(\mathrm{A})t^2_{pd}}{27} \bigg[\frac{J_{\mathrm{H}}}{(\Delta + l \Omega)} \frac{ (t_1 -t_3) \cos \psi^{ml}_0 + t_2 \sin \psi^{ml}_0}{(U-3J_{\mathrm{H}}-n\Omega)(U-J_{\mathrm{H}}-n\Omega)} \bigg], \\
\label{eq.137.4}
h^{\mathrm{eff}}_{ij} & = \sum_{\{n\}} \frac{8 \mathcal{J}_n(\mathrm{A}_0) \mathcal{J}_m(\mathrm{A}) \mathcal{J}_l(\mathrm{A})t^2_{pd}}{27} \frac{\sin \psi_0^{ml}}{\Delta + l \Omega} \bigg[ \frac{t_1 - t_3}{U-3J_{\mathrm{H}}-n\Omega} +  \frac{t_1 - t_3}{U-J_{\mathrm{H}}-n\Omega} \bigg],
\end{align}
\end{subequations}
where $\sum_{\{n\}}$ corresponds to $\sum_{n,m,l = -\infty}^{\infty}$ with the constraint $n+m+l = 0$. Note that the super-exchange coupling $J^{(3)}_{ij}$ vanishes in the absence of the applied circularly polarized light. This result is consistent with the previous theoretical work~\cite{PhysRevLett.102.017205}.



\section{Oblique incidence of the circularly polarized light}

\begin{figure}[t]
\centering
\includegraphics[width=0.3\linewidth]{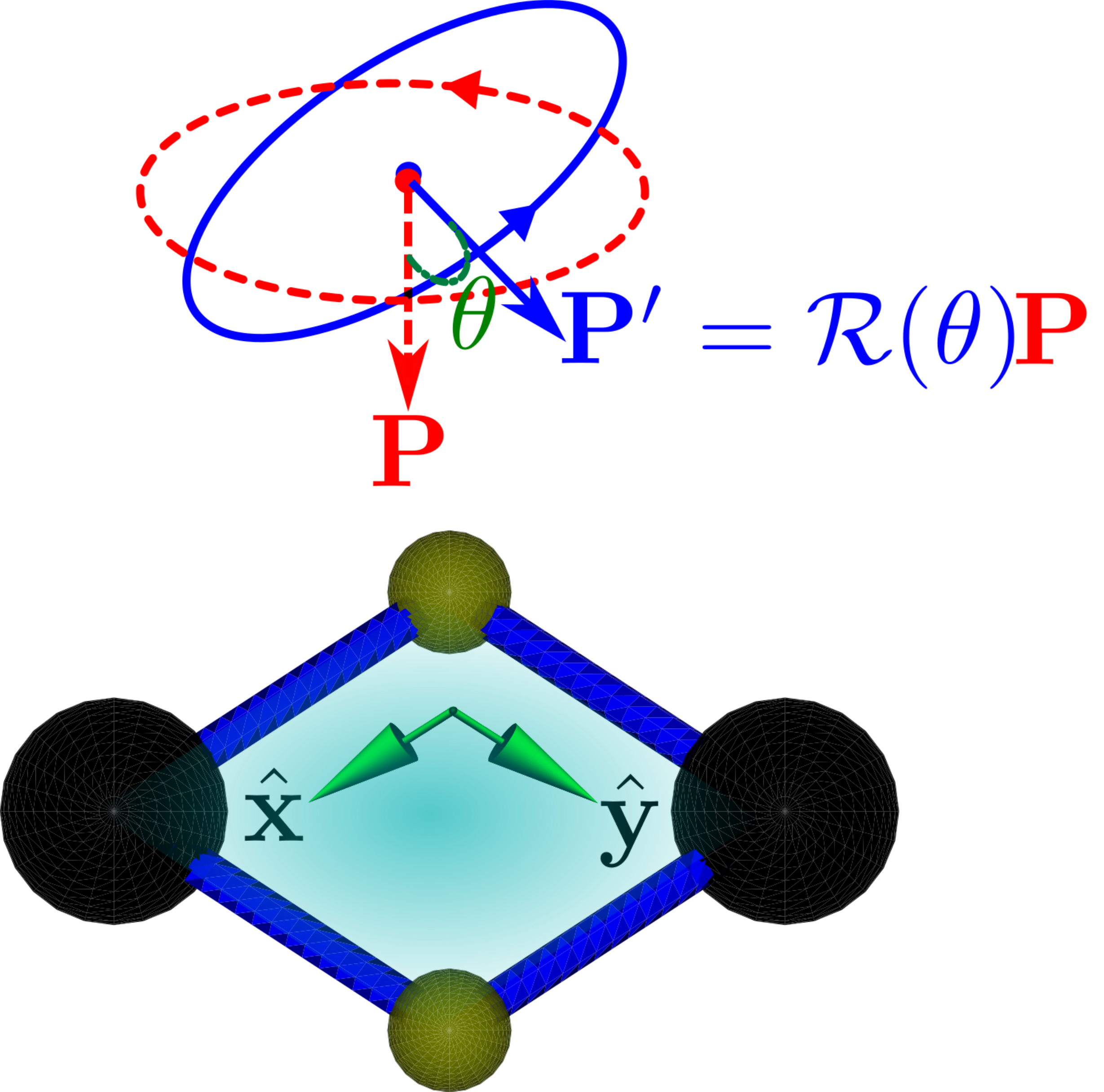} 
\caption{The four-site cluster oriented along the $xy$-plane is irradiated with a circularly polarized light whose polarization axis $\mathbf{P}$ is tilted by an angle $\theta$ along the $x$-axis. The polarization axis -  red-dashed schematic - for the applied laser (previously considered scenario) is oriented along $z$-axis, whereas the tiled polarization of the applied laser (solid blue schematic) is obtained by rotating the red dashed polarization by the rotation matrix $\mathcal{R}(\theta)$.}
\label{fig:Fig2}
\end{figure} 

Previously, we considered a circularly polarized light whose polarization axis $\mathbf{P}$ is perpendicular to the plane of the four-site cluster. However, in real materials, the laser can be shined at an angle to the plane of the four-site cluster (see Fig.~\ref{fig:Fig2}). We adopt a geometry where the polarization axis of the laser is rotated with respect to the $x$-axis by an angle $\theta$. If the gauge potential $\mathbf{A}(t)$ for the perpendicular incidence is given by $\mathbf{A}(t) = E_0/\Omega (\hat{\mathbf{i}} \sin\Omega t  + \hat{\mathbf{j}} \cos\Omega t)$ then we can use the local coordinate system to rotate the vector $\mathbf{A}(t)$ by a rotation matrix $\mathcal{R}(\theta)$ as 
\begin{align}\label{eq.139}
\mathbf{A}'(t) & = \mathcal{R}(\theta) \mathbf{A}(t) \nonumber \\
& = \begin{pmatrix}
1 & 0 & 0 \\
0 & \cos \theta & -\sin \theta \\
0 & \sin \theta & \cos \theta
\end{pmatrix} \begin{pmatrix}
\sin \Omega t\\
\cos \Omega t \\
0
\end{pmatrix} \nonumber \\
& = \begin{pmatrix}
\sin \Omega t \\
\cos \theta \cos \Omega t \\
\sin \theta \cos \Omega t
\end{pmatrix},
\end{align}
consequently, the rotated gauge potential can be thought as an elliptical polarization compared to the previously incident circularly polarized light as $\mathbf{A}'(t) = E_0/\Omega (\hat{\mathbf{i}} \sin\Omega t  + \hat{\mathbf{j}} \cos \theta \cos\Omega t+ \hat{\mathbf{k}} \sin \theta \cos \Omega t)$. With this modification, the results for third-order exchange couplings and the emergent magnetic magnetic field for both the single- and multi-orbital analysis is modified as following: in Eqs.~(\ref{seq.8.1})-(\ref{seq.8.2}) and Eq. (\ref{eq.137.4}) the argument of the Bessel functions and the angular dependence are changed as following
\begin{subequations}
\begin{align}
\label{eq.140.1}
\mathcal{J}_{m,l}(\mathrm{A}) \rightarrow \mathcal{J}_{m,l}(\mathrm{A'}); \quad \mathrm{A}' = r \mathrm{A}, \quad r = \sqrt{\cos^2 \psi_0 + \sin^2 \psi_0 \cos^2 \theta}, \\
\label{eq.140.2}
\sin \psi_0^{ml} \rightarrow \sin \beta_0^{ml}; \quad \beta_0^{ml} = (m-l) \beta_0, \quad  \tan \beta_0 = \tan \psi_0 \cos \theta,
\end{align}
\end{subequations}
since both the argument of the Bessel functions and the angular variations now depends on the angle of the oblique incidence of the applied laser, the latter can be utilized to tune the relative strengths and the orientations of the exchange couplings and the magnetic fields.

\section{Deriving effective Hamiltonian using Exact Diagonalization}

We consider a two-site problem along with a ligand. The Hamiltonian is consequently written as 
\begin{subequations}
\begin{align}
\label{eq.141.1}
\mathcal{H}(t) & = \mathcal{H}_0 + \mathcal{H}_1(t) + \mathcal{H}_2(t), \\
\label{eq.141.2}
\mathcal{H}_0 & = U n_{1\uparrow}n_{1\downarrow}+ U n_{2\uparrow}n_{2\downarrow} + \Delta \left( p^{\dagger}_{\uparrow}p_{\uparrow} + p^{\dagger}_{\downarrow}p_{\downarrow} \right),  \\
\label{eq.141.3} 
\mathcal{H}_1(t) & = -e^{i \phi_{12}(t)} t^{12}_{\sigma \sigma'} d_{1\sigma}^\dagger d_{2\sigma'}  + \mathrm{h.c.}, \\
\label{eq.141.4}
\mathcal{H}_2(t) & = -t_{pd}  \big( e^{i \theta_{1p} (t)} d_{1\sigma}^\dagger p_{\sigma}^{\phantom\dagger}+  e^{i \theta_{2p} (t)} d_{2\sigma}^\dagger p_{\sigma}^{\phantom\dagger}\big)+ \text{h.c.}, 
\end{align}
\end{subequations}
where the hopping elements in Eq.~(\ref{eq.141.3}) are written in a matrix form as 
\begin{subequations}
\begin{align}
\label{eq.142}
t^{12}_{\sigma \sigma'} & = \big[ t_{dd} \mathbb{I}_2 + i \bm{\alpha}_{12} \cdot \bm{\tau} \big]_{\sigma \sigma'} = \begin{pmatrix}
t_{dd} + i \alpha^3_{12} & i \alpha^1_{12} + \alpha^2_{12}\\
i \alpha^1_{12} - \alpha^2_{12} &  t_{dd} - i \alpha^3_{12} \\
\end{pmatrix},
\end{align}
\end{subequations}
where we assumed the spin-orbit coupling vector $\bm{\alpha}_{12} = (\alpha^1_{12}, \alpha^2_{12},\alpha^3_{12})$. Note that the above hopping term breaks the inversion symmetry while being Hermitian. The Hamiltonian in the basis set \{$|s_1, s_2, s_{\mathcal{L}} \rangle$, $|D, 0, 0\rangle$ \} without the time-dependence can be written as (here $s_1$, $s_2$ and $s_{\mathcal{L}}$ are the spins at sites 1,2 and ligand, respectively and $D$ is a double occupancy on a site) 
\begin{equation}\label{eq.143}
\mathcal{H} = \begin{pmatrix}
0 & 0 & 0 & 0 & -i \alpha^1_{12} - \alpha^2_{12}  &  t_{pd} &  0 &  0 &  0 \\
0 & 0 & 0 & 0 &  t_{dd} + i\alpha^2_{12}    &  0  &  t_{pd} &  0 &  0 \\
0 & 0 & 0 & 0 &  -t_{dd} + i \alpha^3_{12}     &  0 &  0 &  t_{pd} &  0 \\
0 & 0 & 0 & 0 &  i \alpha^1_{12} - \alpha^2_{12}  &  0  &  0 &  0 &  t_{pd}  \\
i \alpha^1_{12} - \alpha^2_{12}  &  t_{dd} - i \alpha^2_{12} &  -t_{dd} - i \alpha^3_{12} & -i \alpha^1_{12} - \alpha^2_{12}  &U  &0  &  t_{pd} &  -t_{pd} &  0 \\
t_{pd} &  0&  0  & 0 &0 & \Delta   &  0 &  0 &  0 \\
0 &  t_{pd} &  0  & 0 &t_{pd} & 0   &  \Delta &  0 &  0 \\
0 &  0&  t_{pd}  & 0 &- t_{pd} & 0   &  0 &  \Delta &  0 \\
0 &  0&  0  & t_{pd} &0 & 0   &  0 &  0 &  \Delta 
\end{pmatrix} \begin{pmatrix}
|\uparrow ,\uparrow, 0 \rangle \\
|\uparrow ,\downarrow, 0\rangle \\ 
|\downarrow, \uparrow, 0\rangle\\
|\downarrow, \downarrow, 0\rangle\\
|\uparrow \downarrow,0, 0\rangle  \\
|\uparrow, 0 ,  \uparrow \rangle  \\
|\uparrow, 0 ,  \downarrow \rangle \\
|\downarrow, 0 ,  \uparrow \rangle \\
|\downarrow, 0 , \downarrow \rangle 
\end{pmatrix}.
\end{equation} 
The time-dependence is restricted to the hopping part, $\mathcal{H}_1(t) + \mathcal{H}_2(t)$. The time-independent matrix in Bloch structure form is given by~\cite{Eckardt_2015, Mentink2015}
\begin{equation}\label{eq.floquet}
(\epsilon_\alpha+l\omega)  \ket{\psi_{\alpha, m}} = \sum_{m'} H_{m-m'} \ket{\psi_{\alpha, m'}}.
\end{equation}
Here $\epsilon_\alpha$ is the quasi-energy, $H_m = \int_{0}^{T}dt~e^{- im\omega t} H(t)$, and $ \alpha \in \{ \ket{s_1, s_2, s_{\mathcal{L}}},~\ket{D, 0, 0} \}$  and $m$ is index for the $m^\text{th}$ Floquet sector. Since, we are interested in evaluating parameters for the spin Hamiltonian, we restrict the basis to $m=0$ sector and singly occupied basis using projector; $P_{s, 0}= P_{s} P_{m=0} $, where, $P_{s}  \in \ket{s_1, s_2, 0} \bra{s_1, s_2, 0}$. We make use of the eigen-decomposed Hamiltonian to evaluate the effective spin Hamiltonian $\mathcal{H}_\text{eff}$ and can be evaluated using the relation
\begin{equation}\label{eq.144}
\mathcal{H}^{\mathrm{ed}}_\text{eff} =  P_{s', 0} \sum_{n} E_n \ket{\psi_n} \bra{\psi_n} P_{s, 0},
\end{equation} 
where $E_n$ and $|\psi_{n}\rangle$ are the eigen-energy and eigenvector of Eq.~\eqref{eq.floquet}. We know that the circularly polarized light breaks the time-reversal symmetry and the generic effective spin Hamiltonian can be written in the form
\begin{equation}\label{eq.145}
\mathcal{H}_\text{eff} = \begin{bmatrix} S_1^x&  	S_1^y &   	S_1^z \end{bmatrix}
\begin{bmatrix}
J+\Gamma_\text{xx} & \Gamma_\text{yz} +D^z & \Gamma_\text{xz}-D^y  \\
\Gamma_\text{yz} -D^z  & J+\Gamma_\text{yy} & \Gamma_\text{yz}+D^x  \\
\Gamma_\text{zx}+D^y  & \Gamma_\text{zy}-D^x & J+  \Gamma_\text{zz}
\end{bmatrix} 
\begin{bmatrix} S_2^x\\ S_2^y\\ S_2^z \end{bmatrix}
+ \textbf{h}\cdot (\textbf{S}_1 - \textbf{S}_2). 
\end{equation}
Further the Hamiltonian can be written as
\begin{equation}\label{eq.146}
\scalebox{0.85}[1]{$\mathcal{H}_{\mathrm{eff}}  = \Psi^{\mathrm{T}}_1  \begin{pmatrix}
	\frac{1}{4} \left(\Gamma _{\text{xx}}-\Gamma _{\text{yy}}\right)  - \frac{i}{2}  \Gamma _{\text{xy}}&\frac{1}{2} \left(J+i D_z\right)+ \frac{1}{4} \left(\Gamma _{\text{xx}}+ \Gamma _{\text{yy}}\right) & \frac{1}{2} \left(\Gamma _{\text{zx}} -i \Gamma _{\text{yz}}\right)- \frac{1}{2} \left( D_y+iD_x \right)\\
\frac{1}{2} \left(J-i D_z\right)+ \frac{1}{4} \left(\Gamma _{\text{xx}}+ \Gamma _{\text{yy}}\right) & 	\frac{1}{4} \left(\Gamma _{\text{xx}}-\Gamma _{\text{yy}}\right)  + \frac{i}{2}  \Gamma _{\text{xy}}& \frac{1}{2} \left(\Gamma _{\text{zx}} + i \Gamma _{\text{yz}}\right) -\frac{1}{2} \left( D_y-iD_x \right)\\
\frac{1}{2} \left(\Gamma _{\text{zx}} -i \Gamma _{\text{yz}}\right)+ \frac{1}{2} \left( D_y+iD_x \right) & \frac{1}{2} \left(\Gamma _{\text{zx}} + i \Gamma _{\text{yz}}\right) +\frac{1}{2} \left( D_y-iD_x \right) & J+\Gamma _{\text{zz}} 
\end{pmatrix} 
\Psi_2 + \mathbf{h} \cdot (\mathbf{S}_1 - \mathbf{S}_2),$}
\end{equation}
where the spinor $\Psi_i$ is defined as $\Psi_i = \left( S^+_i, S^-_i, S^z_i\right)^{\mathrm{T}}$. In the last term, $\textbf{h} = (h_x, h_y, h_z)$ depends on the plaquette orientation. For the plaquette in the $xy-$plane, we have $h_x = h_y =0 $ and the last term simplifies to $h_z(S_1^z - S_2^z)$. One can evaluate the various exchange parameters as written in Eq.~\eqref{eq.146} by comparing it with Eq.~(\ref{eq.144}). For example, one can make use of the relations
\begin{subequations}
\begin{align}
\label{eq.147.1}
\Braket{\uparrow, \downarrow, 0 | \mathcal{H}^{\mathrm{ed}}_{\mathrm{eff}} | \uparrow, \downarrow,0}  = -\frac{1}{4}(J+\Gamma_{zz}) + h_z, \\ 
\label{eq.147.2}
\Braket{\downarrow, \uparrow, 0 | \mathcal{H}^{\mathrm{ed}}_{\mathrm{eff}} | \downarrow, \uparrow,0} = -\frac{1}{4}(J+\Gamma_{zz}) - h_z,
\end{align}
\end{subequations}
to evaluate the $h_z$ for the situation discussed above. Similar exercise is carried out for the multi-orbital Hubbard model for evaluating the various spin exchange parameters. 

\bibliographystyle{apsrev4-1}
\bibliography{MS_qsl}